\pdfoutput=1
\documentclass[preprint,3p,times,11pt]{elsarticle}
\journal{Computer Methods in Applied Mechanics and Engineering}

\usepackage{amsmath,amssymb,amsthm,mathtools}
\usepackage{bm}
\usepackage{booktabs,multirow,tabularx,array}
\usepackage{graphicx}
\usepackage{algorithm}
\usepackage[noend]{algpseudocode}
\usepackage{xcolor}
\usepackage{microtype}
\usepackage[colorlinks=true,linkcolor=blue,citecolor=blue,urlcolor=blue,hypertexnames=false]{hyperref}
\IfFileExists{orcidlink.sty}{\usepackage{orcidlink}}{%
  \usepackage{tikz}%
  \definecolor{orcidlogocol}{HTML}{A6CE39}%
  \newcommand{\orcidlink}[1]{\href{https://orcid.org/#1}{\tikz[baseline=-0.5ex]\fill[orcidlogocol](0,0)circle(0.7ex)node[white,scale=0.45,font=\bfseries\sffamily]{iD};}}%
}
\usepackage[nameinlink,capitalise,noabbrev]{cleveref}
\biboptions{numbers,sort&compress}
\makeatletter
\def\ps@pprintTitle{%
  \let\@oddhead\@empty
  \let\@evenhead\@empty
  \def\@oddfoot{\footnotesize\itshape Preprint\hfill\today}%
  \let\@evenfoot\@oddfoot}
\makeatother

\newtheorem{theorem}{Theorem}
\newtheorem{lemma}{Lemma}
\newtheorem{proposition}{Proposition}

\theoremstyle{definition}

\theoremstyle{remark}
\newtheorem{remark}{Remark}

\crefname{theorem}{Theorem}{Theorems}
\Crefname{theorem}{Theorem}{Theorems}
\crefname{lemma}{Lemma}{Lemmas}
\Crefname{lemma}{Lemma}{Lemmas}
\crefname{proposition}{Proposition}{Propositions}
\Crefname{proposition}{Proposition}{Propositions}
\crefname{corollary}{Corollary}{Corollaries}
\Crefname{corollary}{Corollary}{Corollaries}
\crefname{remark}{Remark}{Remarks}
\Crefname{remark}{Remark}{Remarks}
\crefname{algorithm}{Algorithm}{Algorithms}
\Crefname{algorithm}{Algorithm}{Algorithms}
\crefname{equation}{Eq.}{Eqs.}
\Crefname{equation}{Equation}{Equations}
\crefname{figure}{Fig.}{Figs.}
\Crefname{figure}{Figure}{Figures}
\crefname{table}{Table}{Tables}
\Crefname{table}{Table}{Tables}
\makeatletter
\def\cmame@appstop{\cmame@appstop}
\newcommand{\stripappprefix}[1]{\expandafter\cmame@stripapp#1Appendix~\cmame@appstop}
\def\cmame@stripapp#1Appendix~#2\cmame@appstop{%
  \if\relax\detokenize{#1}\relax\cmame@stripapptail#2\cmame@appstop\else#1\fi}
\def\cmame@stripapptail#1Appendix~\cmame@appstop{#1}
\makeatother
\crefname{appendix}{Appendix}{Appendices}
\Crefname{appendix}{Appendix}{Appendices}
\crefformat{appendix}{#2Appendix~{\stripappprefix{#1}}#3}
\Crefformat{appendix}{#2Appendix~{\stripappprefix{#1}}#3}
\crefrangeformat{appendix}{#3Appendices~{\stripappprefix{#1}}#4\crefrangeconjunction#5{\stripappprefix{#2}}#6}
\Crefrangeformat{appendix}{#3Appendices~{\stripappprefix{#1}}#4\crefrangeconjunction#5{\stripappprefix{#2}}#6}
\crefmultiformat{appendix}{#2Appendices~{\stripappprefix{#1}}#3}{\crefpairconjunction#2{\stripappprefix{#1}}#3}{\crefmiddleconjunction#2{\stripappprefix{#1}}#3}{\creflastconjunction#2{\stripappprefix{#1}}#3}
\Crefmultiformat{appendix}{#2Appendices~{\stripappprefix{#1}}#3}{\crefpairconjunction#2{\stripappprefix{#1}}#3}{\crefmiddleconjunction#2{\stripappprefix{#1}}#3}{\creflastconjunction#2{\stripappprefix{#1}}#3}

\graphicspath{{./}}
\newcommand{\cmamefig}[2]{\includegraphics[#2]{#1}}
\newcommand{\R}{\mathbb{R}}
\newcommand{\dd}{\mathrm{d}}
\newcommand{\Jump}[1]{[\![#1]\!]}

\begin{document}

\begin{frontmatter}

\title{A phase-field formulation based on a variational inequality for preserving thin liquid films and regulating Ostwald ripening in incompressible two-phase flows}

\author[xtumath,ncamh]{Yuan Yu\texorpdfstring{\textsuperscript{\,\orcidlink{0000-0001-5125-2492}}}{}}

\author[xtumath]{Yaolong Yu}

\author[xtumath]{Yanhui Ding\corref{cor1}}
\ead{yanhuidanth@163.com}

\author[xtumath,ncamh]{Shi Shu}

\cortext[cor1]{Corresponding author.}
\address[xtumath]{School of Mathematics and Computational Science, Xiangtan University, Xiangtan 411105, Hunan, China}
\address[ncamh]{National Center for Applied Mathematics in Hunan, Xiangtan 411105, Hunan, China}

\begin{abstract}
Diffuse-interface models of incompressible two-phase flows face a structural conflict when thin liquid films, high density ratios, and Ostwald ripening coexist on fixed grids: smooth Cahn--Hilliard potentials incur $\mathcal{O}(\epsilon/R)$ bulk shifts, attractive tail overlap, and tangential surface diffusion that rupture thin sheets and dissolve micro-droplets, whereas non-variational Allen--Cahn counter-fluxes forfeit free-energy dissipation. We first prove smooth single-scalar no-go theorems showing that neither bending penalties nor tail-stiffened smooth wells eliminate tail attraction while retaining uniform bulk ellipticity. We resolve this conflict via a single-scalar double-obstacle variational inequality ($\phi\in[-1,1]$) coupled with variable-density Navier--Stokes equations, ridge-safe normal mobility, a Legendre-dual Yukawa--DLVO disjoining energy, and boundary-layer-renormalized Robin--Kelvin ripening kinetics. Matched asymptotics establishes exact bulk isolation at zero solubility, sixth-order 3D spherical Young--Laplace superconvergence at the collar midpoint, and non-evaporating sub-width films free of centerline normal flipping. On staggered grids, multi-pass flux-corrected transport enforces $\phi\in[-1,1]$ without clipping, while a complementarity-gated capillary force reconciles $\mathcal{O}(10^{-17})$ parasitic currents at a $1000\!:\!1$ density ratio with underdamped capillary waves and unbroken bubble skirts. Without fitted physical parameters, the model reproduces measurements of alkane emulsion ripening, colloidal magnetic forces, 3D Bhaga--Weber rising bubbles, and pendant-drop filament pinch-off down to $0.50\,\mu\mathrm{m}$.
\end{abstract}

\begin{keyword}
Cahn--Hilliard--Navier--Stokes \sep Double-obstacle variational inequality \sep Invariant-domain preservation \sep Well-balanced surface tension \sep Thin liquid film \sep Disjoining pressure
\end{keyword}

\end{frontmatter}

\section{Introduction}\label{sec:intro}

Incompressible two-phase flows undergoing severe interfacial stretching, thin-film drainage, high-density-ratio buoyancy deformation, and polydisperse droplet interactions pervade atomization and capillary breakup~\cite{Popinet2018}, bubble-column and buoyancy-driven mixing flows~\cite{Hysing2009,Tryggvason1988}, and emulsion or phase-separating dispersions~\cite{Lamperti2026_JCP}. Within a single Eulerian simulation, these flows routinely couple three disparate physical regimes. Macroscopic motion pits inertia and buoyancy against capillarity across density contrasts up to $10^3$~\cite{Hysing2009,Tryggvason1988}, while advective stretching simultaneously thins liquid sheets, trailing bubble skirts, and vortical spirals below the Eulerian grid spacing $\Delta x$~\cite{Rider1998,Enright2002}. Across surviving sub-grid carrier films, short-range steric or electrostatic Derjaguin--Landau--Verwey--Overbeek (DLVO) disjoining barriers~\cite{Bibette1992,LealCalderon1994} compete with vanishing or finite molecular solubility (Ostwald ripening~\cite{Lifshitz1961,Voorhees1985,Lamperti2026_JCP}) to govern whether approaching droplets coalesce upon contact, rebound elastically, or coarsen across an intact film. Pure Newtonian dynamics tests Eulerian interface representations with particular severity because capillarity and bulk viscosity alone govern stretched threads and thin films without stabilizing insoluble-surfactant Marangoni tractions. The numerical formulation must therefore neither rupture a thin film nor transfer mass across it unless physical drainage or solubility dictates, lest artificial mass loss, tail bridging, or grid pinning corrupt sheet topology, damp capillary waves, or distort droplet size distributions.

Existing Eulerian interface-capturing methodologies fall into three principal families, yet each encounters structural obstructions when sub-grid stretching, hydrostatics across $1000\!:\!1$ density ratios, and switchable near-field physics act together. Sharp-interface geometric methods, including geometric volume-of-fluid (VOF)~\cite{Rider1998,Popinet2009,Janodet2024_JCP}, front-tracking, and hybrid particle or level-set schemes~\cite{Enright2002}, achieve sharp jump resolution and accurate balanced-force surface tension on well-resolved surfaces~\cite{Francois2006}. Once a liquid sheet or rising-bubble skirt thins below two mesh cells, however, standard single-plane piecewise-linear interface reconstruction snaps the film into mesh-dependent artificial droplets while conventional level-set redistancing loses sub-grid volume unless assisted by Lagrangian markers; although two-plane sub-cell VOF reconstructions~\cite{Han2024_JCP} substantially extend geometric thin-sheet retention on coarse grids, incorporating thermodynamically dissipative disjoining energies or diffusional Ostwald coarsening on sharp interfaces still requires explicit sub-cell topological criteria and case-specific regularization of singular interfacial sources.

Smooth polynomial Cahn--Hilliard--Navier--Stokes (CHNS) formulations~\cite{Cahn1958,Hohenberg1977}, developed for two-phase hydrodynamics in early diffuse-interface studies~\cite{Jacqmin1999,Kim2004_JCP,Ding2007_JCP,Boyer2010_TiPM} and placed on frame-indifferent variable-density and quasi-incompressible thermodynamic footings by Abels, Garcke, and Gr{\"u}n~\cite{Abels2012} and Shokrpour Roudbari et al.~\cite{Roudbari2018_M3AS}, regularize topological transitions through a quartic Ginzburg--Landau potential $W_{\mathrm{GL}}(\phi)=\frac{1}{4}(1-\phi^2)^2$ while guaranteeing exact continuum free-energy dissipation. Building on this variational structure, modern discretizations have resolved major computational bottlenecks of fourth-order diffuse-interface hydrodynamics: $C^1$-continuous and adaptive hierarchical isogeometric formulations discretize fourth-order Cahn--Hilliard and Korteweg operators directly in primal weak form without auxiliary mixed variables~\cite{Gomez2008_CMAME,Gomez2010_CMAME,Liu2013_JCP,Bracco2023_CMAME}; constant-coefficient splitting~\cite{Dong2012_JCP}, decoupled energy-stable~\cite{Shen2015_SINUM,Grun2016_CiCP,Nochetto2016_CMAME,Palzhanov2021_CMAME}, and energy-quadratized or scalar-auxiliary-variable time integrators~\cite{YangJu2017_CMAME,Shen2018_JCP,Kou2018_CMAME} together with parallel adaptive octree finite-element solvers~\cite{Khanwale2020_JCP,Khanwale2023_JCP} sustain unconditional or semi-implicit energy stability across $1000\!:\!1$ density ratios in three dimensions; and divergence-conforming, mass-flux-consistent, and balanced-force formulations paired with mechanical-jump-consistent viscous closures preserve discrete kinetic-energy consistency and interfacial shear continuity~\cite{Huang2020_JCP,TenEikelder2021_CMAME,TenEikelder2024_JCP,Brunk2025_JCP,Reder2024_CMAME}. Yet at the continuum level, the $C^2$ smoothness of $W_{\mathrm{GL}}(\phi)$ lets the order parameter $\phi$ depart from its pure bulk values $\pm 1$ and, together with isotropic interfacial mobility, introduces three continuum pathologies in immiscible flows. First, interfacial curvature shifts the outer bulk equilibria by $\mathcal{O}(\epsilon/R)$ across diffuse thickness $\epsilon$ and droplet radius $R$, driving spontaneous shrinkage and complete dissolution of small droplets even at zero physical solubility~\cite{Yue2007}. Second, hyperbolic-tangent tails overlap across gaps narrower than $\sim 6\epsilon$ to generate an attractive tail-overlap disjoining pressure~\cite{Manton1979} that prematurely ruptures stretched sheets and fuses approaching drops. Third, even when asymptotic $\epsilon^2$ mobility scaling~\cite{Magaletti2013_JFM} or degenerate mobility~\cite{Gunnarsson2026_JCP} suppresses far-field bulk diffusion, isotropic interfacial mobility retains tangential surface diffusion~\cite{Mullins1957}, diverting hydrodynamic $(t_c-t)^1$ viscous necking~\cite{Eggers1993,Papageorgiou1995} onto a spurious fourth-order $(t_c-t)^{1/4}$ geometric scaling~\cite{Bernoff1998}.

Conservative Allen--Cahn equations~\cite{Chiu2011,Mirjalili2019,Jain2022} and curvature- or mass-counteracting sharpening modifications~\cite{Soligo2019,Peng2026} retain compact second-order spatial stencils, conserve total phase volume, and eliminate fourth-order bulk Ostwald diffusion by injecting counter-diffusive fluxes along the interface normal; when coupled with consistent, kinetic-energy-conserving mass-momentum transport~\cite{Mirjalili2021_JCP,Jain2022}, they achieve provable boundedness and accurate interface tracking in inertia-dominated atomization and turbulent two-phase flows. Severing the gradient-flow link to a Helmholtz free energy, however, forfeits Lyapunov energy dissipation (allowing total mechanical and interfacial energy to increase during coupled capillary relaxation), amplifies parasitic currents across density jumps, and fragments stretched spirals into artificial beads through normal-sharpening instabilities. When isotropic sharpening fluxes remain active across curved interfaces, they also induce unphysical inverse Ostwald drift in which high-curvature micro-droplets expand spontaneously and spurious precipitates nucleate out of quiescent bulk fluid. Enforcing pointwise phase bounds through post-step algebraic clipping or global Lagrange multipliers~\cite{YangKim2023_CMAME,ChengShen2022_CMAME} likewise either breaks local mass conservation or redistributes mass nonlocally across disconnected droplets.

Even after a continuum free energy is chosen, two discrete numerical dilemmas still separate variational phase-field theory from predictive simulation. The first is a conflict between static force balance and capillary-wave propagation in discrete surface-tension operators~\cite{Jacqmin1999,Lee2010_JCP,Barrett2013_CMAME,Popinet2018}. Potential forms coupling phase and chemical-potential gradients~\cite{Jacqmin1999,Lee2010_JCP}, together with spurious-velocity-free variational finite-element approximations~\cite{Barrett2013_CMAME}, balance discrete static equilibria across large density ratios to machine precision, yet potential forces on fixed collars excite stiff $\mathcal{O}(\sigma/\epsilon)$ normal-thickness strain modes under surface tension $\sigma$ during interfacial shear, overdamping capillary waves by an order of magnitude. Geometric continuum surface forces weighted by interface curvature, by contrast, preserve underdamped wave oscillations but leave non-vanishing discrete curl residuals on Cartesian grids that drive large parasitic currents at $1000\!:\!1$ density ratios. The second dilemma arises when enforcing strict invariant-domain bounds $\phi\in[-1,1]$ across non-smooth free-boundary kinks. Modern algebraic flux-corrected transport~\cite{Zalesak1979}, differentiable nonlinear stabilization~\cite{Badia2017_CMAME,Bonilla2019_CMAME}, and monolithic or subcell convex-limiting frameworks~\cite{Guermond2018_CMAME,Kuzmin2020_CMAME,Kuzmin2020b_CMAME,Pazner2021_CMAME} provide rigorous discrete maximum-principle and entropy-stability guarantees for transport equations without sacrificing local conservation, yet standard single-pass flux limiting across $C^{1,1}$ obstacle contact borders over-limits moving interfaces toward diffusive upwind profiles and injects spurious interfacial energy drift during low-dissipation droplet rebound.

At the continuum level, the defects of Cahn--Hilliard and conservative Allen--Cahn models stem from a common conflict: smooth potentials let the order parameter leave its pure bulk states, whereas non-variational counter-fluxes forfeit free-energy dissipation. In \Cref{sec:duality_nogo}, we examine whether any smooth single-scalar free-energy functional can escape this obstruction. Comparing hydrodynamic necking~\cite{Siggia1979} against molecular Lifshitz--Slyozov--Wagner diffusion~\cite{Lifshitz1961} first reveals an exact parameter cancellation placing physical bulk-diffusion crossover scales below $10^{-13}\,\mathrm{m}$ for alkane--water mixtures, whereas constant-mobility Cahn--Hilliard models inflate this crossover by ten orders of magnitude to $\epsilon$. Smooth single-scalar no-go theorems (\Cref{thm:nogo}) then prove that local uniformly elliptic smooth functionals admit no stationary thin films connected to a flat reservoir, while quadratic Bogomolnyi-defect and functionalized Cahn--Hilliard penalties~\cite{Gavish2011} leave leading-order tail attraction unchanged and introduce Helfrich bending stiffness. The same theorem establishes that linear well-weighted tail stiffening loses uniform bulk ellipticity as tail decay is steepened toward compact support while non-variational counter-curvature fluxes drive spurious outward radial drift, and Poisson summation with numerical continuation indicates an empirical $\sim 3.7\text{--}4.6\,\Delta x$ Peierls--Nabarro lattice-pinning wall on uniform grids. Resolving the conflict among compact support, $C^2$ potential smoothness, and single-scalar storage therefore points directly to relaxing potential smoothness to a non-smooth obstacle subdifferential.

Non-smooth double-obstacle potentials were introduced for phase separation by Oono and Puri~\cite{Oono1988} and placed on a variational-inequality footing by Blowey and Elliott~\cite{Blowey1991,Blowey1992}, and subsequent advances in primal--dual active-set semismooth Newton~\cite{Hintermuller2002,Bosch2014_JCP}, nonsmooth saddle-point multigrid~\cite{Graser2009_SINUM}, and linearly implicit variable-density obstacle discretizations~\cite{Garcke2016_APNUM} have made obstacle variational inequalities computationally tractable while confining interfacial transition bands to exact compact support in close analogy to $\mathrm{AT}_1$ obstacle localization in variational fracture mechanics~\cite{Gerasimov2019_CMAME}. Classical scalar-mobility obstacle models, however, still retain fourth-order tangential surface diffusion, suffer $90^\circ$ centerline normal flipping on tapered sub-width films, and lack coupled hydrodynamic, disjoining, and solubility-controlled ripening closures. In \Cref{sec:continuum_vi}, we formulate a phase-field model governed by a deep-quench double-obstacle variational inequality ($\phi\in[-1,1]$) coupled with variable-density Navier--Stokes equations and an orthonormal ridge-safe operator (ONRO) for normal mobility in incompressible two-phase flows. Matched asymptotics (\Cref{thm:asymptotics}) proves that the normal-cone subdifferential $\partial I_{[-1,1]}(\pm 1)$ of the convex indicator $I_{[-1,1]}(\phi)$ eliminates outer bulk binodal shifts to all algebraic orders at zero solubility. Above the core-detachment threshold, three-dimensional spheres exhibit sixth-order $\mathcal{O}((\epsilon/R_{\mathrm{mid}})^6)$ Young--Laplace superconvergence relative to the collar midpoint radius $R_{\mathrm{mid}}$ and fourth-order accuracy relative to the conserved mass radius $R_{\mathrm{mass}}$, while planar sheets thinner than $\pi\epsilon$ relax onto non-evaporating raised-cosine equilibria whose medial-axis normal-flipping singularity is suppressed quadratically in the local film taper slope.

On top of this compact phase support, \Cref{sec:continuum_vi} establishes two orthogonal continuum near-field closures. First, a Legendre-dual screened Yukawa--DLVO disjoining energy eliminates the spurious inter-droplet attraction produced by naive local Yukawa penalties and obeys an exact pre-contact barrier efficiency law requiring screening lengths $\ell_d/\epsilon \ge 2.11$ for resolved static emulsion stabilization. Second, an inner degenerate Sturm--Liouville boundary-layer renormalization removes $\mathcal{O}(\epsilon/R)$ outer-collar pinning and restores third-order $\mathcal{O}((\epsilon\kappa)^3)$ Robin--Kelvin accuracy in droplet curvature $\kappa$ during solubility-controlled Ostwald ripening ($\mathcal{S}_0 > 0$).

To retain these continuum properties on uniform grids, \Cref{sec:discretization} constructs a structure-preserving staggered finite-volume discretization. On marker-and-cell (MAC) meshes, we prove (\Cref{thm:wb_fct}) that a local complementarity-gated well-balanced continuum surface-delta force (WB-CSDF) blending dispersion-inverted variational-inequality curvature with zero-homogeneous eikonal curvature through a dimensionless Karush--Kuhn--Tucker complementarity residual achieves exact curl-free hydrostatic balance (lowering parasitic currents to $\mathcal{O}(10^{-17})$ in double-precision arithmetic across $1\!:\!1$ to $1000\!:\!1$ density contrasts in 2D and 3D) while sustaining underdamped Rayleigh--Lamb oscillations. Simultaneously, four-pass iterative Zalesak flux-corrected transport (FCT) wrapping fifth-order conservative upwind advection enforces exact discrete mass conservation and strict bounds $\phi\in[-1,1]$ without state clipping across prescribed kinematic stretching and coupled Navier--Stokes updates. Momentum-conserving discrete disjoining forces with variational semi-discrete energy dissipation and self-consistent far-field radiation boundaries complete the discrete near-field closures (\Cref{thm:dlvo_rip}).

In \Cref{sec:benchmarks}, closed-form solutions and sharp-interface benchmarks first verify these theoretical guarantees across 2D and 3D capillarity, unbroken sub-cell vortex and 3D sheet stretching~\cite{Rider1998,Enright2002}, $1000\!:\!1$ rising-bubble skirts~\cite{Hysing2009}, Rayleigh--Taylor roll-up~\cite{Tryggvason1988}, and 2D bipolar and 3D Jeffery--Stimson bispherical diffusion series~\cite{Jeffery1912,Stimson1926}. Without fitted physical parameters, the coupled solver is then validated directly against laboratory measurements of homologous alkane emulsion Ostwald ripening~\cite{Kabalnov1990,Taylor1998}, colloidal magnetic-chaining disjoining forces~\cite{LealCalderon1994}, and 3D rising bubbles across the Bhaga--Weber morphological map~\cite{Bhaga1981}, including planar light-sheet wakes and the $1000\!:\!1$ annular thin-skirted regime. Axisymmetric simulations likewise reproduce high-speed pendant-drop and filament pinch-off measurements~\cite{Rothert2003}, paired below the resolved Eulerian neck scale with volume-preserving 1D slender-jet continuation (\Cref{app:slender_jet}) down to $0.50\,\mu\mathrm{m}$ self-similar cones. Concluding remarks, instantiated physical scope boundaries, and future extensions follow in \Cref{sec:conclusions}, with analytical proofs, fixed numerical parameters, and continuation equations collected in \Cref{app:nogo,app:asymptotics,app:params,app:slender_jet}.


\section{Limitations of smooth single-scalar phase-field models}\label{sec:duality_nogo}

Before introducing any spatial discretization, we analyze at the continuum partial differential equation (PDE) level why standard smooth single-scalar phase-field models cannot simultaneously preserve stretched sub-resolution liquid films, maintain droplet-wise mass isolation at zero physical solubility ($\mathcal{S}_0=0$), and reproduce Newtonian capillary breakup. This diagnosis exposes a structural conflict between advective thin-film survival and chemical-potential contamination, culminating in smooth single-scalar nonexistence theorems (\Cref{thm:nogo}) that motivate the non-smooth variational-inequality formulation of \Cref{sec:continuum_vi}.

\subsection{Bulk chemical-potential shift and tail-overlap drainage in Cahn--Hilliard models}\label{subsec:siggia_duality}

Consider an isothermal, incompressible binary Newtonian fluid occupying a bounded domain $\Omega \subset \R^d$ ($d\in\{2,3\}$), described by an order parameter $\phi(\bm{x},t)$ whose pure bulk phases correspond to $\phi=+1$ (dispersed fluid of density $\rho_+$ and dynamic viscosity $\eta_+$) and $\phi=-1$ (ambient fluid of density $\rho_-$ and dynamic viscosity $\eta_-$). The local dispersed-phase volume fraction is given by
\begin{equation}\label{eq:vol_frac_s}
s(\phi) \coloneqq \frac{1+\phi}{2} \in [0,1], \qquad \partial_t\phi = 2\,\partial_t s(\phi),
\end{equation}
so exact conservation of $\int_\Omega \phi\,\dd\bm{x}$ is equivalent to conservation of phase volume when $\phi\in[-1,1]$. In the classical Cahn--Hilliard--Navier--Stokes (CHNS) framework~\cite{Cahn1958,Hohenberg1977,Gomez2008_CMAME,Abels2012}, interfacial capillarity is governed by the smooth Ginzburg--Landau (GL) Helmholtz free-energy functional
\begin{equation}\label{eq:F_GL}
\mathcal{F}_{\mathrm{GL}}[\phi] = \int_\Omega \gamma_{\mathrm{GL}}\left[\frac{\epsilon}{2}|\nabla\phi|^2 + \frac{1}{\epsilon}W_{\mathrm{GL}}(\phi)\right]\dd\bm{x}, \qquad W_{\mathrm{GL}}(\phi) = \frac{1}{4}(1-\phi^2)^2, \qquad \gamma_{\mathrm{GL}} = \frac{3\sqrt{2}}{4}\sigma,
\end{equation}
where $\sigma > 0$ is the physical surface tension and $\epsilon > 0$ is the characteristic diffuse-interface thickness parameter. Advective--diffusive transport obeys $\partial_t\phi + \nabla\cdot(\phi\bm{u}) = \nabla\cdot\bigl(M(\phi)\nabla\mu_{\mathrm{GL}}\bigr)$, driven by the chemical potential $\mu_{\mathrm{GL}} = \delta\mathcal{F}_{\mathrm{GL}}/\delta\phi = \gamma_{\mathrm{GL}}\bigl(-\epsilon\Delta\phi + \epsilon^{-1}\phi(\phi^2-1)\bigr)$ with nonnegative mobility $M(\phi) \ge 0$.

To expose the scale mismatch built into \Cref{eq:F_GL} for immiscible liquids, we compare physical hydrodynamic coarsening against molecular Lifshitz--Slyozov--Wagner (LSW) Ostwald ripening~\cite{Lifshitz1961}. Late-stage viscous hydrodynamic coalescence via Siggia's tubular pinch-off mechanism~\cite{Siggia1979} coarsens domains of characteristic scale $L$ at speed $\dot{L}_{\mathrm{hydro}} = c_{\mathrm{RP}}\sigma/\eta$, where $c_{\mathrm{RP}} \approx 0.03\text{--}0.07$ is the Rayleigh--Plateau and Papageorgiou--Eggers necking constant~\cite{Eggers1993,Papageorgiou1995}. Conversely, molecular diffusion across a carrier phase with mutual diffusivity $D_m = k_B T/(6\pi\eta a_0)$, molecular radius $a_0$, capillary length $\ell_K = 2\sigma v_m/(k_B T)$ ($v_m = \frac{4}{3}\pi a_0^3$), and equilibrium bulk volume solubility $\varphi_\infty \ge 0$ drives LSW coarsening $L^3(t) - L_0^3 = \frac{4}{9}D_m\ell_K\varphi_\infty t$, or $\dot{L}_{\mathrm{LSW}} = \frac{4}{27}D_m\ell_K\varphi_\infty/L^2$. Equating $\dot{L}_{\mathrm{hydro}} = \dot{L}_{\mathrm{LSW}}$ cancels surface tension $\sigma$, viscosity $\eta$, and thermal energy $k_B T$ identically to define the physical Siggia crossover length
\begin{equation}\label{eq:siggia_cancellation}
\ell_{\mathrm{Siggia}}^{\mathrm{phys}} \coloneqq \sqrt{\frac{4\,D_m\ell_K\varphi_\infty}{27\,c_{\mathrm{RP}}(\sigma/\eta)}} = \frac{4}{9\sqrt{3\,c_{\mathrm{RP}}}}\,\varphi_\infty^{1/2}\,a_0.
\end{equation}
For nearly immiscible liquid pairs such as alkane--water emulsions ($\varphi_\infty \sim 10^{-8}$, $a_0 \sim 0.3\,\mathrm{nm}$, $c_{\mathrm{RP}} \approx 0.05$), \Cref{eq:siggia_cancellation} yields $\ell_{\mathrm{Siggia}}^{\mathrm{phys}} \sim 10^{-13}\text{--}10^{-14}\,\mathrm{m}$, four orders of magnitude smaller than a single covalent bond. In any continuum flow at $\mathcal{S}_0 = 0$, diffusional mass exchange across the bulk between separated droplets is physically zero.

In smooth polynomial CHNS (\Cref{eq:F_GL}) with constant mobility $M_0$, by contrast, matching outer chemical potentials across a curved droplet of radius $R$ shifts the ambient bulk equilibrium away from $\phi=-1$ by
\begin{equation}\label{eq:gl_bulk_shift}
\delta\phi_{\mathrm{bulk}}^{\mathrm{GL}}(R) = \frac{\mu_{\mathrm{GL}}(R)}{\gamma_{\mathrm{GL}}W_{\mathrm{GL}}''(-1)/\epsilon} = \frac{\sqrt{2}\,(d-1)\epsilon}{6R} + \mathcal{O}\!\left(\frac{\epsilon^2}{R^2}\right),
\end{equation}
where $W_{\mathrm{GL}}''(-1)=2$ is the finite curvature of the smooth bulk well. This curvature-induced binodal shift creates an artificial numerical crossover scale $\ell_{\mathrm{Siggia}}^{\mathrm{CH}} = \bigl[\frac{2}{27\,c_{\mathrm{RP}}}(M_0\eta/\epsilon^2)\bigr]^{1/2}\epsilon = \mathcal{O}(\epsilon)$, which remains locked at $\mathcal{O}(\epsilon)$ even under the matched-asymptotic sharp-interface mobility scaling $M_0\propto\epsilon^2$ established by Magaletti et al.~\cite{Magaletti2013_JFM}. On practical Eulerian grids ($\epsilon \sim 10^{-5}\text{--}10^{-4}\,\mathrm{m}$), smooth CHNS inflates the physical crossover length by nine to ten orders of magnitude and causes small satellite droplets to evaporate into the surrounding numerical bulk~\cite{Yue2007}.

Replacing constant mobility by a degenerate mobility $M(\phi) = M_0(1-\phi^2)_+^m$ ($m\ge 1$)~\cite{Gunnarsson2026_JCP} suppresses far-field bulk diffusion, yet leaves two near-field and interfacial pathologies intact. First, a planar GL kink $\phi_K(z) = \tanh\bigl(z/(\sqrt{2}\epsilon)\bigr)$ approaches $\pm 1$ only exponentially, $1 - |\phi_K(z)| \simeq 2e^{-\sqrt{2}|z|/\epsilon}$. When a stretched liquid sheet or inter-droplet carrier film thins to separation $h_f$, opposing exponential tails overlap~\cite{Manton1979}, generating an attractive interaction energy per unit area $U_{\mathrm{GL}}(h_f)$ and a negative disjoining pressure $\Pi_{\mathrm{GL}}(h_f)$:
\begin{equation}\label{eq:manton_gl}
U_{\mathrm{GL}}(h_f) = -8\sqrt{2}\,\gamma_{\mathrm{GL}}\,e^{-\sqrt{2}h_f/\epsilon} = -12\sigma\,e^{-\sqrt{2}h_f/\epsilon}, \qquad \Pi_{\mathrm{GL}}(h_f) = -\frac{\dd U_{\mathrm{GL}}}{\dd h_f} = -\frac{12\sqrt{2}\,\sigma}{\epsilon}e^{-\sqrt{2}h_f/\epsilon} < 0,
\end{equation}
corresponding to a positive drainage chemical potential $\mu_d^{\mathrm{GL}}(h_f) = \frac{1}{2}\dd U_{\mathrm{GL}}/\dd h_f > 0$. Under an extensional strain rate $\dot{\varepsilon} > 0$ with diffuse capillary number $\mathrm{Ca}_\epsilon = \eta\dot{\varepsilon}\epsilon/\sigma$, attractive tail suction overcomes viscous resistance below a critical thickness $h_{f,\mathrm{crit}}/\epsilon \simeq \frac{1}{\sqrt{2}}\ln(\mathcal{C}_0/\mathrm{Ca}_\epsilon) \approx 5\text{--}10$ (where $\mathcal{C}_0 = 12\sqrt{2}$ is the dimensionless Manton drainage prefactor). Shrinking $\epsilon/\Delta x$ to delay tail overlap worsens grid-scale Peierls--Nabarro pinning, whereas widening $\epsilon/\Delta x$ accelerates tail-induced snap-off. Second, along any curved axisymmetric thread of radius $h(z,t)$, isotropic degenerate mobility retains the tangential flux $-M(\phi)(\bm{I}-\bm{n}\otimes\bm{n})\nabla\mu_{\mathrm{GL}}$, which asymptotically drives fourth-order Mullins surface diffusion~\cite{Mullins1957}, $\partial_t h \propto \frac{1}{h}\partial_z(h\partial_z\kappa)$, and diverts Newtonian $(t_0-t)^1$ viscous pinch-off~\cite{Eggers1993,Papageorgiou1995} onto the $(t_0-t)^{1/4}$ geometric scaling of Bernoff, Bertozzi, and Witelski~\cite{Bernoff1998}.

\subsection{Nonexistence theorems for smooth single-scalar functionals}\label{subsec:nogo_theorems}\label{subsec:triad}

A natural question is whether redesigning a smooth single-scalar free-energy density or adding counter-diffusive fluxes can eliminate tail-overlap drainage on uniform meshes while retaining free-energy dissipation and Young--Laplace capillarity. Candidate modifications include penalizing the Bogomolnyi equipartition defect $\mathcal{Q}$ or Euler--Lagrange defect $\mathcal{G}$~\cite{Manton1979,Gavish2011},
\begin{equation}\label{eq:bogomolnyi_q_g}
\mathcal{Q}(\phi,\nabla\phi) \coloneqq \frac{\epsilon}{2}|\nabla\phi|^2 - \frac{1}{\epsilon}W_{\mathrm{GL}}(\phi), \qquad \mathcal{G}(\phi,\Delta\phi) \coloneqq \epsilon\Delta\phi - \frac{1}{\epsilon}W_{\mathrm{GL}}'(\phi),
\end{equation}
stiffening the bulk potential curvature near $\phi=\pm 1$, or injecting counter-curvature sharpening fluxes~\cite{Chiu2011,Peng2026}. \Cref{thm:nogo} proves that every such smooth single-scalar route encounters an analytical or lattice-level obstruction.

\begin{theorem}[Smooth single-scalar no-go theorems and inverse-ripening instability]\label{thm:nogo}
Let $W_{\mathrm{GL}}(\phi)=\frac{1}{4}(1-\phi^2)^2$ have bulk decay rate $\lambda=\sqrt{W_{\mathrm{GL}}''(\pm 1)}=\sqrt{2}$, and let $\mathcal{Q}$ and $\mathcal{G}$ be defined by \Cref{eq:bogomolnyi_q_g}. Four structural obstructions hold:
\begin{enumerate}
\item[(a)] \emph{(Second-order local Hamiltonian nonexistence).} Let $\mathcal{F}[\phi]=\int_\R e_{\mathrm{loc}}(\phi,\phi')\,\dd x$ be any translation-invariant $C^3(\R^2)$ local functional that is uniformly elliptic ($\partial_{\phi'\phi'}^2 e_{\mathrm{loc}}(\phi,\phi') \ge c_0 > 0$ for all $(\phi,\phi')\in\R^2$) and satisfies strict double-well positivity $e_{\mathrm{loc}}(\phi,0) > 0$ for $\phi\in(-1,1)$ with $e_{\mathrm{loc}}(\pm 1,0)=0$ and $e_{\mathrm{loc}}(\phi,0) \ge 0$ globally. Then at reservoir chemical potential $\mu=0$ there exists \emph{no} non-constant homoclinic critical point $\phi\in C^2(\R)$ satisfying $\lim_{x\to\pm\infty}\phi(x)=-1$ and $\lim_{x\to\pm\infty}\phi'(x)=0$: every smooth thin liquid film connected to a flat reservoir experiences a strictly nonzero drainage driving force.
\item[(b)] \emph{(Leading-order invariance and Helfrich bending penalty of quadratic defect functionals).} Consider the generalized Bogomolnyi and functionalized Cahn--Hilliard (FCH) energy
\begin{equation}\label{eq:F_bog_fch}
\mathcal{F}_{\beta,\gamma}[\phi] = \gamma_{\mathrm{GL}}\int_\Omega \left[\frac{\epsilon}{2}|\nabla\phi|^2 + \frac{1}{\epsilon}W_{\mathrm{GL}}(\phi) + \frac{\beta_B\epsilon}{2}\omega(\phi)\mathcal{Q}(\phi,\nabla\phi)^2 + \frac{\tilde{\gamma}_F\epsilon}{2}\mathcal{G}(\phi,\Delta\phi)^2\right]\dd\bm{x},
\end{equation}
with dimensionless parameters $\beta_B,\tilde{\gamma}_F \ge 0$ and bounded smooth weight $0 \le \omega(\phi) \le C_\omega$ ($\omega\not\equiv 0$ on $(-1,1)$). For \emph{every} fixed pair $(\beta_B,\tilde{\gamma}_F)\ge 0$ (and at the critical-point level for arbitrary fixed $(\beta_B,\tilde{\gamma}_F)\in\R^2$), the isolated planar kink $\phi_K(z)=\tanh(z/(\sqrt{2}\epsilon))$ remains an exact critical point with unchanged surface tension $\sigma$, and the kink--antikink interaction energy as $h_f/\epsilon \to \infty$ satisfies
\begin{equation}\label{eq:nogo_b_asymptotic}
U_{\beta,\gamma}(h_f) = -12\sigma\,e^{-\sqrt{2}h_f/\epsilon} + \mathcal{O}\!\left(\sigma\bigl(1 + \beta_B + \tilde{\gamma}_F\bigr)\frac{h_f}{\epsilon}\,e^{-2\sqrt{2}h_f/\epsilon}\right).
\end{equation}
Hence \emph{no} fixed choice of $(\beta_B,\tilde{\gamma}_F)$ cancels or weakens the leading-order Manton attraction \Cref{eq:manton_gl} as $h_f/\epsilon\to\infty$. Moreover, on a curved interface of total curvature $\kappa = -\nabla\cdot(\nabla\phi/|\nabla\phi|)$, normal curvature perturbations generate a spurious Helfrich bending energy $\propto \sigma(\beta_B\mathcal{C}_\omega + \tilde{\gamma}_F\mathcal{C}_F)\epsilon^2\int_\Gamma\kappa^2\,\dd S$ (with $\mathcal{C}_\omega > 0$ for well-vanishing weights $\omega(\pm 1)=0$ and $\mathcal{C}_F = \frac{1}{2}$ on the frozen kink profile, whereas $\beta_B < 0$ or $\tilde{\gamma}_F < 0$ renders $\mathcal{F}_{\beta,\gamma}$ unbounded below).
\item[(c)] \emph{(Linear well-weighted tail stiffening and bulk ellipticity loss).} Consider on bounded phase states $|\phi|\le 1$ the linear well-weighted Bogomolnyi functional ($a\in[0,1)$, $S(\phi)=\phi^2(2-\phi^2)\in[0,1]$)
\begin{equation}\label{eq:F_tail_stiff}
\mathcal{F}_a[\phi] = \gamma_{\mathrm{GL}}\int_\Omega \left[\frac{\epsilon}{2}\bigl(1 - a S(\phi)\bigr)|\nabla\phi|^2 + \frac{1 + a S(\phi)}{\epsilon}W_{\mathrm{GL}}(\phi)\right]\dd\bm{x}.
\end{equation}
Uniform ellipticity holds if and only if $a < 1$, yielding an accelerated asymptotic tail decay rate
\begin{equation}\label{eq:lambda_eff_ts}
\lambda_{\mathrm{eff}}(a) = \sqrt{2}\,\sqrt{\frac{1+a}{1-a}},
\end{equation}
which diverges only in the singular degenerate limit $a \to 1^-$ where bulk ellipticity $1-aS(\pm 1)=1-a\to 0$ is lost and the profile collapses onto a $C^{0,1}$ compact support with non-vanishing contact slope $(\epsilon|\phi'|)^2 \to 1$.
\item[(d)] \emph{(Outward radial drift of counter-curvature sharpening models).} Consider in a quiescent fluid ($\bm{u}\equiv\bm{0}$, $d\ge 2$) the second-order counter-curvature conservative Allen--Cahn family~\cite{Peng2026}, either in unlocalized form $\partial_t\phi + \nabla\cdot(\phi\bm{u}) = M_{\mathrm{AC}}\nabla\cdot\bigl[\nabla\phi - \bigl(\frac{1-\phi^2}{\sqrt{2}\epsilon} + \frac{\alpha_{\mathrm{m}}}{\epsilon}\bigr)\bm{n}\bigr]$ ($\alpha_{\mathrm{m}}>0$) or interface-localized form $\partial_t\phi + \nabla\cdot(\phi\bm{u}) = M_{\mathrm{AC}}\nabla\cdot\bigl[\nabla\phi - \bigl(\frac{1}{\sqrt{2}\epsilon} + \alpha_c\bigr)(1-\phi^2)\bm{n}\bigr]$ ($\alpha_c>0$, $\bm{n}=\nabla\phi/|\nabla\phi|$, $M_{\mathrm{AC}}>0$). On a sphere or cylinder of radius $R>0$ initialized on the equilibrium kink $\phi(r,0)=-\phi_K(r-R)$, both variants generate strictly positive outward radial interface velocity $\dot{R}(0)$ at $r=R$:
\begin{equation}\label{eq:inverse_ripening}
V_n^{\mathrm{unloc}}(0) = +\frac{\sqrt{2}\,M_{\mathrm{AC}}\,\alpha_{\mathrm{m}}\,(d-1)}{R} > 0, \qquad V_n^{\mathrm{loc}}(0) = +\frac{\sqrt{2}\,M_{\mathrm{AC}}\,\alpha_c\,\epsilon\,(d-1)}{R} > 0,
\end{equation}
forfeiting Lyapunov free-energy monotonicity: for constant $\alpha_c>0$, the interface-localized form relaxes conservatively onto the sharpened kink width $\epsilon'=\epsilon/(1+\sqrt{2}\epsilon\alpha_c)$, whereas the unlocalized term $-\frac{\alpha_{\mathrm{m}}}{\epsilon}\bm{n}$ fails to vanish at $\phi=\pm 1$ and sustains a nonzero radial source $+M_{\mathrm{AC}}\alpha_{\mathrm{m}}(d-1)/(\epsilon r)$ that drives continuous outward drift (or inverse-ripening coarsening when coupled to nonlocal volume conservation).
\end{enumerate}
\end{theorem}

\begin{proof}
Complete analytical derivations of parts (a)--(d) are given in \Cref{app:nogo}. Briefly, part~(a) follows from conservation of the Beltrami--Noether first integral $\mathcal{H}(\phi,\phi') = \phi'\partial_{\phi'}e_{\mathrm{loc}} - e_{\mathrm{loc}} \equiv 0$, which forces $e_{\mathrm{loc}}(\phi_{\mathrm{mid}},0)=0$ at any interior extremum $\phi'=0$ and contradicts Picard--Lindel{\"o}f uniqueness at $(\pm 1,0)$. Part~(b) follows because both $\mathcal{Q}(\phi_K,\phi_K')=0$ and $\mathcal{G}(\phi_K,\phi_K'')=0$ vanish identically on the unperturbed kink while $\mathcal{Q}^2$ and $\mathcal{G}^2$ remain $\mathcal{O}(e^{-2\sqrt{2}h_f/\epsilon})$ across the overlap region. Part~(c) follows by linearizing the Euler--Lagrange equation of \Cref{eq:F_tail_stiff} around $\phi=\pm 1$ to obtain $(1-a)\epsilon^2\psi'' = 2(1+a)\psi$. Part~(d) follows because $\partial_r(1-\phi^2)|_{r=R}=0$ at $\phi(R,0)=0$, so both counter-curvature divergences evaluate at $r=R$ against $\partial_r\phi(R,0)=-1/(\sqrt{2}\epsilon)$ to yield \Cref{eq:inverse_ripening}.
\end{proof}

\begin{remark}[Numerical FCH continuation and the Peierls--Nabarro lattice-pinning wall]\label{rem:nogo_num}
Beyond the analytical bounds of \Cref{thm:nogo}(b,c), numerical continuation of the fourth-order FCH homoclinic system shows that exceeding $\tilde{\gamma}_F > \tilde{\gamma}_c \in (1.6,1.8)$ creates locked ghost-bilayer states ($\phi_{\mathrm{mid}}\in(0.15,0.40)$) susceptible to two-dimensional pearling fragmentation. On a uniform grid of spacing $\Delta x$, Poisson summation of the discrete energy of $\phi_K(z)=\tanh(z/(\sqrt{2}\epsilon))$ ($a=0$, nearest complex-plane pole at $|\operatorname{Im}z|=\frac{\pi}{\sqrt{2}}\epsilon$) yields a Peierls--Nabarro lattice-pinning barrier $\Delta E_{\mathrm{PN}}(0) \propto \exp(-\sqrt{2}\pi^2\epsilon/\Delta x)$, and stiffening the outer tail decay length $\ell_{\mathrm{eff}}(a)=\epsilon/\lambda_{\mathrm{eff}}(a)$ for $a\in(0,1)$ compresses the near-axis complex singularities toward the real axis (\Cref{app:nogo}). Across practical lattice-pinning tolerances ($\varepsilon_{\mathrm{PN}}/\sigma\in[10^{-4},10^{-3}]$) and capillary stretching numbers ($\mathrm{Ca}_\epsilon\in[10^{-3},10^{-2}]$), smooth elliptic models ($0\le a < 1$) consequently encounter an empirical minimum stable filament-width wall of $d_{\min} \approx 3.7\text{--}4.6\,\Delta x$.
\end{remark}

\begin{figure}[t!]
\centering
\cmamefig{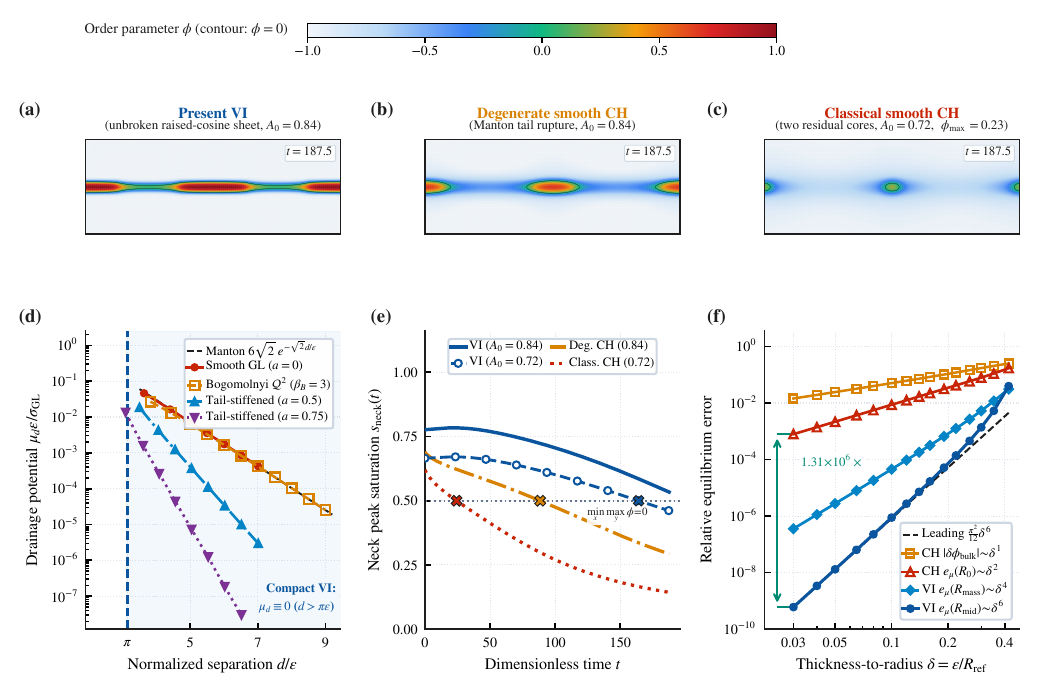}{width=\textwidth}
\caption{Thin-film survival, tail-overlap chemical potential, and three-dimensional spherical Young--Laplace convergence across smooth single-scalar phase-field models and the double-obstacle variational inequality (\Cref{sec:continuum_vi}). (a--c)~Two-dimensional sub-width planar-sheet evolutions in the central $128\times 48$ strip of a $128\times 64$ periodic box under identical extensional pulse forcing ($\epsilon=1.45\Delta x$, $t\le 33.75$, normalized initial amplitude $A_0 = 2h_0/(\pi\epsilon)$): (a)~double-obstacle variational inequality (labelled \texttt{Present VI}, unbroken raised-cosine sheet at $A_0 = 0.84$); (b)~degenerate smooth CH (tail-suction rupture into isolated droplets at $A_0=0.84$); (c)~classical smooth CH (bulk dissolution losing the $\phi=0$ interface at $A_0=0.72$). (d)~One-dimensional tail-overlap drainage chemical potential $\mu_d\epsilon/\sigma_{\mathrm{GL}}$ ($\sigma_{\mathrm{GL}}=\sigma$) versus normalized film separation $d/\epsilon$ ($d \equiv h_f$, $\Delta x = 0.1\epsilon$): Bogomolnyi penalties ($\beta_B=3$, \Cref{thm:nogo}(b)) preserve the leading Manton exponential $6\sqrt{2}\,e^{-\sqrt{2}d/\epsilon}$, whereas smooth tail stiffening ($a\in\{0.5,0.75\}$, \Cref{thm:nogo}(c)) steepens decay only by sacrificing bulk ellipticity toward the compact cutoff $d=\pi\epsilon$ ($\mu_d \equiv 0$ for $d>\pi\epsilon$). (e)~Neck peak saturation $s_{\mathrm{neck}}(t)=\min_x\max_y\frac{1+\phi}{2}$ versus time $t$ ($A_0\in\{0.72,0.84\}$; crosses mark the $\phi=0$ rupture threshold $s_{\mathrm{neck}}=0.50$, whereas the variational inequality labelled \texttt{VI} at $A_0=0.84$ stays unbroken above $0.53$). (f)~Three-dimensional spherical Young--Laplace errors across $\delta=\epsilon/R_{\mathrm{mid}}\in[0.03,0.42]$ ($R_{\mathrm{mid}}=\frac{1}{2}(r_-+r_+)$ and $R_{\mathrm{mass}}$ defined in \Cref{subsec:asymptotics}): smooth CH incurs $\mathcal{O}(\delta^1)$ bulk shifts and $\mathcal{O}(\delta^2)$ chemical-potential error, whereas the double-obstacle VI exhibits sixth-order superconvergence $e_\mu(R_{\mathrm{mid}})\simeq\frac{\pi^2}{12}\delta^6$ (\Cref{thm:asymptotics}(b)), lowering $e_\mu(R_{\mathrm{mid}})$ by $1.31\times 10^6\times$ at $\delta=0.03$.}
\label{fig:duality_nogo}
\end{figure}

Within the smooth functional classes covered by \Cref{thm:nogo}(a)--(c) and illustrated by the smooth baseline curves in \Cref{fig:duality_nogo}(b--e), a local phase-field free energy cannot simultaneously maintain $C^2$ smoothness of its bulk potential $W(\phi)$, evolve a single scalar order parameter $\phi(\bm{x},t)$ without auxiliary interfacial-scale fields, and enforce identically zero tail-overlap interaction ($\Pi(h_f)\equiv 0$) beyond a finite separation $h_f > h_c$. Promoting the local interface half-width to a second evolving scalar field relaxes the single-scalar constraint at the expense of solving an additional parabolic or elliptic PDE and managing cross-coupling stiffness across large density ratios; that multi-field route is not pursued here.

Relaxing $C^2$ potential smoothness instead resolves this conflict within a single scalar order parameter. Rather than stopping at the degenerate $C^{0,1}$ kink limit $a\to 1^-$ of \Cref{thm:nogo}(c), where bulk ellipticity vanishes and contact slopes remain discontinuous, we pass directly to the proper, convex, lower-semicontinuous Blowey--Elliott double-obstacle indicator functional~\cite{Oono1988,Blowey1991,Blowey1992,Hintermuller2002}, whose equilibrium profiles attain $C^{1,1}$ free-boundary regularity with exact tangency $\phi'(\pm r_\pm)=0$, confining the interfacial layer to compact support while retaining constant interior gradient ellipticity.


\section{Continuum variational-inequality formulation and matched asymptotics}\label{sec:continuum_vi}\label{sec:vi_wb}\label{sec:dlvo_rip}

\Cref{thm:nogo} leaves one structural property open to change within a single-scalar variational model: the smoothness of the bulk potential. Building on this observation, we formulate the complete continuum two-phase system in four steps, coupling the deep-quench double-obstacle variational inequality with frame-indifferent variable-density hydrodynamics and ridge-safe normal mobility (\Cref{subsec:gov_vi}), establishing its sharp-interface limit, three-dimensional spherical superconvergence, and planar sub-width film equilibria (\Cref{subsec:asymptotics}), and deriving orthogonal continuum closures for screened Yukawa--DLVO disjoining repulsion (\Cref{subsec:dlvo}) and boundary-layer-renormalized Ostwald ripening (\Cref{subsec:ripening}).

\subsection{Double-obstacle free energy, governing equations, and ridge-safe normal mobility}\label{subsec:gov_vi}

Let $\mathcal{K} \coloneqq \{\psi \in H^1(\Omega) : |\psi(\bm{x})| \le 1 \text{ a.e.\ in }\Omega\}$ denote the closed convex admissible set of bounded phase states. Following Oono and Puri~\cite{Oono1988} and Blowey and Elliott~\cite{Blowey1991,Blowey1992} in the zero-temperature (\emph{deep-quench}) limit, we replace the smooth quartic potential in \Cref{eq:F_GL} by the quadratic concave obstacle potential paired with the convex indicator functional $I_{[-1,1]}$:
\begin{equation}\label{eq:F_DO}
\mathcal{F}_{\mathrm{DO}}[\phi] = \int_\Omega \gamma_{\mathrm{DO}}\left[\frac{\epsilon}{2}|\nabla\phi|^2 + \frac{1-\phi^2}{2\epsilon}\right]\dd\bm{x} + \int_\Omega I_{[-1,1]}(\phi)\,\dd\bm{x}, \qquad \gamma_{\mathrm{DO}} = \frac{2\sigma}{\pi},
\end{equation}
where $I_{[-1,1]}(\phi)=0$ for $|\phi|\le 1$ and $+\infty$ otherwise. Across a planar equilibrium interface $\phi_0(z) = \sin(z/\epsilon)$ for $|z|\le \frac{\pi}{2}\epsilon$ and $\phi_0(z)=\operatorname{sgn}(z)$ for $|z|>\frac{\pi}{2}\epsilon$, equipartition $\frac{\epsilon}{2}|\phi_0'|^2 = \frac{1-\phi_0^2}{2\epsilon} = \frac{1}{2\epsilon}\cos^2(z/\epsilon)$ recovers the exact physical surface tension $\gamma_{\mathrm{DO}}\int_{-\pi\epsilon/2}^{\pi\epsilon/2}\epsilon^{-1}\cos^2(z/\epsilon)\,\dd z = \sigma$.

Because $I_{[-1,1]}$ is proper, convex, and lower semicontinuous, the Fr{\'e}chet derivative of $\mathcal{F}_{\mathrm{DO}}$ generalizes to the multivalued subdifferential inclusion
\begin{equation}\label{eq:mu_subdiff}
\mu \in \frac{\delta\mathcal{F}_{\mathrm{DO}}}{\delta\phi}(\phi) = \mu_{\mathrm{reg}}(\phi) + \partial I_{[-1,1]}(\phi), \qquad \mu_{\mathrm{reg}}(\phi) \coloneqq \gamma_{\mathrm{DO}}\left(-\epsilon\Delta\phi - \frac{\phi}{\epsilon}\right),
\end{equation}
with normal cone $\partial I_{[-1,1]}(-1) = (-\infty,0]$, $\partial I_{[-1,1]}(\phi)=\{0\}$ for $|\phi|<1$, and $\partial I_{[-1,1]}(+1) = [0,+\infty)$. Equivalently, $\mu$ and $\phi\in\mathcal{K}$ satisfy the pointwise Karush--Kuhn--Tucker (KKT) complementarity conditions and weak variational inequality
\begin{equation}\label{eq:kkt_complementarity}
|\phi| \le 1, \qquad (\mu - \mu_{\mathrm{reg}})_{\pm}\,\bigl(1 \mp \phi\bigr) = 0, \qquad \int_\Omega (\mu_{\mathrm{reg}} - \mu)(\psi - \phi)\,\dd\bm{x} \ge 0 \quad \forall\,\psi\in\mathcal{K}.
\end{equation}
Inside the open interfacial transition band $\Omega_{\mathrm{int}}(t) \coloneqq \{\bm{x}\in\Omega : |\phi(\bm{x},t)| < 1\}$, the subdifferential vanishes and $\mu = \mu_{\mathrm{reg}}$ holds classically. In the saturated bulk phases $\phi=\pm 1$, the contact reaction $\xi \coloneqq \mu - \mu_{\mathrm{reg}} \in \partial I_{[-1,1]}(\pm 1)$ absorbs ambient chemical-potential offsets without displacing $\phi$ from $\pm 1$.

We couple \Cref{eq:F_DO,eq:mu_subdiff} to an incompressible Newtonian fluid through the volume-fraction density and viscosity laws
\begin{equation}\label{eq:rho_eta_laws}
\rho(\phi) = \rho_+ s(\phi) + \rho_-\bigl(1 - s(\phi)\bigr), \qquad \eta(\phi) = \eta_+ s(\phi) + \eta_-\bigl(1 - s(\phi)\bigr), \qquad s(\phi) = \frac{1+\phi}{2}.
\end{equation}
Across high-viscosity-contrast gas--liquid collars ($\eta_-/\eta_+ \gg 1$), the arithmetic viscosity in \Cref{eq:rho_eta_laws} may be replaced by the mechanical-jump-consistent harmonic shear viscosity $\eta_{\mathrm{harm}}(\phi) = [s(\phi)/\eta_+ + (1-s(\phi))/\eta_-]^{-1} > 0$~\cite{Reder2024_CMAME} to preserve internal gas circulation (with regime assignments summarized in \Cref{app:params}). Because $\phi\in\mathcal{K}$ enforces $s(\phi)\in[0,1]$ everywhere, both $\rho(\phi)$ and $\eta(\phi)$ remain strictly positive at density and viscosity contrasts exceeding $10^3$ without artificial property clamping. Frame-indifferent Abels--Garcke--Gr{\"u}n (AGG)~\cite{Abels2012} momentum and phase coupling then yields the governing continuum system
\begin{subequations}\label{eq:gov_agg_do}
\begin{align}
\partial_t(\rho\bm{u}) + \nabla\cdot\bigl((\rho\bm{u} + \bm{J}_\rho)\otimes\bm{u}\bigr) + \nabla p &= \nabla\cdot\bigl(2\eta(\phi)\bm{D}(\bm{u})\bigr) + \bm{f}_{\mathrm{cap}} + \bm{f}_{\mathrm{DLVO}} + \rho(\phi)\bm{g}, \label{eq:gov_mom}\\
\nabla\cdot\bm{u} &= 0, \label{eq:gov_div}\\
\partial_t\phi + \nabla\cdot(\phi\bm{u}) &= -\nabla\cdot\bm{J}_\phi + q_\phi, \qquad \bm{J}_\rho = \frac{\rho_+ - \rho_-}{2}\,\bm{J}_\phi, \label{eq:gov_phi}
\end{align}
\end{subequations}
where $\bm{D}(\bm{u})=\frac{1}{2}(\nabla\bm{u}+\nabla\bm{u}^\top)$ is the rate-of-deformation tensor, $\bm{g}$ is gravitational acceleration, $\bm{f}_{\mathrm{cap}} = -\phi\nabla\mu$ is the reversible capillary force, and $\bm{f}_{\mathrm{DLVO}}$ (controlled by disjoining energy scale $A_{\mathrm{rep}}\ge 0$) and $q_\phi$ (controlled by dimensionless carrier solubility number $\mathcal{S}_0\ge 0$) denote the screened disjoining force and interfacial ripening source derived in \Cref{subsec:dlvo,subsec:ripening} (both vanishing identically in baseline immiscible flows $A_{\mathrm{rep}}=\mathcal{S}_0=0$). Including the relative diffusive mass flux $\bm{J}_\rho$ in \Cref{eq:gov_mom} maintains exact consistency between total mass continuity, $\partial_t\rho + \nabla\cdot(\rho\bm{u} + \bm{J}_\rho) = \frac{1}{2}(\rho_+-\rho_-)q_\phi$, and kinetic-energy transport across large density ratios~\cite{Abels2012,Garcke2016_APNUM,Khanwale2020_JCP,TenEikelder2024_JCP}.

Closing the relaxation flux $\bm{J}_\phi = -\bm{\mathcal{M}}_{\mathrm{ONRO}}(\phi,\nabla\phi,\bm{D}(\bm{u}))\nabla\mu_{\mathrm{tot}}$ (with $\mu_{\mathrm{tot}}=\mu + \mu_{\mathrm{DLVO}}$) must avoid both fourth-order tangential Mullins surface diffusion (\Cref{subsec:siggia_duality}) and medial-axis axial drainage on sub-width thin films. A naive unit-normal projector $\bm{P}_\perp^{\mathrm{naive}} = (\nabla\phi\otimes\nabla\phi)/(|\nabla\phi|^2+\delta_{\mathrm{reg}}^2)$ becomes singular along the symmetry plane $y=0$ of any weakly tapered sub-width sheet ($|\phi(x,0)| < 1$): there $\partial_y\phi(x,0)=0$ while axial taper gives $\partial_x\phi(x,0)\neq 0$, rotating $\nabla\phi/|\nabla\phi|$ by $90^\circ$ into the axial direction ($P_{xx}^{\mathrm{naive}}|_{y=0}\to 1$) and draining mass along the sheet. We eliminate this singularity through an orthonormal ridge-safe operator (ONRO) for normal projection normalized by the double-obstacle equipartition invariant:
\begin{equation}\label{eq:P_ONRO}
\bm{P}_\perp^{\mathrm{ONRO}}(\phi,\nabla\phi) \coloneqq \frac{2\epsilon^2\,\nabla\phi\otimes\nabla\phi}{\epsilon^2|\nabla\phi|^2 + (1-\phi^2)_+ + \delta_{\mathrm{reg}}^2},
\end{equation}
where $\delta_{\mathrm{reg}} > 0$ is a small dimensionless regularization constant (fixed at $\delta_{\mathrm{reg}}^2 = 10^{-12} \ll (\partial_x h)^2$; \Cref{app:params}). On any resolved single interface where $\epsilon^2|\nabla\phi|^2 = 1-\phi^2$, the denominator of \Cref{eq:P_ONRO} equals $2\epsilon^2|\nabla\phi|^2 + \delta_{\mathrm{reg}}^2$, so the factor of $2$ in the numerator makes $\bm{P}_\perp^{\mathrm{ONRO}} = \bm{n}\otimes\bm{n} + \mathcal{O}(\delta_{\mathrm{reg}}^2)$ a unit-trace rank-one normal projector satisfying $\bm{P}_\perp^{\mathrm{ONRO}}\bm{\tau}=\bm{0}$ for every tangent vector $\bm{\tau}\perp\nabla\phi$. On an unsaturated sub-width ridge where $|\nabla\phi|\to 0$ while $(1-\phi^2)_+ = \mathcal{O}(1) > 0$, by contrast, the denominator remains bounded below by $(1-\phi^2)_+$ and suppresses $\bm{P}_\perp^{\mathrm{ONRO}}$ quadratically in the taper slope. Pairing \Cref{eq:P_ONRO} with the strain-adaptive bulk-lockout scalar mobility $M(\phi,|\bm{D}(\bm{u})|) \coloneqq (1-\phi^2)_+\bigl(\gamma_0\epsilon^2 U_{\mathrm{ref}} + \beta\epsilon^3|\bm{D}(\bm{u})|\bigr)/\gamma_{\mathrm{DO}}$ ($\gamma_0 > 0$, $\beta \ge 0$, reference speed $U_{\mathrm{ref}}$) and a $C^1$ medial-axis sensor $\chi_{\mathrm{ma}} = H_3(\mathcal{D}_{\mathrm{ma}}; s_{\mathrm{lo}}, s_{\mathrm{hi}}) \in [0,1]$ built from the cubic Hermite ramp $H_3(s;a,b)=3\hat{s}^2-2\hat{s}^3$ ($\hat{s}=\operatorname{clip}((s-a)/(b-a),0,1)$) defines the symmetric positive semi-definite mobility tensor
\begin{equation}\label{eq:M_ONRO}
\bm{\mathcal{M}}_{\mathrm{ONRO}}(\phi,\nabla\phi,\bm{D}(\bm{u})) \coloneqq M\bigl(\phi,|\bm{D}(\bm{u})|\bigr)\bigl(1 - \chi_{\mathrm{ma}}(\phi,\nabla\phi)\bigr)\bm{P}_\perp^{\mathrm{ONRO}}(\phi,\nabla\phi), \qquad \mathcal{D}_{\mathrm{ma}} \coloneqq \frac{\bigl[(1-\phi^2)_+ - \epsilon^2|\nabla\phi|^2\bigr]_+}{(1-\phi^2)_+ + \delta_{\mathrm{ma}}},
\end{equation}
where $\delta_{\mathrm{ma}} > 0$ and $0 < s_{\mathrm{lo}} < s_{\mathrm{hi}} < 1$ are dimensionless ridge-sensor constants (fixed at $\delta_{\mathrm{ma}}=0.18$ and $(s_{\mathrm{lo}},s_{\mathrm{hi}})=(0.30,0.62)$ in 3D, with $\chi_{\mathrm{ma}}\equiv 0$ in 2D and axisymmetric runs; \Cref{sec:discretization,app:params}). Differentiating the total energy $\mathcal{E}_{\mathrm{tot}}[\bm{u},\phi] = \int_\Omega \frac{1}{2}\rho(\phi)|\bm{u}|^2\,\dd\bm{x} + \mathcal{F}_{\mathrm{DO}}[\phi]$ at $A_{\mathrm{rep}}=\mathcal{S}_0=0$ and $\bm{g}=\bm{0}$ under periodic or no-slip, zero-flux boundary conditions ($\bm{u}=\bm{0}$, $\partial_n\phi = \bm{n}\cdot\bm{J}_\phi = 0$ on $\partial\Omega$) yields the exact Clausius--Duhem dissipation law
\begin{equation}\label{eq:clausius_duhem}
\frac{\dd\mathcal{E}_{\mathrm{tot}}}{\dd t} = -\int_\Omega 2\eta(\phi)|\bm{D}(\bm{u})|^2\,\dd\bm{x} - \int_\Omega \nabla\mu\cdot\bm{\mathcal{M}}_{\mathrm{ONRO}}(\phi,\nabla\phi,\bm{D}(\bm{u}))\nabla\mu\,\dd\bm{x} \le 0.
\end{equation}

\subsection{Sharp-interface limit, spherical superconvergence, and sub-width thin-film equilibria}\label{subsec:asymptotics}

We next determine the sharp-interface limit and finite-thickness equilibrium profiles of \Cref{eq:F_DO,eq:mu_subdiff,eq:gov_agg_do,eq:P_ONRO}, establishing exact outer bulk isolation, sixth-order Young--Laplace superconvergence for three-dimensional spheres, and non-evaporating sub-width planar thin films.

\begin{theorem}[Matched asymptotics, sixth-order 3D spherical superconvergence, and planar sub-width equilibria]\label{thm:asymptotics}
Consider the deep-quench double-obstacle variational inequality \Cref{eq:F_DO,eq:mu_subdiff,eq:gov_agg_do} with mobility \Cref{eq:M_ONRO} at $\mathcal{S}_0=0$ and $A_{\mathrm{rep}}=0$. Three properties hold:
\begin{enumerate}
\item[(a)] \emph{(Sharp-interface limit and exact bulk isolation).} For any smooth moving interface $\Gamma(t)$ with principal curvatures $|\kappa_i|\epsilon \ll 1$ separated from other interface segments by gap $H > \pi\epsilon + \mathcal{O}(\epsilon^3\kappa^2)$, the transition layer has compact half-widths $r_\pm(\bm{x}_s) = \frac{\pi}{2}\epsilon + \mathcal{O}(\epsilon^3\kappa^2)$, outside which $\phi \equiv \pm 1$ and $\bm{J}_\phi \equiv \bm{0}$ hold \emph{identically} to all algebraic orders in $\epsilon$. As $\epsilon \to 0^+$, $(\bm{u},p,\Gamma(t))$ converges formally to the classical sharp-interface two-phase incompressible Navier--Stokes equations with pure advective interface motion $V_n = \bm{u}\cdot\bm{n}$ (zero fourth-order tangential surface diffusion) and Young--Laplace traction jump $\Jump{-p\bm{I} + 2\eta\bm{D}(\bm{u})}\bm{n} = -\sigma\kappa\bm{n}$.
\item[(b)] \emph{(Dimensional superconvergence of radial equilibria).} For a stationary radial droplet in $\R^d$ with inner and outer compact obstacle radii $0 < r_- < r_+$, define the collar midpoint radius $R_{\mathrm{mid}} \coloneqq \frac{1}{2}(r_- + r_+)$, zero-level radius $R_{\phi=0}$ ($\phi(R_{\phi=0})=0$), mass radius $R_{\mathrm{mass}}$, and thickness ratio $\delta \coloneqq \epsilon/R_{\mathrm{mid}}$:
\begin{itemize}
\item In $d=2$ (circular cylinder), the exact Bessel $J_0/Y_0$ free-boundary solution satisfies
\begin{equation}\label{eq:yl_2d_asym}
2\mu_{\mathrm{eq}}^{\mathrm{2D}} = \frac{\sigma}{R_{\mathrm{mid}}}\left[1 + c_2^{(2)}\delta^2 + \mathcal{O}(\delta^4)\right], \qquad c_2^{(2)} = \frac{\pi^2 - 6}{16} \approx 0.241850.
\end{equation}
\item In $d=3$ (sphere), let $\xi_1 \approx 4.49340946$ be the smallest positive root of $\tan\xi = \xi$. For every droplet above the exact core-detachment threshold $R_{\mathrm{mid}} > R_{\mathrm{det,mid}}^{\mathrm{3D}} \coloneqq \frac{1}{2}\xi_1\epsilon \approx 2.246705\,\epsilon$ (corresponding to $R_{\phi=0} > 2.145116\,\epsilon$ and $R_{\mathrm{mass}} > 2.529844\,\epsilon$), both the $\mathcal{O}(\delta^2)$ and $\mathcal{O}(\delta^4)$ Young--Laplace errors vanish identically relative to $R_{\mathrm{mid}}$:
\begin{align}
2\mu_{\mathrm{eq}}^{\mathrm{3D}} &= \frac{2\sigma}{R_{\mathrm{mid}}}\left[1 - \frac{\pi^2}{12}\delta^6 - \frac{\pi^4}{16}\delta^8 + \mathcal{O}(\delta^{10})\right], \label{eq:yl_3d_superconv}\\
\frac{R_{\mathrm{mass}}^{\mathrm{3D}}}{R_{\mathrm{mid}}} &= 1 + \frac{\pi^2(12 - \pi^2)}{48}\delta^4 + \frac{\pi^2}{4}\delta^6 + \mathcal{O}(\delta^8), \qquad \frac{R_{\phi=0}^{\mathrm{3D}}}{R_{\mathrm{mid}}} = 1 - \frac{\pi - 2}{2}\delta^2 + \mathcal{O}(\delta^4), \label{eq:rmass_3d_superconv}
\end{align}
where $R_{\mathrm{mass}}^{\mathrm{3D}} = \bigl(\frac{3}{4\pi}\int_\Omega s(\phi)\,\dd\bm{x}\bigr)^{1/3}$, yielding fourth-order Laplace accuracy $\mathcal{O}(\delta^4)$ relative to $R_{\mathrm{mass}}^{\mathrm{3D}}$. For $R_{\mathrm{mass}}^{\mathrm{3D}} < R_{\mathrm{det,mass}}^{\mathrm{3D}}$, the inner core detaches from $+1$ onto a non-evaporating spherical Bessel profile $\phi_{\mathrm{sub}}^{\mathrm{3D}}(r) = -1 + C_{\mathrm{sph}}\bigl(j_0(r/\epsilon) - \cos\xi_1\bigr)$ with spherical Bessel function $j_0(\xi)=\sin\xi/\xi$, locked outer support $r_+ = \xi_1\epsilon$, and mass-determined amplitude $C_{\mathrm{sph}}\in\bigl(0, 2/(1-\cos\xi_1)\bigr)$.
\end{itemize}
\item[(c)] \emph{(Non-evaporating sub-width sheets and $\mathcal{O}((\partial_x h)^2)$ suppression of $90^\circ$ normal flipping).} Below the single-interface compact support width $\pi\epsilon$, a uniform planar liquid sheet or bubble skirt of mass-equivalent thickness $2h \in (0,\pi\epsilon)$ admits the exact one-dimensional cross-sectional raised-cosine obstacle equilibrium ($A = \frac{2h}{\pi\epsilon} \in (0,1)$):
\begin{equation}\label{eq:subwidth_rc}
\phi_{\mathrm{rc}}(y;h) = \begin{cases} -1 + A\bigl(1 + \cos(y/\epsilon)\bigr), & |y| \le \pi\epsilon, \\ -1, & |y| > \pi\epsilon, \end{cases} \qquad \mu_{\mathrm{rc}}(h) = \frac{2\sigma}{\pi\epsilon}(1 - A).
\end{equation}
Along the medial ridge $y=0$ of any weakly tapered sub-width sheet $\phi(x,y)=\phi_{\mathrm{rc}}(y;h(x))$ with $(\partial_x h)^2 \ll A(x)(1-A(x))$, the naive normal projector gives $P_{xx}^{\mathrm{naive}}|_{y=0} = 1 + \mathcal{O}(\delta_{\mathrm{reg}}^2)$, whereas \Cref{eq:P_ONRO} yields
\begin{equation}\label{eq:onro_pxx_bound}
P_{xx}^{\mathrm{ONRO}}\big|_{y=0} = \frac{8\,(\partial_x h)^2}{\pi^2 A(x)\bigl(1 - A(x)\bigr)} + \mathcal{O}\bigl((\partial_x h)^4, \delta_{\mathrm{reg}}^2\bigr).
\end{equation}
\end{enumerate}
\end{theorem}

\begin{proof}
Complete derivations are given in \Cref{app:asymptotics}. Physically, the sixth-order spherical superconvergence \Cref{eq:yl_3d_superconv} arises from the radial Liouville transformation $u(r) = r(\phi(r) + \tilde{\mu})$: in $d=3$, the centrifugal curvature potential $\frac{(d-1)(d-3)}{4r^2}$ in the radial Helmholtz operator $\Delta_r + \epsilon^{-2}$ vanishes identically. Consequently, $\phi(r)$ solves the constant-coefficient harmonic equation $u'' + \epsilon^{-2}u = 0$, reducing the inner and outer free-boundary contact conditions at $r_\pm$ to trigonometric identities whose $\mathcal{O}(\delta^2)$ and $\mathcal{O}(\delta^4)$ series coefficients cancel algebraically.
\end{proof}

These analytical properties explain the contrast observed in \Cref{fig:duality_nogo}(a,e,f) and summarized in \Cref{tab:arch_compare}: under identical extensional forcing, the double-obstacle sheet at $A_0=0.84$ relaxes onto an unbroken raised-cosine ligament without bulk loss or Manton snap-off, while in three dimensions \Cref{eq:yl_3d_superconv} lowers the spherical Young--Laplace chemical-potential error at $\delta=0.03$ by six orders of magnitude relative to smooth CH.

\begin{remark}[Volume-fraction diagnostics on sub-width thin films]\label{rem:subwidth_diag}
Equation~\eqref{eq:subwidth_rc} carries a direct consequence for thin-film post-processing. Whenever a sub-width liquid sheet or trailing skirt thins into the range $2h \in (0, \frac{\pi}{2}\epsilon)$ ($A < \frac{1}{2}$), its centerline peak satisfies $\phi_{\mathrm{rc}}(0) = 2A - 1 \in (-1,0)$, so the $\phi=0$ zero-level contour disappears even though the physical sheet thickness $2h = \int_{-\pi\epsilon}^{\pi\epsilon} s(\phi_{\mathrm{rc}})\,\dd y > 0$ remains $100\%$ conserved inside $|y|\le \pi\epsilon$. Sheet-thickness and mass diagnostics must therefore integrate the conserved volume fraction $s(\phi)=\frac{1+\phi}{2}$ rather than thresholding $\phi > 0$.
\end{remark}

\begin{table}[t!]
\centering
\caption{Mathematical comparison of Eulerian diffuse-interface formulations across free-energy dissipation, bulk chemical-potential isolation, thin-film tail interaction, three-dimensional spherical Young--Laplace scaling ($\delta = \epsilon/R_{\mathrm{mid}}$), and sub-width sheet equilibria (\Cref{sec:duality_nogo,sec:continuum_vi}). Within the single-scalar formulations covered by \Cref{thm:nogo,thm:asymptotics}, replacing smooth polynomial wells with the double-obstacle indicator $I_{[-1,1]}(\phi)$ eliminates outer bulk shifts and far-field tail suction identically while retaining free-energy decay and sixth-order spherical midpoint accuracy.}
\label{tab:arch_compare}
\small
\setlength{\tabcolsep}{4.2pt}
\renewcommand{\arraystretch}{1.16}
\resizebox{\textwidth}{!}{%
\begin{tabular}{@{}lcccccc@{}}
\toprule
Formulation & Scalar fields & Variational $\dd\mathcal{E}/\dd t\le 0$ & Bulk shift $\delta\phi_{\mathrm{bulk}}$ ($\mathcal{S}_0=0$) & Tail suction $\Pi(h_f)$ ($h_f>\pi\epsilon$) & 3D sphere YL error & Sub-width $2h<\pi\epsilon$ \\
\midrule
Classical smooth CH~\cite{Cahn1958,Abels2012} & $1$ ($\phi$) & Yes & $\mathcal{O}(\epsilon/R)$ & $-12\sqrt{2}\frac{\sigma}{\epsilon}e^{-\sqrt{2}h_f/\epsilon}$ & $\mathcal{O}(\delta^2)$ & Evaporates \\
Degenerate smooth CH~\cite{Gunnarsson2026_JCP} & $1$ ($\phi$) & Yes & Suppressed & $-12\sqrt{2}\frac{\sigma}{\epsilon}e^{-\sqrt{2}h_f/\epsilon}$ & $\mathcal{O}(\delta^2)$ & Mullins / snaps \\
Bogomolnyi / FCH~\cite{Gavish2011} & $1$ ($\phi$) & Yes ($\beta_B>0$) & $\mathcal{O}(\epsilon/R)$ & Unchanged $\mathcal{O}(e^{-\sqrt{2}h_f/\epsilon})$ & $\mathcal{O}(\delta^2)$ & Helfrich / pearls \\
Smooth tail-stiffened (\Cref{eq:F_tail_stiff}) & $1$ ($\phi$) & Yes ($a<1$) & $\mathcal{O}\bigl(\epsilon/[(1+a)R]\bigr)$ & $\propto e^{-\lambda_{\mathrm{eff}}(a)h_f/\epsilon}$ & $\mathcal{O}(\delta^2)$ & PN wall $\sim 4\Delta x$ \\
Local conservative AC~\cite{Chiu2011,Jain2022} & $1$ ($\phi$) & \textbf{No} ($\Delta\mathcal{E}>0$) & Exp.\ tail & Normal bead split & $\mathcal{O}(\delta^2)$ & Fragments \\
Counter-diffusive mcAC~\cite{Peng2026} & $1$ ($\phi$) & \textbf{No} ($\Delta\mathcal{E}>0$) & Exp.\ tail / nucl. & Outward $\dot{R}(0)>0$ & Sharpens / drifts & Fragments \\
Two-field dynamic-width model & $2$ ($\phi,\ell$) & Yes & Exp.\ tail & Compressed $\ell(\bm{x})$ & $\mathcal{O}(\delta^2)$ & Extra PDE cost \\
\textbf{Present double-obstacle VI} & $\mathbf{1}$ ($\bm{\phi}$) & \textbf{Yes (exact)} & $\bm{0}$ \textbf{(exact)} & $\bm{0}$ \textbf{(exact)} & $\bm{\mathcal{O}(\delta^6)}$ ($\bm{R_{\mathrm{mid}}}$) & \textbf{Exact cosine} \\
\bottomrule
\end{tabular}%
}
\end{table}

\subsection{Legendre-dual screened disjoining energy and pre-contact barrier efficiency}\label{subsec:dlvo}

In emulsions, microfluidic dispersions, and colloidal suspensions, baseline immiscible hydrodynamics is often accompanied by short-range steric or electrostatic DLVO disjoining barriers that stabilize approaching droplets across an intact carrier film~\cite{Bibette1992,LealCalderon1994}. Adding an empirical non-variational body force $\bm{f}_{\mathrm{disj}} = -\Pi_{\mathrm{emp}}(h)\nabla s(\phi)$ directly to \Cref{eq:gov_mom} breaks Lyapunov energy decay during droplet rebound and violates momentum conservation on unequal droplets.

Even within an auxiliary-field variational formulation, the sign of the variational principle requires care. When interfaces with screening length $\ell_d > 0$ interact via a Yukawa kernel, a seemingly natural local functional is the infimum penalty
\begin{equation}\label{eq:F_dlvo_naive}
\mathcal{F}_{\mathrm{naive}}[\phi] \coloneqq \inf_{\psi\in H^1(\Omega)} \frac{A_{\mathrm{rep}}}{2}\int_\Omega \Bigl[\ell_d^2|\nabla\psi|^2 + \bigl(\psi - s(\phi)\bigr)^2\Bigr]\dd\bm{x}, \qquad A_{\mathrm{rep}} > 0,
\end{equation}
where $A_{\mathrm{rep}}$ is the disjoining energy scale. Although \Cref{eq:F_dlvo_naive} is nonnegative, its minimizer solves $(I - \ell_d^2\Delta)\psi = s(\phi)$ under periodic or homogeneous Neumann boundary conditions, whence integration by parts gives $\mathcal{F}_{\mathrm{naive}}[\phi] = \frac{A_{\mathrm{rep}}}{2}\int_\Omega s(\phi)\bigl(s(\phi)-\psi\bigr)\,\dd\bm{x}$. For two disjoint droplets $s = s_1 + s_2$ ($\operatorname{supp}s_1 \cap \operatorname{supp}s_2 = \emptyset$, so $\int_\Omega s_1 s_2\,\dd\bm{x}=0$), the cross-interaction energy of \Cref{eq:F_dlvo_naive} is strictly negative:
\begin{equation}\label{eq:naive_cross_attr}
\mathcal{E}_{\mathrm{int}}^{\mathrm{naive}}(h) = -A_{\mathrm{rep}}\int_\Omega s_1\,(I - \ell_d^2\Delta)^{-1}s_2\,\dd\bm{x} < 0,
\end{equation}
becoming more negative as the surface-to-surface separation $h$ decreases. Minimizing \Cref{eq:F_dlvo_naive} therefore produces spurious inter-droplet attraction that accelerates coalescence rather than preventing it.

To obtain unconditionally dissipative short-range repulsion, we pass to the convex Legendre--Fenchel dual maximization principle on $H^1(\Omega)$:
\begin{equation}\label{eq:F_DLVO_cont}
\mathcal{F}_{\mathrm{DLVO}}[\phi] \coloneqq \sup_{\psi\in H^1(\Omega)} \int_\Omega \left[A_{\mathrm{rep}}\,s(\phi)\psi - \frac{A_{\mathrm{rep}}}{2}\Bigl(\psi^2 + \ell_d^2|\nabla\psi|^2\Bigr)\right]\dd\bm{x} = \frac{A_{\mathrm{rep}}}{2}\int_\Omega s(\phi)\,(I - \ell_d^2\Delta)^{-1}s(\phi)\,\dd\bm{x}.
\end{equation}
Letting $\psi_d = (I - \ell_d^2\Delta)^{-1}s(\phi) \in [0,1]$ denote the unique screened potential, the first variation of $\mathcal{F}_{\mathrm{DLVO}}$ with respect to $\phi$ is $\mu_{\mathrm{DLVO}} = \delta\mathcal{F}_{\mathrm{DLVO}}/\delta\phi = \frac{1}{2}A_{\mathrm{rep}}\psi_d$, generating the conservative Korteweg disjoining force $\bm{f}_{\mathrm{DLVO}} = \mu_{\mathrm{DLVO}}\nabla\phi$ in \Cref{eq:gov_mom}. Driving $\bm{J}_\phi = -\bm{\mathcal{M}}_{\mathrm{ONRO}}\nabla(\mu + \mu_{\mathrm{DLVO}})$ at $\mathcal{S}_0=0$ preserves the exact Clausius--Duhem dissipation law \Cref{eq:clausius_duhem} for the augmented total energy $\mathcal{E}_{\mathrm{tot}} + \mathcal{F}_{\mathrm{DLVO}}$.

Once the outer supports of two approaching double-obstacle collars touch at surface-to-surface separation $h = \pi\epsilon$, the intervening carrier film forms the $\phi\mapsto-\phi$ counterpart of the cross-sectional equilibrium \Cref{eq:subwidth_rc}, whose areal energy $e(A)=2\sigma A(2-A)$ exerts a finite-range capillary suction $\dd e/\dd(2h)=\frac{4\sigma}{\pi\epsilon}(1-A)$ for $2h<\pi\epsilon$ while vanishing identically for $h>\pi\epsilon$; static repulsion must therefore arrest approach across $h \ge \pi\epsilon$. Convolving two planar double-obstacle profiles against the screened Green's kernel determines the exact pre-contact barrier efficiency $\eta_\Pi(\alpha)$ relative to the sharp-interface ceiling $A_{\mathrm{rep}}/2$ as a function of $\alpha \coloneqq \ell_d/\epsilon$:
\begin{equation}\label{eq:eta_pi_exact}
\eta_\Pi(\alpha) \coloneqq \frac{\Pi_{\mathrm{DLVO}}(\pi\epsilon)}{A_{\mathrm{rep}}/2} = \left[\frac{\alpha^2}{1+\alpha^2}\,\frac{1 + e^{-\pi/\alpha}}{2}\right]^2 = \begin{cases} \frac{1}{4}\alpha^4 + \mathcal{O}(\alpha^6), & \alpha \to 0^+, \\ 1 - \frac{\pi}{\alpha} + \mathcal{O}(\alpha^{-2}), & \alpha \to \infty. \end{cases}
\end{equation}
Specifically, $\eta_\Pi(\alpha) \ge 25\%$ holds if and only if $\alpha = \ell_d/\epsilon \ge \alpha_{25\%} \approx 2.1073$ ($\eta_\Pi = 33.3\%$ at $\alpha = 2.70$), whereas for under-resolved screening $\alpha < 1$, $\eta_\Pi(\alpha) < 6.80\%$ collapses quartically so capillary meniscus suction overwhelms static repulsion upon collar contact.

\begin{proposition}[Continuum Yukawa--DLVO sign duality and pre-contact barrier efficiency]\label{prop:dlvo_cont}
For any two disjoint droplets $s = s_1 + s_2$ separated by surface-to-surface gap $h > \pi\epsilon$, the Legendre-dual interaction energy of \Cref{eq:F_DLVO_cont} satisfies $\mathcal{E}_{\mathrm{int}}^{\mathrm{DLVO}}(h) = -\mathcal{E}_{\mathrm{int}}^{\mathrm{naive}}(h) = +A_{\mathrm{rep}}\int_\Omega s_1(I-\ell_d^2\Delta)^{-1}s_2\,\dd\bm{x} > 0$. For radially symmetric droplets in $\R^d$ translated along their line of centers, $\partial_h\mathcal{E}_{\mathrm{int}}^{\mathrm{DLVO}}(h) < 0$ holds strictly for all $h > \pi\epsilon$ (up to $\mathcal{O}(e^{-L/\ell_d})$ periodic image tails on $[-L/2,L/2]^d$), and across planar collars at contact onset $h = \pi\epsilon$ the pre-contact disjoining pressure obeys \Cref{eq:eta_pi_exact}.
\end{proposition}

\begin{proof}
Positivity and sign duality follow from the positive Green's kernel of $(I-\ell_d^2\Delta)^{-1}$ and $\mathcal{F}_{\mathrm{naive}}[s] + \mathcal{F}_{\mathrm{DLVO}}[s] = \frac{1}{2}A_{\mathrm{rep}}\int_\Omega s^2\,\dd\bm{x}$; the explicit Green's convolution yielding \Cref{eq:eta_pi_exact} is given in \Cref{app:asymptotics}.
\end{proof}

\begin{remark}[Physical regime demarcation and Rayleigh self-energy threshold]\label{rem:dlvo_scope}
Equation~\eqref{eq:eta_pi_exact} establishes a sharp physical scope boundary for $\mathcal{F}_{\mathrm{DLVO}}[\phi]$. As a conservative equilibrium free energy depending solely on $\phi$, it captures adsorbed steric and electrical double-layer stabilization in liquid--liquid emulsions and microfluidic dispersions whenever the screening length is resolved on the Eulerian grid ($\ell_d/\epsilon \ge 2.11 \implies \eta_\Pi \ge 25\%$), arresting capillary coalescence and preserving multi-droplet satellite necklaces against secondary fusion. Because the volumetric Yukawa self-energy carries an electrostatic self-surface-tension reduction $\Delta\sigma_{\mathrm{DLVO}} = -c_{\mathrm{col}}(\epsilon/\ell_d)\frac{1}{4}A_{\mathrm{rep}}\ell_d$ across a single collar ($c_{\mathrm{col}}\to 1$ as $\epsilon/\ell_d\to 0$, $c_{\mathrm{col}}\approx 0.72$ at $\ell_d/\epsilon=2.2$), spherical droplets remain below Lord Rayleigh's fissility instability whenever $\sigma_{\mathrm{eff}} = \sigma + \Delta\sigma_{\mathrm{DLVO}} > 0$. Conversely, for uncharged sub-millimeter droplets colliding in ambient gas~\cite{Qian1997}, physical van der Waals and steric scales satisfy $\ell_d \sim 1\,\mathrm{nm} \ll \epsilon$, where the quartic law $\eta_\Pi(\alpha) = \frac{1}{4}\alpha^4 + \mathcal{O}(\alpha^6)$ predicts a vanishing static barrier; aerodynamic rebound in that regime is governed instead by rate-dependent viscous Stefan--Reynolds--Knudsen compression of a sub-micron gas film ($\propto -\eta_g\dot{h}_f/h_f^3$).
\end{remark}

\subsection{Boundary-layer-renormalized Robin--Kelvin closure for Ostwald ripening}\label{subsec:ripening}

To complement short-range repulsion when weak molecular solubility is active ($\mathcal{S}_0 > 0$, where the dimensionless solubility number $\mathcal{S}_0$ scales the equilibrium carrier volume solubility $\varphi_\infty$ and vanishes identically with $q_\phi\equiv 0$ in immiscible flows), dissolved monomer supersaturation $u_c(\bm{x},t) = (c - c_\infty)/c_\infty$ in the carrier phase $1-s(\phi)$ obeys the quasi-steady degenerate diffusion--reaction system
\begin{equation}\label{eq:ripening_pde}
-D_m\nabla\cdot\Bigl(\bigl(1 - s(\phi)\bigr)\nabla u_c\Bigr) + K(\kappa)\,\delta_s(\phi)\bigl(u_c - u_K(\kappa)\bigr) = 0, \qquad q_\phi = 2\,\mathcal{S}_0\,K(\kappa)\,\delta_s(\phi)\bigl(u_c - u_K(\kappa)\bigr),
\end{equation}
where $D_m > 0$ is the ambient monomer diffusivity, $u_K(\kappa) = \exp(\frac{1}{2}\ell_K\kappa) - 1$ is the nonlinear Kelvin equilibrium supersaturation in terms of total curvature $\kappa = \frac{d-1}{R}$ ($d\in\{2,3\}$) and capillary length $\ell_K$, $K(\kappa) > 0$ is the interfacial kinetic transfer coefficient, and $\delta_s(\phi)$ is the normalized algebraic surface-area density
\begin{equation}\label{eq:delta_s_alg}
\delta_s(\phi) \coloneqq \frac{8}{3\pi\epsilon}\bigl(1 - \phi^2\bigr)_+^2, \qquad \int_{-\pi\epsilon/2}^{\pi\epsilon/2}\delta_s\bigl(\phi_0(n)\bigr)\,\dd n = \frac{8}{3\pi}\int_{-\pi/2}^{\pi/2}\cos^4\zeta\,\dd\zeta = 1.
\end{equation}
Because $\delta_s(\phi)$ vanishes quadratically as $(1-\phi^2)_+^2 \to 0$ at both obstacle bounds $\phi=\pm 1$, the source $q_\phi$ is strictly localized to the open transition collar $|\phi|<1$.

Recent pore-network~\cite{Mehmani2022_JCP} and variationally consistent phase-field dissolution and precipitation models~\cite{Lamperti2026_JCP} establish thermodynamically consistent frameworks for curvature-driven diffusive exchange and its sharp-interface convergence; in diffuse-interface implementations, a large penalty stiffness $K \gg D_m/\epsilon$ is commonly chosen to enforce $u_c \approx u_K(\kappa)$ across the transition layer~\cite{Lamperti2026_JCP}. Yet on a compact double-obstacle collar as $K\epsilon/D_m \to \infty$, $u_c$ becomes pinned at the outer edge $\zeta = +\frac{\pi}{2}$ of the compact collar ($r_+ = R_0 + \frac{\pi}{2}\epsilon$) rather than at the physical dividing surface $\zeta=0$ ($r=R_0$), shifting the effective droplet radius by $+\frac{\pi}{2}\epsilon$ and introducing an $\mathcal{O}(\epsilon/R)$ flux bias on practical meshes.

We remove this extrapolation bias analytically through matched boundary-layer asymptotics. Measuring the stretched outward normal coordinate $\zeta = (r - R_0)/\epsilon \in (-\frac{\pi}{2},\frac{\pi}{2})$ across the radial equilibrium profile $\phi_0(\zeta) = -\sin\zeta$ (where $\zeta \le -\frac{\pi}{2}$ is the saturated droplet core $\phi=+1$ and $\zeta \ge +\frac{\pi}{2}$ is the pure carrier $\phi=-1$), one has $1 - s(\phi_0(\zeta)) = \frac{1+\sin\zeta}{2}$ and $\delta_s(\phi_0) = \frac{8}{3\pi\epsilon}\cos^4\zeta$. Setting $K(\kappa) = \Lambda(\epsilon\kappa)\,D_m/\epsilon$ and expanding $u_c - u_K(\kappa) = \epsilon\partial_r u_c^{\mathrm{ext}}(R_0)\,W(\zeta)$ reduces the leading-order inner boundary layer to the singular degenerate Sturm--Liouville problem
\begin{equation}\label{eq:sturm_liouville_inner}
-\frac{\dd}{\dd\zeta}\!\left(\frac{1 + \sin\zeta}{2}\frac{\dd W_0}{\dd\zeta}\right) + \frac{8\Lambda_0}{3\pi}\cos^4\zeta\,W_0(\zeta) = 0, \qquad \zeta\in\left(-\frac{\pi}{2},\frac{\pi}{2}\right),
\end{equation}
subject to regularity at the degenerate inner core edge $\zeta \to -\frac{\pi}{2}^+$ (where $1+\sin\zeta \simeq \frac{1}{2}(\zeta+\frac{\pi}{2})^2$) and unit outward flux $\frac{\dd W_0}{\dd\zeta}(\frac{\pi}{2}) = 1$ at $\zeta = +\frac{\pi}{2}$. Matching $W_0(\frac{\pi}{2})$ to the outer harmonic continuation $W^{\mathrm{ext}}(\zeta) = \zeta + b_{\mathrm{ext}}(\Lambda_0)/\epsilon$ and enforcing zero extrapolation shift $b_{\mathrm{ext}}(\Lambda_0) \equiv 0$ ($W_0(\frac{\pi}{2}) = \frac{\pi}{2}$) uniquely determines the universal eigenvalue $\Lambda_\star$, while expanding both the curved diffusion operator and the curved double-obstacle profile (\Cref{eq:yl_2d_asym,eq:yl_3d_superconv}) through $\mathcal{O}(\epsilon^2\kappa^2)$ yields its second-order curvature renormalization:
\begin{equation}\label{eq:lambda_star_betas}
\begin{aligned}
K(\kappa) &= \frac{D_m}{\epsilon}\,\Lambda_\star\max\!\Bigl(0.15,\, 1 + \beta_1\,\tilde{\kappa} + \beta_2^{(d)}\,\tilde{\kappa}^2\Bigr), \qquad \tilde{\kappa} \coloneqq \min(\epsilon\kappa, 0.85), \\
\Lambda_\star &= 3.880110546586, \quad \beta_1 = +0.18435619, \quad \beta_2^{(2)} = -0.54020129, \quad \beta_2^{(3)} = -0.29394123,
\end{aligned}
\end{equation}
where the positivity floor and curvature cap safeguard terminal droplet vanishing ($R\to 0$). Evaluating the curved radial operator on a frozen planar $-\sin\zeta$ ansatz without the Laplace-pressure obstacle shift $-m$ would instead give $(-0.273883, +0.218392, +0.136350)$, showing why the inner expansion must be matched against the true curved obstacle profile.

\begin{proposition}[Third-order boundary-layer-renormalized Robin--Kelvin accuracy]\label{prop:ripening_sl}
With $\Lambda_\star$ and $(\beta_1,\beta_2^{(d)})$ given by \Cref{eq:lambda_star_betas} in the asymptotic regime $\epsilon\kappa \ll 1$, the inner boundary-layer solution of \Cref{eq:ripening_pde} around any droplet of total curvature $\kappa = \frac{d-1}{R}$ ($d\in\{2,3\}$) matches the sharp-interface Gibbs--Thomson boundary condition $u_c^{\mathrm{ext}}(R_{\mathrm{mass}}) = u_K(\kappa)$ with residual error $\mathcal{O}\bigl((\epsilon\kappa)^3\bigr)$, reducing $\mathcal{O}(\epsilon/R)$ outer-collar penalty pinning by two asymptotic orders, while any closed or zero-net-flux outer boundary condition satisfies $\int_\Omega q_\phi\,\dd\bm{x} = 0$.
\end{proposition}

\begin{proof}
Frobenius series selection at the regular singular point $\zeta = -\frac{\pi}{2}$ and curved matched-asymptotic expansions establishing $(\Lambda_\star,\beta_1,\beta_2^{(d)})$ are given in \Cref{app:asymptotics}; integrating \Cref{eq:ripening_pde} over $\Omega$ gives $\int_\Omega q_\phi\,\dd\bm{x} = 2\mathcal{S}_0 D_m\int_{\partial\Omega}(1-s(\phi))\partial_n u_c\,\dd S = 0$.
\end{proof}


\section{Structure-preserving staggered finite-volume discretization}\label{sec:discretization}

Discretizing the non-smooth continuum system of \Cref{sec:continuum_vi} on a fixed Cartesian mesh must preserve four structural invariants that standard finite differences lose: curl-free hydrostatic capillary balance across large density jumps without overdamping dynamic capillary waves, strict phase-bound invariance $\phi_{\bm{i}}\in[-1,1]$ without mass-violating state clipping, positive-definite screened repulsion with zero net self-force, and mass-conserving Ostwald ripening on bounded domains. We construct this discretization on a uniform staggered Marker-and-Cell (MAC) grid in three parts, introducing isotropic grid operators and a complementarity-gated capillary force (\Cref{subsec:exact_wb}), pairing fifth-order upwind reconstruction with multi-pass flux-corrected transport (\Cref{subsec:fct_cisl5}), and closing the discrete disjoining and radiative ripening solves within a four-stage time-stepping cycle (\Cref{subsec:disc_nearfield}).

\subsection{Staggered grid operators and complementarity-gated capillary force}\label{subsec:exact_wb}

We discretize \Cref{eq:gov_agg_do} on a uniform staggered MAC grid of spacing $\Delta x$, storing scalars $(\phi_{\bm{i}}, \mu_{\bm{i}}, p_{\bm{i}})$ at cell centers $\bm{x}_{\bm{i}}$ ($\bm{i}\in\mathbb{Z}^d$, $d\in\{2,3\}$) and normal velocity and force components $(u_f, F_f)$ on cell faces $f = (\bm{i}+\frac{1}{2}\bm{e}_\alpha)$ ($\alpha\in\{1,\dots,d\}$). Two-point face gradients, cell-centered divergences, and face arithmetic averages are defined by
\begin{equation}\label{eq:mac_grad_div}
(\nabla_f \psi)_f \coloneqq \frac{\psi_{\bm{i}+\bm{e}_\alpha} - \psi_{\bm{i}}}{\Delta x}, \qquad (\nabla_h\cdot\bm{u})_{\bm{i}} \coloneqq \frac{1}{\Delta x}\sum_{\alpha=1}^d \bigl(u_{\bm{i}+\frac{1}{2}\bm{e}_\alpha} - u_{\bm{i}-\frac{1}{2}\bm{e}_\alpha}\bigr), \qquad \overline{\psi}^f \coloneqq \frac{\psi_{\bm{i}} + \psi_{\bm{i}+\bm{e}_\alpha}}{2}.
\end{equation}
On periodic or no-penetration domains, these operators satisfy discrete summation-by-parts $\langle \bm{u}, \nabla_f\psi\rangle_F = -\langle \nabla_h\cdot\bm{u}, \psi\rangle_C$ and annihilate every discrete edge or plaquette circulation identically, $(\nabla_{\mathrm{MAC}}\times \nabla_f\psi)_e \equiv 0$. To suppress grid-orientation anisotropy in cell-centered interfacial derivatives within a single-cell halo, $\mu_{\mathrm{reg}}$ and $\bm{P}_\perp^{\mathrm{ONRO}}$ use the compact 9-point ($d=2$) and 19-point ($d=3$) isotropic stencils $\nabla_{\mathrm{ICS}}$ and $\Delta_{\mathrm{ICS}}$ (with a separable 27-point isotropic variant documented in \Cref{app:params}):
\begin{align}\label{eq:ics_laplacian}
(\Delta_{\mathrm{ICS}}\phi)_{\bm{i}}^{\mathrm{2D}} &= \frac{1}{\Delta x^2}\left(\frac{2}{3}\sum_{|\bm{s}|_1=1}\phi_{\bm{i}+\bm{s}} + \frac{1}{6}\sum_{|\bm{s}|_1=2,|\bm{s}|_\infty=1}\phi_{\bm{i}+\bm{s}} - \frac{10}{3}\phi_{\bm{i}}\right), \nonumber\\
(\Delta_{\mathrm{ICS}}\phi)_{\bm{i}}^{\mathrm{3D}} &= \frac{1}{\Delta x^2}\left(\frac{1}{3}\sum_{|\bm{s}|_1=1}\phi_{\bm{i}+\bm{s}} + \frac{1}{6}\sum_{|\bm{s}|_1=2,|\bm{s}|_\infty=1}\phi_{\bm{i}+\bm{s}} - 4\phi_{\bm{i}}\right).
\end{align}
Fourier expansion of \Cref{eq:ics_laplacian} gives $-\widehat{\Delta_{\mathrm{ICS}}}\phi(\bm{k}) = |\bm{k}|^2\bigl(1 - \frac{1}{12}|\bm{k}|^2\Delta x^2\bigr)\hat{\phi}(\bm{k}) + \mathcal{O}(|\bm{k}|^6\Delta x^4)$, or $\Delta_{\mathrm{ICS}}\phi = \Delta\phi + \frac{1}{12}\Delta x^2\Delta^2\phi + \mathcal{O}(\Delta x^4)$, alongside $\widehat{\nabla_{\mathrm{ICS}}}\phi(\bm{k}) = \mathrm{i}\bm{k}\bigl(1 - \frac{1}{6}|\bm{k}|^2\Delta x^2\bigr)\hat{\phi}(\bm{k}) + \mathcal{O}(|\bm{k}|^5\Delta x^4)$: both operators are second-order accurate, and because their $\mathcal{O}(\Delta x^2)$ truncation terms are purely radial, directional anisotropy enters only at $\mathcal{O}(\Delta x^4)$. When evaluating \Cref{eq:M_ONRO} on the mesh, $\mathcal{D}_{\mathrm{ma}}$ employs the central-stencil dispersion length $\epsilon_{\mathrm{disc}} = \epsilon/\operatorname{sinc}(\Delta x/\epsilon)$ with unnormalized $\operatorname{sinc}(q) \coloneqq \sin(q)/q$, and $\chi_{\mathrm{ma}}$ acts on the single-hop face-neighbor maximum $\mathcal{D}_{\mathrm{ma}}^{\mathrm{dil}}$ across three-dimensional collars (\Cref{app:params}).

Discretizing the capillary force $\bm{f}_{\mathrm{cap}}$ for a non-smooth obstacle potential exposes a conflict between static equilibrium and dynamic wave propagation. Directly forming $-\phi\nabla_f\mu_{\mathrm{reg}}$ or $\overline{\mu_{\mathrm{reg}}}^f\nabla_f\phi$ fails across the free boundary $|\phi|=1$, where the regular chemical potential jumps to $\mu_{\mathrm{reg}}(\pm 1) = \mp\gamma_{\mathrm{DO}}/\epsilon \neq 0$ as the contact multiplier $\xi \in \partial I_{[-1,1]}(\pm 1)$ activates. Restricting $\mu_{\mathrm{reg}}$ to a feathered collar potential restores static balance on quiescent droplets, yet penalizes transient thickness compression during capillary shear oscillations with artificial $\mathcal{O}(\sigma/\epsilon)$ normal stiffness that heavily damps wave motion. Conversely, geometric continuum surface forces $\sigma\overline{\kappa_{\mathrm{eik}}}^f\nabla_f s(\phi)$ preserve oscillatory motion because the eikonal curvature $\kappa_{\mathrm{eik}} = -\nabla\cdot(\nabla\phi/|\nabla\phi|)$ is homogeneous of degree zero in $|\nabla\phi|$, but leave $\mathcal{O}(\Delta x^2/\epsilon^3)$ discrete curl residuals that drive persistent parasitic currents across large density ratios.

\Cref{lem:collar_coarea} provides the curl-free equilibrium building block needed to resolve both halves of this conflict.

\begin{lemma}[$C^1$-feathered collar potential and coarea MAC curl annihilation]\label{lem:collar_coarea}
Let $\epsilon \ge \Delta x$, $\delta_b \in (0,1)$, and $n_{\mathrm{pass}} \ge 1$ be an integer, and let $\mu_{\mathrm{reg},\bm{i}} = \gamma_{\mathrm{DO}}(-\epsilon\Delta_{\mathrm{ICS}}\phi_{\bm{i}} - \phi_{\bm{i}}/\epsilon)$. Define the $C^1$ interior Hermite weight $w_{\mathrm{in}}(\phi) \coloneqq H_3\bigl(1-|\phi|; 0, \delta_b\bigr) \in [0,1]$, the interior stencil weight $w_{\mathrm{ext}}(\phi) \coloneqq \bigl[(1-\delta_b)^2 - \phi^2\bigr]_+^2$, and the feathered collar potential $\tilde{\mu}_{\bm{i}} = w_{\mathrm{in}}(\phi_{\bm{i}})\mu_{\mathrm{reg},\bm{i}} + \bigl(1 - w_{\mathrm{in}}(\phi_{\bm{i}})\bigr)\mu_{\mathrm{col},\bm{i}}$, where $\mu_{\mathrm{col}}$ is propagated outward across $n_{\mathrm{pass}}$ neighbor shells by normalized isotropic convex averaging of $w_{\mathrm{ext}}\mu_{\mathrm{reg}}$ over $w_{\mathrm{ext}}$ with nonnegative stencil weights. Furthermore, define the normalized double-obstacle Korteweg antiderivative
\begin{equation}\label{eq:G_DO}
G_{\mathrm{DO}}(\phi) \coloneqq \frac{2}{\pi}\left(\phi\sqrt{(1-\phi^2)_+} + \arcsin\bigl(\operatorname{proj}_{[-1,1]}(\phi)\bigr)\right) \in [-1,1], \qquad G_{\mathrm{DO}}(\pm 1) = \pm 1,
\end{equation}
which satisfies $\frac{1}{2}\nabla G_{\mathrm{DO}}(\phi) = \frac{2}{\pi}\sqrt{(1-\phi^2)_+}\nabla\phi = \delta_\Gamma^\epsilon\bm{n}$ in terms of the regularized interfacial Dirac density $\delta_\Gamma^\epsilon \coloneqq \frac{2}{\pi}\sqrt{(1-\phi^2)_+}|\nabla\phi|$, with unit jump $\int_{-1}^{1}\frac{1}{2}G_{\mathrm{DO}}'(\phi)\,\dd\phi = 1$. Suppose that at a resolved discrete VI equilibrium every active face $(\nabla_f G_{\mathrm{DO}}(\phi))_f \neq 0$ lies within $n_{\mathrm{pass}}$ stencil hops of at least one interior collar cell ($|\phi_{\bm{j}}| < 1-\delta_b$, so the propagated denominator weight is strictly positive) and $\mu_{\mathrm{reg},\bm{j}} \equiv \mu_{\mathrm{eq}}$ across $|\phi_{\bm{j}}| < 1-\delta_b$. Then $\tilde{\mu}_{\bm{i}} \equiv \mu_{\mathrm{eq}}$ and $(\nabla_f\tilde{\mu})_f \equiv 0$ hold identically across every active face, and for any constant $\kappa_0\in\R$ on a connected interfacial collar, $F_f = \sigma\kappa_0\,\nabla_f(\frac{1}{2}G_{\mathrm{DO}}(\phi))$ satisfies $(\nabla_{\mathrm{MAC}}\times \bm{F})_e \equiv 0$ identically while $(1/\rho_f)F_f = (1/\rho_f)\nabla_f p_{\mathrm{cap}}$ lies in the exact range of the variable-density pressure gradient for every $\rho_+/\rho_- \in [1,\infty)$.
\end{lemma}

\begin{proof}
Because $\mu_{\mathrm{reg},\bm{i}} = \mu_{\mathrm{eq}}$ wherever $w_{\mathrm{ext}}(\phi_{\bm{i}}) > 0$ ($|\phi_{\bm{i}}| < 1-\delta_b$), normalized convex averaging yields $\mu_{\mathrm{col},\bm{i}} \equiv \mu_{\mathrm{eq}}$ and hence $\tilde{\mu}_{\bm{i}} = w_{\mathrm{in}}\mu_{\mathrm{eq}} + (1-w_{\mathrm{in}})\mu_{\mathrm{eq}} \equiv \mu_{\mathrm{eq}}$ across the entire $n_{\mathrm{pass}}$-shell halo straddling $|\phi|=1$. On any connected collar where $\overline{\kappa}^f\equiv\kappa_0$ wherever $\nabla_f G_{\mathrm{DO}}(\phi)\neq 0$, $F_f = \nabla_f p_{\mathrm{cap},\bm{i}}$ is the exact MAC gradient of $p_{\mathrm{cap},\bm{i}} = \frac{1}{2}\sigma\kappa_0(G_{\mathrm{DO}}(\phi_{\bm{i}})+1)$, which equals $\sigma\kappa_0$ in the interior $\phi_{\bm{i}}=+1$ and $0$ in the exterior $\phi_{\bm{i}}=-1$; hence $F_f/\rho_f = (\nabla_f p_{\mathrm{cap}})_f/\rho_f$ is balanced identically by $p_{\bm{i}}=p_{\mathrm{cap},\bm{i}}$ and $\bm{u}_f\equiv\bm{0}$ regardless of how $\rho_f > 0$ is interpolated.
\end{proof}

Combining \Cref{lem:collar_coarea} (evaluated at $\delta_b=10^{-3}$ and $n_{\mathrm{pass}}=3$ for capillary force assembly; \Cref{app:params}) with \Cref{eq:yl_2d_asym,eq:yl_3d_superconv} defines our complementarity-gated well-balanced continuum surface-delta force (WB-CSDF). Let $q \coloneqq \min(\Delta x/\epsilon, 1.2)$ denote the mesh-to-thickness ratio, and introduce the discrete collar tension calibration
\begin{equation}\label{eq:sigma_disc}
\sigma_{\mathrm{disc}}(q) \coloneqq \sigma\left(1 - \frac{q^2}{18} - \frac{q^4}{45}\right),
\end{equation}
together with the unique real root $\kappa_{\mathrm{VI},\bm{i}}^{(d)}$ of the $d$-dimensional matched-asymptotic Young--Laplace relation
\begin{equation}\label{eq:kappa_vi_d}
\kappa_{\mathrm{VI},\bm{i}}^{(d)}\left(1 + c_2^{(d)}\epsilon^2\bigl(\kappa_{\mathrm{VI},\bm{i}}^{(d)}\bigr)^2\right) = \frac{2\tilde{\mu}_{\bm{i}}}{\sigma_{\mathrm{disc}}(q)}, \qquad c_2^{(2)} = \frac{\pi^2 - 6}{16}, \quad c_2^{(3)} = 0.
\end{equation}
Here \Cref{thm:asymptotics}(b) simplifies \Cref{eq:kappa_vi_d} directly: in two dimensions, inverting the depressed cubic removes the $+\frac{\pi^2-6}{16}\delta^2$ cylindrical bias, whereas in three dimensions sixth-order midpoint superconvergence sets $c_2^{(3)} \equiv 0$, reducing \Cref{eq:kappa_vi_d} to the linear relation $\kappa_{\mathrm{VI},\bm{i}}^{(3)} = 2\tilde{\mu}_{\bm{i}}/\sigma_{\mathrm{disc}}(q)$. On each MAC face $f$, departure from local KKT phase equilibrium is measured through the dimensionless active-link chemical-potential jump
\begin{equation}\label{eq:kkt_gate}
\mathcal{R}_{\mathrm{KKT},f} \coloneqq \frac{\epsilon}{\gamma_{\mathrm{DO}}}\,\bigl|\Delta_{\mathrm{act}}\tilde{\mu}\bigr|_f^{\mathrm{dil}}, \qquad g_{\mathrm{eq},f} \coloneqq 1 - H_3\bigl(\mathcal{R}_{\mathrm{KKT},f}; \eta_{\mathrm{lo}}, \eta_{\mathrm{hi}}\bigr) \in [0,1],
\end{equation}
where $|\Delta_{\mathrm{act}}\tilde{\mu}|^{\mathrm{dil}}$ is the Moore-dilated maximum of $|\tilde{\mu}_{\bm{j}}-\tilde{\mu}_{\bm{i}}|$ across active neighbor links ($|\phi|<1-10^{-9}$) and $0 < \eta_{\mathrm{lo}} < \eta_{\mathrm{hi}}$ are the equilibrium-to-wave transition thresholds (baseline $[\eta_{\mathrm{lo}}, \eta_{\mathrm{hi}}] = [10^{-5}, 2\times 10^{-4}]$, degenerating to the purely dynamic eikonal branch $g_{\mathrm{eq},f}=0$ in axisymmetric pinch-off runs; in three-dimensional rising bubbles with under-resolved skirt tips, the outer ridge weight $\chi_{\mathrm{ma}}$ in \Cref{eq:exact_wb_force} is replaced by $\chi_{\mathrm{eff}}=\max(\chi_{\mathrm{ma}},\chi_{\mathrm{BKZ}})$ using the Brackbill--Kothe--Zemach~\cite{Brackbill1992} corner-normal coherence sensor $\chi_{\mathrm{BKZ}}$; \Cref{app:params,app:slender_jet}). Blending $\kappa_{\mathrm{VI}}^{(d)}$ with the zero-homogeneous narrow-band eikonal curvature $\kappa_{\mathrm{eik},\bm{i}} = -\nabla_{\mathrm{ICS}}\cdot\bigl(\nabla_{\mathrm{ICS}}\psi/(|\nabla_{\mathrm{ICS}}\psi|+\delta_{\mathrm{reg}})\bigr)$ (where $\psi$ is the signed-distance coordinate extended off $\epsilon\arcsin(\operatorname{proj}_{[-1,1]}(\phi))$) yields the complementarity-gated MAC face force
\begin{equation}\label{eq:exact_wb_force}
F_{\mathrm{cap},f} \coloneqq \bigl(1 - \overline{\chi_{\mathrm{ma}}}^f\bigr)\,\sigma\Bigl[g_{\mathrm{eq},f}\,\overline{\kappa_{\mathrm{VI}}^{(d)}}^f + \bigl(1 - g_{\mathrm{eq},f}\bigr)\overline{\kappa_{\mathrm{eik}}}^f\Bigr]\nabla_f\!\left(\frac{1}{2}G_{\mathrm{DO}}(\phi)\right) + \overline{\chi_{\mathrm{ma}}}^f\,\overline{\tilde{\mu}}^f\,\nabla_f\phi.
\end{equation}

\begin{remark}[Planar stencil dispersion and curved-collar polynomial calibration]\label{rem:dispersion}
Along an axis-aligned one-dimensional profile $\phi(x) = \sin(k_{\mathrm{disc}}x)$, exact annihilation of $-\epsilon\Delta_{\mathrm{ICS}}\phi - \phi/\epsilon = 0$ gives $k_{\mathrm{disc}}\epsilon = \frac{\arccos(1-q^2/2)}{q} = 1 + \frac{q^2}{24} + \frac{3q^4}{640} + \mathcal{O}(q^6)$, corresponding to a planar discrete tension factor $1 - q^2/24$. On two-dimensional circular and three-dimensional spherical collars sampled across oblique grid orientations and integrated against the two-point face difference $\nabla_f(\frac{1}{2}G_{\mathrm{DO}}(\phi))$, solid-angle orientation averaging combined with midpoint face quadrature shifts the effective collar tension slightly below $1 - q^2/24$. The fourth-degree polynomial $\sigma_{\mathrm{disc}}(q) = \sigma(1 - \frac{q^2}{18} - \frac{q^4}{45})$ in \Cref{eq:sigma_disc} is a fixed numerical-dispersion correction pre-calibrated on static curved-collar equilibria across $q=\min(\Delta x/\epsilon,1.2)$ and held identically constant across every 2D and 3D simulation in this work.
\end{remark}

\subsection{Bound-preserving flux-corrected transport and discrete baseline guarantees}\label{subsec:fct_cisl5}

Maintaining the obstacle constraint $\phi_{\bm{i}}\in[-1,1]$ during advection $\partial_t\phi + \nabla\cdot(\phi\bm{u}) = 0$ without post-step algebraic clipping ($\phi\leftarrow\operatorname{clip}(\phi,-1,1)$) is essential for local mass conservation and positivity of $\rho(\phi)$ and $\eta(\phi)$. Post-step clipping alters total phase mass whenever dispersive overshoots occur near $|\phi|=1$, while dimension-split one-dimensional semi-Lagrangian updates omit multi-dimensional cross-derivatives $-\Delta t\,u_x u_y\partial_{xy}\phi$ in vortical flows and single-pass flux limiters over-diffuse the $C^{1,1}$ obstacle kinks at $|\phi|=1$.

We therefore combine fifth-order conservative upwind spatial reconstruction with four-pass multi-dimensional Zalesak flux-corrected transport (FCT)~\cite{Zalesak1979,Kuzmin2020_CMAME,Kuzmin2020b_CMAME}. For $u_{i+1/2}\ge 0$, the high-order flux $F_f^{\mathrm{high}}$ uses the zero-Courant upwind-biased weights $(\frac{1}{30}, -\frac{13}{60}, \frac{47}{60}, \frac{9}{20}, -\frac{1}{20})$ on the five-cell stencil $(\phi_{i-2},\allowbreak\phi_{i-1},\allowbreak\phi_i,\allowbreak\phi_{i+1},\allowbreak\phi_{i+2})$, augmented by fourth-degree Courant polynomial terms. Prescribed kinematic stretching benchmarks (\Cref{subsec:bench_adv_enright}) embed this spatial operator inside a three-stage strong-stability-preserving Runge--Kutta (SSPRK3) integrator~\cite{Shu1988} with zero-Courant weights, whereas coupled Navier--Stokes steps execute a single solenoidal split MAC update.

\Cref{thm:wb_fct} unifies the discrete hydrostatic, capillary-wave, and invariant-domain guarantees of the baseline immiscible discretization (\Cref{subsec:exact_wb,subsec:fct_cisl5}).

\begin{theorem}[Discrete hydrostatic balance, capillary-wave release, and clipping-free bound preservation]\label{thm:wb_fct}
Let $\bm{u}_f$ satisfy discrete MAC incompressibility $(\nabla_h\cdot\bm{u})_{\bm{i}} = 0$ on $\R^d$ ($d\in\{2,3\}$) with positive face densities $\rho_f > 0$. Three discrete properties hold in exact arithmetic:
\begin{enumerate}
\item[(a)] \emph{(Null-tension consistency and arbitrary-density-ratio $M$-droplet hydrostatics).} When $\sigma = 0$, $F_{\mathrm{cap},f} \equiv 0$ identically on every face. For $\sigma > 0$, $\epsilon \ge \Delta x$, and $M\ge 1$ static droplets of collar midpoint radii $R_{\mathrm{mid},m} > \frac{\pi}{2}\epsilon + 2\Delta x$ separated by disjoint-halo gaps $H_{mn} > \pi\epsilon + 2\Delta x$ at exact discrete VI equilibrium ($\mathcal{R}_{\mathrm{KKT},f}=0$, whence $\mathcal{R}_{\mathrm{KKT},f} \le \eta_{\mathrm{lo}}$ locks $g_{\mathrm{eq},f}=1$ alongside $\overline{\chi_{\mathrm{ma}}}^f=0$ on all interfacial faces), $(\nabla_{\mathrm{MAC}}\times\bm{F}_{\mathrm{cap}})_e \equiv 0$ vanishes identically and $\bm{u}_f \equiv \bm{0}$ is an exact discrete steady state simultaneously across all $M$ droplets for every $\rho_+/\rho_- \in [1,\infty)$ (whereas under a finite pre-relaxation residual $0<\mathcal{R}_{\mathrm{KKT},f}\le\eta_{\mathrm{lo}}$, the residual discrete curl is bounded linearly as $\mathcal{O}(\sigma\mathcal{R}_{\mathrm{KKT},f}/(\epsilon\Delta x^2))$ independently of $\rho_+/\rho_-$). Moreover, the discrete pressure jump satisfies $\Delta p_m = (d-1)\sigma/R_{\mathrm{mid},m} + \mathcal{O}(\Delta x^2/\epsilon^2, \epsilon^4/R_{\mathrm{mid},m}^5)$ in $d=2$ and $\Delta p_m = (d-1)\sigma/R_{\mathrm{mid},m} + \mathcal{O}(\Delta x^2/\epsilon^2, \epsilon^6/R_{\mathrm{mid},m}^7)$ in $d=3$ with fourth-order directional isotropy.
\item[(b)] \emph{(Zero-homogeneous capillary-wave release).} Under any capillary-wave perturbation satisfying $\mathcal{R}_{\mathrm{KKT},f} \ge \eta_{\mathrm{hi}}$, $g_{\mathrm{eq},f}=0$ switches \Cref{eq:exact_wb_force} onto $\sigma\overline{\kappa_{\mathrm{eik}}}^f\nabla_f(\frac{1}{2}G_{\mathrm{DO}}(\phi))$, whose underlying continuous eikonal curvature functional satisfies $\partial_\varepsilon \kappa_{\mathrm{eik}}|_{n=0}=0$ identically under normal thickness strain $\phi(n)\mapsto\phi((1+\varepsilon)n)$ (leaving only $\mathcal{O}(\Delta x^2/\epsilon^2)$ discrete truncation error on the mesh) and eliminates artificial $\mathcal{O}(\sigma/\epsilon)$ normal-strain damping.
\item[(c)] \emph{(Solenoidal convex-hull and clipping-free multi-pass Zalesak FCT).} Under the sharp outflow Courant condition
\begin{equation}\label{eq:cfl_outflow}
\nu_{\mathrm{out},\bm{i}} \coloneqq \frac{\Delta t}{\Delta x}\sum_{f\in\partial\mathcal{C}_{\bm{i}}} (u_f\,\sigma_{\bm{i},f})_+ \le 1 \qquad \forall\,\bm{i},
\end{equation}
where $\sigma_{\bm{i},f}\in\{\pm 1\}$ is the outward face orientation sign of cell $\mathcal{C}_{\bm{i}}$, the first-order upwind step $\phi_{\bm{i}}^{(0)} = \phi_{\bm{i}}^* - \frac{\Delta t}{\Delta x}\sum_{f\in\partial\mathcal{C}_{\bm{i}}}\sigma_{\bm{i},f}F_f^{\mathrm{upw}}$ is a convex combination of $\phi_{\bm{i}}^*$ and its $2d$ face neighbors in $[\phi_{\bm{i}}^{\min},\phi_{\bm{i}}^{\max}]\subseteq[-1,1]$. Iterating $p=1,\dots,P$ ($P=4$) Zalesak passes from $A_f^{(0)} = F_f^{\mathrm{high}} - F_f^{\mathrm{upw}}$,
\begin{equation}\label{eq:fct_multipass}
\phi_{\bm{i}}^{(p)} = \phi_{\bm{i}}^{(p-1)} - \frac{\Delta t}{\Delta x}\sum_{f\in\partial\mathcal{C}_{\bm{i}}}\sigma_{\bm{i},f}\,\alpha_f^{(p)}A_f^{(p-1)}, \qquad A_f^{(p)} = \bigl(1 - \alpha_f^{(p)}\bigr)A_f^{(p-1)},
\end{equation}
conserves discrete mass $\sum_{\bm{i}}\phi_{\bm{i}}^{(P)}\Delta x^d = \sum_{\bm{i}}\phi_{\bm{i}}^*\Delta x^d$, enforces $\phi_{\bm{i}}^{n+1}\in[-1,1]$ across SSPRK3 with \emph{zero} state clipping, and monotonically reduces residual antidiffusion $|A_f^{(P)}|\le\cdots\le|A_f^{(1)}|$ around non-smooth $C^{1,1}$ obstacle kinks.
\end{enumerate}
\end{theorem}

\begin{proof}
Part~(a) follows by applying \Cref{lem:collar_coarea} componentwise to each disjoint droplet collar $m$, yielding $F_{\mathrm{cap},f} = \nabla_f(\sum_{m=1}^M p_{\mathrm{cap},m})$ even when $\mu_{\mathrm{eq},1}\neq\cdots\neq\mu_{\mathrm{eq},M}$, while accuracy of $\Delta p_m$ follows from \Cref{eq:sigma_disc,eq:kappa_vi_d,eq:yl_2d_asym,eq:yl_3d_superconv}. Part~(b) follows because normal rescaling $\tilde{\phi}(\bm{x}) = \phi(\bm{x}_s + (1+\varepsilon)n\bm{n})$ scales $\nabla\tilde{\phi} = (1+\varepsilon)\nabla\phi + \mathcal{O}(n)$, leaving $\tilde{\bm{n}} = \nabla\tilde{\phi}/|\nabla\tilde{\phi}| = \bm{n} + \mathcal{O}(n)$ and $\kappa_{\mathrm{eik}}|_{n=0} = -\nabla_s\cdot\bm{n}$ invariant in $\varepsilon$. For part~(c), $(\nabla_h\cdot\bm{u})_{\bm{i}}=0$ implies $\sum_{f\in\partial\mathcal{C}_{\bm{i}}}(u_f\sigma_{\bm{i},f})_+ = \sum_{f\in\partial\mathcal{C}_{\bm{i}}}(-u_f\sigma_{\bm{i},f})_+$, so
\begin{equation}\label{eq:upwind_convex}
\phi_{\bm{i}}^{(0)} = \bigl(1 - \nu_{\mathrm{out},\bm{i}}\bigr)\phi_{\bm{i}}^* + \sum_{f\in\partial\mathcal{C}_{\bm{i}}} \frac{\Delta t}{\Delta x}(-u_f\sigma_{\bm{i},f})_+\,\phi_{\bm{i}(f)}^*,
\end{equation}
whose weights are nonnegative under \Cref{eq:cfl_outflow} and sum to $1$. Skew-symmetry of $\sigma_{\bm{i},f}\alpha_f^{(p)}A_f^{(p-1)}$ telescopes $\sum_{\bm{i}}\phi_{\bm{i}}^{(p)}$ identically, while setting $\alpha_f^{(p)} = \min(R_{\bm{i}_{\mathrm{up}}}^{-,(p)}, R_{\bm{i}_{\mathrm{dn}}}^{+,(p)})\in[0,1]$ (with donor and acceptor cells $\bm{i}_{\mathrm{up}},\bm{i}_{\mathrm{dn}}$ oriented along the sign of $A_f^{(p-1)}$) from nodal capacity ratios $R_{\bm{i}}^{\pm,(p)} = \min\bigl(1, Q_{\bm{i}}^{\pm,(p)}/(P_{\bm{i}}^{\pm,(p)}+\delta_{\mathrm{tiny}})\bigr)$---where $Q_{\bm{i}}^{+,(p)} = \phi_{\bm{i}}^{\max}-\phi_{\bm{i}}^{(p-1)}\ge 0$ and $Q_{\bm{i}}^{-,(p)} = \phi_{\bm{i}}^{(p-1)}-\phi_{\bm{i}}^{\min}\ge 0$ denote the remaining admissible nodal capacities, $P_{\bm{i}}^{\pm,(p)}\ge 0$ accumulate inward and outward antidiffusive face fluxes, and $\delta_{\mathrm{tiny}}=10^{-30}$ guards zero-flux division---guarantees $\phi_{\bm{i}}^{\min}\le \phi_{\bm{i}}^{(p)}\le \phi_{\bm{i}}^{\max}\subseteq[-1,1]$ by induction on $p$, extending to SSPRK3 by convexity of $[-1,1]$.
\end{proof}

\subsection{Discrete near-field closures and coupled time-stepping algorithm}\label{subsec:disc_nearfield}

To preserve the repulsive sign duality of \Cref{prop:dlvo_cont} on the staggered MAC grid, we discretize \Cref{eq:F_DLVO_cont} through the discrete Legendre--Fenchel maximization principle
\begin{equation}\label{eq:F_DLVO_dual}
\mathcal{F}_{\mathrm{DLVO},h}[\phi] \coloneqq \sup_{\psi_h} \sum_{\bm{i}}\Delta x^d\left[A_{\mathrm{rep}}\,s(\phi_{\bm{i}})\psi_{\bm{i}} - \frac{A_{\mathrm{rep}}}{2}\Bigl(\psi_{\bm{i}}^2 - \ell_d^2\psi_{\bm{i}}(\Delta_{\mathrm{ICS}}\psi)_{\bm{i}}\Bigr)\right] = \frac{A_{\mathrm{rep}}}{2}\bigl\langle s(\phi), (I - \ell_d^2\Delta_{\mathrm{ICS}})^{-1}s(\phi)\bigr\rangle_h,
\end{equation}
where $\langle a,b\rangle_h = \sum_{\bm{i}}a_{\bm{i}}b_{\bm{i}}\Delta x^d$. Under periodic or homogeneous Neumann boundary conditions, $-\Delta_{\mathrm{ICS}} \succeq 0$ is symmetric positive semi-definite and all off-diagonal neighbor weights in \Cref{eq:ics_laplacian} are nonnegative; hence $\mathcal{A}_{\ell_d} \coloneqq I - \ell_d^2\Delta_{\mathrm{ICS}}$ is a symmetric irreducible Stieltjes (M-) matrix for every $\ell_d > 0$, ensuring strict positive definiteness $\mathcal{A}_{\ell_d} \succ 0$ and entrywise positivity $\mathcal{A}_{\ell_d}^{-1} > 0$. Setting $\psi_{d,\bm{i}} = (\mathcal{A}_{\ell_d}^{-1}s(\phi))_{\bm{i}} \in [0,1]$, the discrete Gateaux derivative of $\mathcal{F}_{\mathrm{DLVO},h}$ with respect to $\phi_{\bm{i}}$ is
\begin{equation}\label{eq:mu_dlvo}
\mu_{\mathrm{DLVO},\bm{i}} \coloneqq \frac{1}{\Delta x^d}\frac{\partial\mathcal{F}_{\mathrm{DLVO},h}}{\partial\phi_{\bm{i}}} = \frac{A_{\mathrm{rep}}}{2}\,\psi_{d,\bm{i}}, \qquad (I - \ell_d^2\Delta_{\mathrm{ICS}})\psi_d = s(\phi) = \frac{1+\phi}{2}.
\end{equation}
On MAC faces, one may form either the potential-gradient force $F_{\mathrm{DLVO},f}^{(1)} \coloneqq -(1 + \overline{\phi}^f)\nabla_f\mu_{\mathrm{DLVO}}$ or the Korteweg dual force $F_{\mathrm{DLVO},f}^{(2)} \coloneqq +\overline{\mu_{\mathrm{DLVO}}}^f\nabla_f\phi$; at any finite $\Delta x$, their difference is an exact discrete gradient:
\begin{align}\label{eq:mac_product_rule}
F_{\mathrm{DLVO},f}^{(2)} - F_{\mathrm{DLVO},f}^{(1)} &= \frac{\mu_{\mathrm{DLVO},\bm{i}} + \mu_{\mathrm{DLVO},\bm{i}+\bm{e}_\alpha}}{2}\frac{\phi_{\bm{i}+\bm{e}_\alpha}-\phi_{\bm{i}}}{\Delta x} + \left(1 + \frac{\phi_{\bm{i}}+\phi_{\bm{i}+\bm{e}_\alpha}}{2}\right)\frac{\mu_{\mathrm{DLVO},\bm{i}+\bm{e}_\alpha}-\mu_{\mathrm{DLVO},\bm{i}}}{\Delta x} \nonumber\\
&\equiv \nabla_f\Bigl((1+\phi)\mu_{\mathrm{DLVO}}\Bigr)_f.
\end{align}

When molecular solubility is active ($\mathcal{S}_0 > 0$), \Cref{eq:ripening_pde} is discretized in conservation form on the MAC grid by regularizing the degenerate carrier diffusivity inside $\phi=+1$ cores as $\max(1-s(\phi_{\bm{i}}), \omega_{\mathrm{floor}})$ ($\omega_{\mathrm{floor}}\in\{10^{-6},10^{-4}\}$; \Cref{app:params}), keeping the linear system nonsingular while quadratic vanishing of $\delta_s(\phi_{\bm{i}})$ at $\phi_{\bm{i}}=\pm 1$ preserves $\phi_{\bm{i}}\in[-1,1]$ without clipping under semi-implicit reaction updates (or explicit substeps satisfying $\Delta t\,\frac{16\mathcal{S}_0|K(\kappa)|\,|u_c-u_K|}{3\pi\epsilon} \le \frac{27}{32}$). On bounded boxes $\Omega = [-L/2,L/2]^d$, periodic boundary conditions introduce an $\mathcal{O}((d_{\mathrm{pair}}/L)^d)$ image-dipole confinement bias. We eliminate this bias without analytical boundary data by imposing the projected box-normal first-order Bayliss--Gunzburger--Turkel (BGT1)~\cite{Bayliss1982} dipole radiation condition $\partial_n(u_c - u_\infty) + \frac{(d-1)(\bm{x}\cdot\bm{n})}{|\bm{x}|^2}(u_c - u_\infty) = 0$ on $\partial\Omega$ (where $\bm{n}$ is the outward unit normal to $\partial\Omega$ and $\int_\Omega q_\phi\,\dd\bm{x}=0$ cancels the far-field monopole so the dipole mode dominates) and determining the far-field constant $u_\infty$ self-consistently via Sherman--Morrison rank-one enforcement of zero net boundary radiative flux. On multi-droplet fields, shell-averaged curvature ($\kappa_{\mathrm{VI}}^{(2)}$ in 2D and geometric normal divergence in 3D, with spherical-volume fallback below $R_k < 3.2\Delta x$; \Cref{app:params}) is computed on each connected $\delta_s(\phi)>0$ collar and propagated across the collar by steepest-ascent watershed dilation along $+\nabla_{\mathrm{ICS}}\phi$, preventing curvature label bleeding across narrow inter-droplet gaps.

\Cref{thm:dlvo_rip} establishes the discrete positivity, momentum-conservation, energy-dissipation, and mass-conservation guarantees of these near-field closures.

\begin{theorem}[Discrete Yukawa--DLVO positivity, momentum conservation, energy decay, and mass-conserving ripening]\label{thm:dlvo_rip}
For the discrete Yukawa--DLVO functional \Cref{eq:F_DLVO_dual,eq:mu_dlvo} and conservative MAC ripening closure of \Cref{eq:ripening_pde,eq:delta_s_alg,eq:lambda_star_betas}, three discrete properties hold:
\begin{enumerate}
\item[(a)] \emph{(Discrete sign duality and strict repulsive positivity).} On any connected periodic or homogeneous Neumann MAC grid with $\ell_d > 0$, $\mathcal{A}_{\ell_d} = I - \ell_d^2\Delta_{\mathrm{ICS}}$ has an entrywise strictly positive inverse $\mathcal{A}_{\ell_d}^{-1} > 0$. Consequently, for any two disjoint discrete droplets $s = s_1 + s_2$ ($s_{1,\bm{i}}s_{2,\bm{i}}=0$ for all $\bm{i}$, $s_1,s_2\not\equiv 0$), the discrete Legendre-dual interaction energy satisfies $\mathcal{E}_{\mathrm{int},h}^{\mathrm{DLVO}} = +A_{\mathrm{rep}}\langle s_1, \mathcal{A}_{\ell_d}^{-1}s_2\rangle_h > 0$, whereas the discrete infimum counterpart $\mathcal{F}_{\mathrm{naive},h}[\phi] = \frac{1}{2}A_{\mathrm{rep}}\langle s, s - \mathcal{A}_{\ell_d}^{-1}s\rangle_h$ yields $\mathcal{E}_{\mathrm{int},h}^{\mathrm{naive}} = -\mathcal{E}_{\mathrm{int},h}^{\mathrm{DLVO}} < 0$.
\item[(b)] \emph{(Discrete MAC equivalence, Noether momentum conservation, and semi-discrete energy decay).} Under exact discrete Helmholtz projection $(\nabla_h\cdot\bm{u}^{n+1}=0)$ on a periodic MAC grid, $F_{\mathrm{DLVO},f}^{(1)}$ and $F_{\mathrm{DLVO},f}^{(2)}$ produce identical projected velocity updates $\bm{u}^{n+1}$ on constant-density steps (with pressures differing by $(1+\phi)\mu_{\mathrm{DLVO}} + C(t)$ via \Cref{eq:mac_product_rule}) and satisfy exact zero net domain force $\sum_{f\parallel\bm{e}_\alpha} F_{\mathrm{DLVO},f}^{(2)}\Delta x^d = 0$ for arbitrary asymmetric multi-droplet configurations. Moreover, when paired with centered solenoidal phase advection $\overline{\phi}^f\bm{u}_f$, variational collar capillarity $\overline{\tilde{\mu}}^f\nabla_f\phi$, and augmented relaxation flux $-\bm{\mathcal{M}}_{\mathrm{ONRO}}\nabla_f(\tilde{\mu}+\mu_{\mathrm{DLVO}})$, they dissipate total semi-discrete energy $\mathcal{E}_{\mathrm{tot},h} = \mathcal{E}_{\mathrm{kin},h} + \mathcal{F}_{\mathrm{DO},h} + \mathcal{F}_{\mathrm{DLVO},h}$ monotonically ($\dd\mathcal{E}_{\mathrm{tot},h}/\dd t \le 0$) at $\mathcal{S}_0=0$.
\item[(c)] \emph{(Nonsingular discrete ripening and two-phase volume conservation).} For any core diffusivity floor $\omega_{\mathrm{floor}} > 0$ and nonnegative surface reactivity $K(\kappa_{\bm{i}})\delta_s(\phi_{\bm{i}}) \ge 0$ not identically zero, the discrete diffusion--reaction system for $u_c$ is nonsingular, and under periodic, homogeneous Neumann, or Sherman--Morrison zero-net-flux BGT1 outer boundary conditions the discrete source satisfies $\sum_{\bm{i}}q_{\phi,\bm{i}}\Delta x^d = 0$ identically, conserving total discrete two-phase volume.
\end{enumerate}
\end{theorem}

\begin{proof}
In part~(a), symmetric positive semi-definiteness of $-\Delta_{\mathrm{ICS}}$ with nonnegative off-diagonal weights and connected neighbor graph makes $\mathcal{A}_{\ell_d} = I - \ell_d^2\Delta_{\mathrm{ICS}}$ an irreducible symmetric Stieltjes matrix, whence $\mathcal{A}_{\ell_d}^{-1} > 0$ entrywise and $\mathcal{E}_{\mathrm{int},h}^{\mathrm{DLVO}} = -\mathcal{E}_{\mathrm{int},h}^{\mathrm{naive}} = A_{\mathrm{rep}}\langle s_1, \mathcal{A}_{\ell_d}^{-1}s_2\rangle_h > 0$. In part~(b), \Cref{eq:mac_product_rule} gives $F_{\mathrm{DLVO},f}^{(2)} - F_{\mathrm{DLVO},f}^{(1)} = \nabla_f\bigl((1+\phi)\mu_{\mathrm{DLVO}}\bigr)_f$; introducing the cell-centered central difference $(\nabla_{c,\alpha}\psi)_{\bm{i}} \coloneqq (\psi_{\bm{i}+\bm{e}_\alpha}-\psi_{\bm{i}-\bm{e}_\alpha})/(2\Delta x)$, summing $F_{\mathrm{DLVO},f}^{(2)}$ over faces aligned with $\bm{e}_\alpha$ gives $\sum_{f\parallel\bm{e}_\alpha} F_{\mathrm{DLVO},f}^{(2)}\Delta x^d = -\langle \phi, \nabla_{c,\alpha}\mu_{\mathrm{DLVO}}\rangle_h = -A_{\mathrm{rep}}\langle s, \nabla_{c,\alpha}\mathcal{A}_{\ell_d}^{-1}s\rangle_h = 0$ because $\nabla_{c,\alpha}^* = -\nabla_{c,\alpha}$ is skew-adjoint and $\mathcal{A}_{\ell_d}^{-1}$ is self-adjoint and translation-invariant. Semi-discrete energy decay follows from convexity of $\mathcal{F}_{\mathrm{DLVO},h}[s]$ and exact MAC summation-by-parts cancellation $\langle \bm{u}, \overline{\mu_{\mathrm{DLVO}}}^f\nabla_f\phi\rangle_F - \langle \mu_{\mathrm{DLVO}}, \nabla_h\cdot(\overline{\phi}^f\bm{u})\rangle_C = -\langle \mu_{\mathrm{DLVO}}\phi, \nabla_h\cdot\bm{u}\rangle_C = 0$ on $\nabla_h\cdot\bm{u}=0$. In part~(c), strict diagonal dominance on active collar cells makes the discrete diffusion--reaction matrix nonsingular for $\omega_{\mathrm{floor}}>0$, and summing the conservative divergence over $\Omega_h$ cancels interior face fluxes identically while the boundary condition enforces zero net outer flux, yielding $\sum_{\bm{i}}q_{\phi,\bm{i}}\Delta x^d = 0$.
\end{proof}

\Cref{alg:mac_step} summarizes the four-stage staggered-MAC update $(\bm{u}^n, p^n, \phi^n) \mapsto (\bm{u}^{n+1}, p^{n+1}, \phi^{n+1})$. Initial static droplet profiles in hydrostatic benchmarks are relaxed beforehand via projected FISTA active-set iteration enforcing \Cref{eq:kkt_complementarity} (playing the role of semismooth Newton and nonsmooth multigrid obstacle solvers~\cite{Hintermuller2002,Graser2009_SINUM,Bosch2014_JCP} on uniform Cartesian grids), while linear-solver tolerances for geometric multigrid preconditioned conjugate gradients (GMG-PCG) and auxiliary elliptic solves are documented in \Cref{app:params}.

\begin{algorithm}[t!]
\caption{Structure-preserving staggered-MAC time step ($\bm{u}^n, p^n, \phi^n \mapsto \bm{u}^{n+1}, p^{n+1}, \phi^{n+1}$)}
\label{alg:mac_step}
\small
\begin{algorithmic}[1]
\Require Divergence-free MAC velocity $\bm{u}_f^n$ ($\nabla_h\cdot\bm{u}^n=0$), bounded phase state $\phi_{\bm{i}}^n \in [-1,1]$, physical parameters $(\sigma,\epsilon,\rho_\pm,\eta_\pm,A_{\mathrm{rep}},\mathcal{S}_0)$.
\State \textbf{Stage 1 (Bound-preserving advection):} Reconstruct fifth-order upwind face fluxes $F_f^{\mathrm{high}}$ and advance $\phi^n \mapsto \phi^{\mathrm{adv}} \in [-1,1]$ via $P=4$ Zalesak FCT passes (\Cref{eq:fct_multipass}, wrapped in SSPRK3 for prescribed velocity fields).
\State \textbf{Stage 2 (Obstacle relaxation and optional ripening):} Evaluate $\nabla_{\mathrm{ICS}}\phi^{\mathrm{adv}}$, $\Delta_{\mathrm{ICS}}\phi^{\mathrm{adv}}$, $\bm{P}_\perp^{\mathrm{ONRO}}$ (\Cref{eq:P_ONRO}), and $\chi_{\mathrm{ma}}$ (\Cref{eq:M_ONRO}); if $A_{\mathrm{rep}}>0$, solve $(I-\ell_d^2\Delta_{\mathrm{ICS}})\psi_d = s(\phi^{\mathrm{adv}})$ for $\mu_{\mathrm{DLVO}}$ (\Cref{eq:mu_dlvo}). Advance FCT-limited degenerate-mobility substeps driven by $\nabla_f(\tilde{\mu}+\mu_{\mathrm{DLVO}})$ (plus $\Lambda_\star$-renormalized source $q_\phi$ if $\mathcal{S}_0>0$, \Cref{eq:ripening_pde,eq:lambda_star_betas}) to obtain $\phi^{n+1}\in[-1,1]$ and $\bm{J}_\rho^{n+1}$.
\State \textbf{Stage 3 (Capillary and disjoining forces):} Construct feathered collar potential $\tilde{\mu}^{n+1}$ (\Cref{lem:collar_coarea}), invert $\kappa_{\mathrm{VI}}^{(d)}$ (\Cref{eq:kappa_vi_d}), evaluate KKT gate $g_{\mathrm{eq},f}$ (\Cref{eq:kkt_gate}), and assemble $F_{\mathrm{cap},f}$ (\Cref{eq:exact_wb_force}) together with $F_{\mathrm{DLVO},f}^{(2)}$ when $A_{\mathrm{rep}}>0$.
\State \textbf{Stage 4 (Variable-density momentum predictor and projection):} Update $\rho_f^{n+1}$ and $\eta^{n+1}$ from $s(\phi^{n+1})$, advance viscous predictor $\bm{u}_f^*$ with consistent drift $\bm{J}_\rho^{n+1}\otimes\bm{u}^n$, solve $-\nabla_h\cdot\bigl(\frac{\Delta t}{\rho_f^{n+1}}\nabla_f p^{n+1}\bigr) = -\nabla_h\cdot\bm{u}^*$ via GMG-PCG, and project $\bm{u}_f^{n+1} = \bm{u}_f^* - \frac{\Delta t}{\rho_f^{n+1}}\nabla_f p^{n+1}$.
\end{algorithmic}
\end{algorithm}


\section{Numerical verification and experimental validation}\label{sec:benchmarks}

\Cref{thm:asymptotics,thm:wb_fct,thm:dlvo_rip} and \Cref{prop:dlvo_cont,prop:ripening_sl} yield quantitative predictions that we test first against closed-form analytical solutions and sharp-interface numerical benchmarks (\Cref{subsec:bench_wb,subsec:bench_adv_enright,subsec:bench_hysing_rt,subsec:bench_rip_dlvo}), and then against laboratory measurements without fitted physical parameters (\Cref{subsec:bench_experiments}). All computations are performed in IEEE-754 double-precision arithmetic with numerical parameters held fixed per \Cref{app:params}. Every unforced ($\bm{g}=\bm{0}$) hydrodynamic test includes a null-tension control ($\sigma=0$, and $\mathcal{S}_0=0$ where applicable) confirmed to maintain identically zero velocity and mass drift, while setting $\sigma=0$ in buoyancy-driven runs vanishes the discrete capillary force identically.

\subsection{Static equilibria, capillary waves, and sub-width sheet convergence}\label{subsec:bench_wb}

\begin{figure}[t!]
\centering
\cmamefig{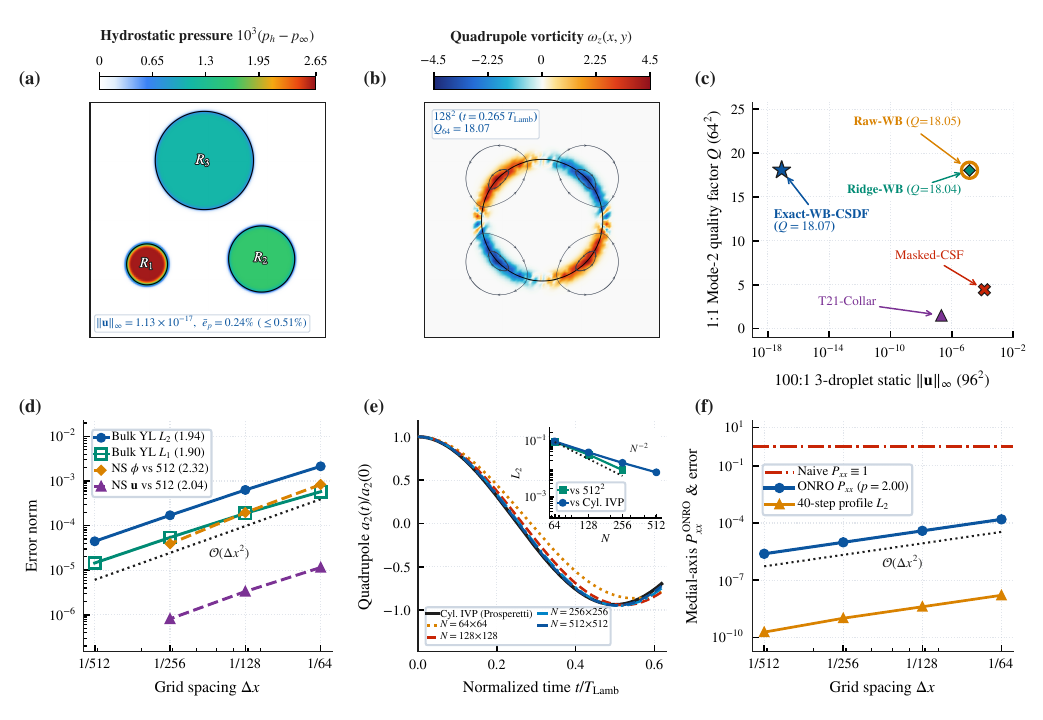}{width=\textwidth}
\caption{Resolution of the two-dimensional static-versus-wave capillary dilemma and four-grid sharp-interface convergence ($64^2 \to 512^2$). (a)~Exact $100\!:\!1$ three-unequal-droplet hydrostatic pressure field $10^3(p_h - p_\infty)$ on $128^2$ ($R\in\{11.5,18.5,27.5\}\Delta x$) under the complementarity-gated capillary force (labelled \texttt{Exact-WB-CSDF}), resolving three sharp Young--Laplace terraced plateaus $\sigma/R_m$ ($\bar{e}_p = 0.24\%$, $\max e_p = 0.51\%$) at machine-zero parasitic velocity ($\|\bm{u}\|_\infty = 1.13\times 10^{-17}$). (b)~Mode-2 Rayleigh--Lamb quadrupole vorticity $\omega_z(x,y)$ and closed streamlines on $128^2$ at $t=0.265\,T_{\mathrm{Lamb}}$ ($64^2$ full-period $Q=18.07$). (c)~Phase-plane comparison of five discrete surface-tension operators pairing $100\!:\!1$ three-droplet static $\|\bm{u}\|_\infty$ ($96^2$, $R\in\{8.5,13.5,20.5\}\Delta x$) against Mode-2 quality factor $Q$ ($64^2$). (d)~Second-order convergence of $100\!:\!1$ bulk Young--Laplace pressure ($p_{L_2}=1.94$, $p_{L_1}=1.90$) under Cahn scaling ($\epsilon=1.90\Delta x$) and coupled $5\!:\!1$ Navier--Stokes velocity and phase errors ($p_{\bm{u}}=2.04$, $p_\phi=2.32$) at fixed $\epsilon$. (e)~Four-grid Mode-2 cylindrical quadrupole waveforms $a_2(t)/a_2(0)$ against the exact viscous initial-value solution of Prosperetti~\cite{Prosperetti1980} ($\epsilon=1.80\Delta x$; inset: $L_2$ waveform convergence vs.\ $N$, self-order $p=1.82$). (f)~Sub-width raised-cosine thin-sheet verification ($2h(x)=1.60\Delta x(1+0.08\cos 2\pi x)$, $|\partial_x h|=\mathcal{O}(\Delta x)$) confirming $\mathcal{O}(\Delta x^2)$ ($p=2.00$) suppression of $90^\circ$ medial-axis normal flipping under \Cref{eq:P_ONRO} (labelled \texttt{ONRO}, $P_{xx,\max}^{\mathrm{ONRO}} = 2.42\times 10^{-6}$ at $512^2$ versus $P_{xx}^{\mathrm{naive}}\equiv 1$, alongside uniform-sheet $40$-step profile $L_2$ convergence).}
\label{fig:wb2d}
\end{figure}

Taken alone, neither a potential form nor a geometric form of the capillary force maintains static balance across large density jumps without overdamping dynamic capillary waves. \Cref{fig:wb2d}(a--c) and \Cref{tab:wb_2d3d} compare five discrete capillary operators across three coexisting unequal droplets at rest ($\rho_+/\rho_- = 100$, $96^2$ in \Cref{fig:wb2d}(c) and \Cref{tab:wb_2d3d}; $128^2$ in \Cref{fig:wb2d}(a)) and a damped density-matched ($\rho_+/\rho_-=1$) Mode-2 Rayleigh--Lamb oscillation ($64^2$ full-period run in \Cref{fig:wb2d}(c) and \Cref{tab:wb_2d3d}; $128^2$ vorticity snapshot at $t=0.265\,T_{\mathrm{Lamb}}$ in \Cref{fig:wb2d}(b), reporting the first-period envelope quality factor $Q \coloneqq \pi/\ln(a_2(0)/|a_2(T_1)|)$ against $Q_{\mathrm{IVP}}=24.76$ for Prosperetti's analytical initial-value solution~\cite{Prosperetti1980} including quiescent boundary-layer spin-up, and $Q_{\mathrm{visc}}=52.4$ for the late-time asymptotic normal mode). Using an unfeathered boolean-masked collar potential with naive normal projection (labelled \texttt{Masked-CSF} in the figure legends, $\overline{\mu_{\mathrm{bool}}}^f\nabla_f\phi$) incurs both large coupled parasitic currents ($\|\bm{u}\|_\infty = 1.31\times 10^{-4}$) and strong wave damping ($Q=4.43$). Replacing that force with eikonal curvature, either without (\texttt{Raw-WB-CSDF}, $Q=18.05$) or with medial-axis blending (\texttt{Ridge-WB-CSDF}, $Q=18.04$), restores underdamped oscillations by zero-homogeneity (\Cref{thm:wb_fct}(b)), yet stalls at $\|\bm{u}\|_\infty = 1.41\times 10^{-5}$ alongside an initial Cartesian curl residual of $9.37\times 10^{-5}$. Conversely, the $C^1$-feathered ungated collar potential (\texttt{T21-Collar}) eliminates the curl residual and lowers static currents to $2.06\times 10^{-7}$, but its $\mathcal{O}(\sigma/\epsilon)$ normal-thickness stiffness crushes the oscillation to $Q=1.51$. Blending $\kappa_{\mathrm{VI}}^{(2)}$ and $\kappa_{\mathrm{eik}}$ through the local KKT gate $\mathcal{R}_{\mathrm{KKT},f}$ in \Cref{eq:exact_wb_force} (\texttt{Exact-WB-CSDF}) resolves both failures simultaneously: parasitic velocities drop by twelve to thirteen orders of magnitude to $\|\bm{u}\|_\infty = 8.16\times 10^{-18}$ on $96^2$ ($1.13\times 10^{-17}$ on $128^2$, with mean Young--Laplace errors of $0.46\%$ on $96^2$ and $0.24\%$ on $128^2$), while the Mode-2 quality factor remains $Q=18.07$ on $64^2$ ($Q=26.73\text{--}31.25$ on $128^2\text{--}512^2$ at $\epsilon=1.80\,\Delta x$).

Grid refinement across $N\in\{64,128,256,512\}$ (\Cref{fig:wb2d}(d--f)) confirms the asymptotic convergence orders predicted in \Cref{sec:continuum_vi,sec:discretization}. Under Cahn scaling ($\epsilon=1.90\,\Delta x$), bulk $100\!:\!1$ Young--Laplace pressure converges at second order ($p_{L_1}=1.90$, $p_{L_2}=1.94$), while fixed-$\epsilon$ coupled $5\!:\!1$ Navier--Stokes evolutions converge at $p_{\bm{u}}=2.04$ and $p_\phi=2.32$. Simultaneously, Mode-2 capillary waveforms converge toward Prosperetti's analytical initial-value solution~\cite{Prosperetti1980} ($L_2$ error $7.48\times 10^{-3}$ at $512^2$, self-order $p=1.82$), and weakly tapered sub-width raised-cosine sheets follow the quadratic $P_{xx}^{\mathrm{ONRO}} = \mathcal{O}((\partial_x h)^2) = \mathcal{O}(\Delta x^2)$ bound (\Cref{eq:onro_pxx_bound}) down to $2.42\times 10^{-6}$ at $512^2$.

\begin{figure}[t!]
\centering
\cmamefig{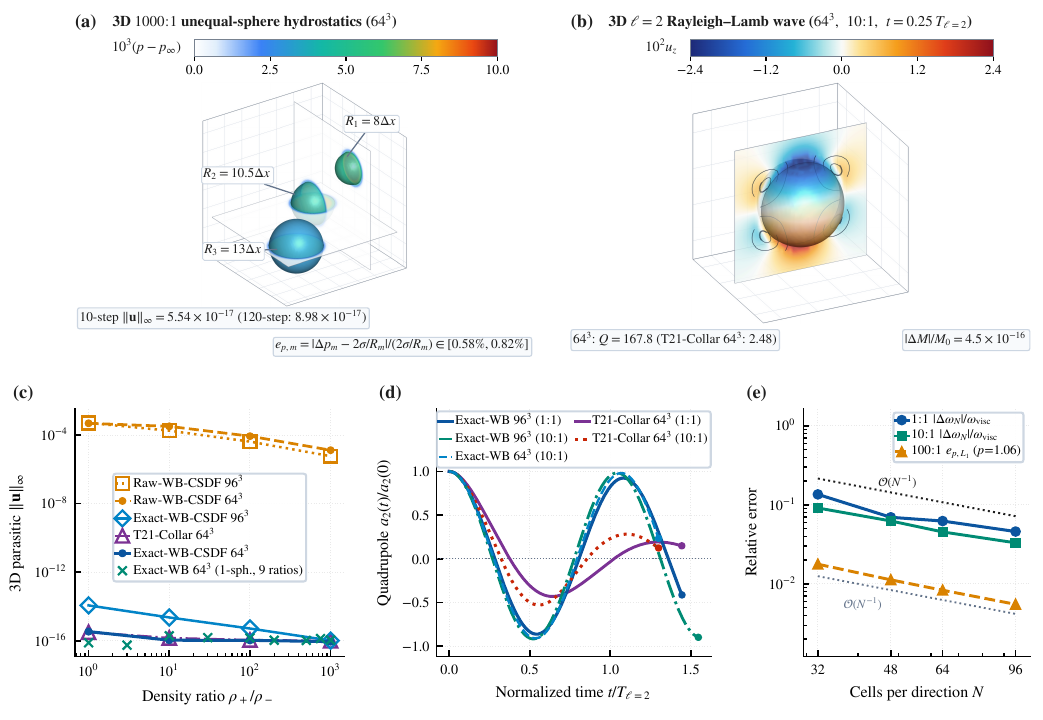}{width=\textwidth}
\caption{Three-dimensional verification of the complementarity-gated capillary force (labelled \texttt{Exact-WB-CSDF}) across $1\!:\!1$ to $1000\!:\!1$ density ratios ($32^3 \to 96^3$). (a)~Three-dimensional $1000\!:\!1$ unequal-sphere hydrostatics on $64^3$ ($R\in\{8,10.5,13\}\Delta x$): 3D quadrant cutaway of $\phi=0$ spheres and orthogonal slices colored by $10^3(p - p_\infty)$ ($10$-step direct stasis $\|\bm{u}\|_\infty = 5.54\times 10^{-17}$, $120$-step cold-start max $8.98\times 10^{-17}$; $\Delta p_m = 2\sigma/R_m$ within $0.58\%\text{--}0.82\%$ at $10$ steps and $\le 0.794\%$ at $120$ steps). (b)~Three-dimensional $\ell=2$ spheroidal Rayleigh--Lamb wave on $64^3$ ($10\!:\!1$, $t=0.25\,T_{\ell=2}$): meridional $10^2 u_z$ slice and quadrupole streamline loops ($Q=167.8$ vs.\ $2.48$ for the ungated collar potential \texttt{T21-Collar} on $64^3$, $|\Delta M|/M_0 = 4.5\times 10^{-16}$). (c)~Peak parasitic velocity $\|\bm{u}\|_\infty$ for three coexisting 3D spheres ($64^3$ and $96^3$) and a nine-ratio single-sphere sweep ($\rho_+/\rho_-\in[1,1000]$), reducing spurious currents by $10^{11}\text{--}10^{12}\times$ relative to \texttt{Raw-WB-CSDF}. (d)~Three-dimensional $\ell=2$ quadrupole oscillations $a_2(t)/a_2(0)$ at $\rho_+/\rho_-\in\{1,10\}$: the gated force achieves $Q=39.86$ ($96^3$, $1\!:\!1$; $Q=26.82$ on $64^3$) and $Q=186.03$ ($96^3$, $10\!:\!1$; $Q=167.77$ on $64^3$), whereas the ungated collar potential (\texttt{T21-Collar}) on $64^3$ damps oscillations by $14.0\text{--}67.6\times$ ($Q=1.91$ and $2.48$). (e)~Four-grid ($N\in\{32,48,64,96\}$) convergence of 3D $\ell=2$ viscous Lamb frequency error at $1\!:\!1$ and $10\!:\!1$ ($\mathcal{O}(N^{-1})$ diffuse viscosity-jump rate) alongside $100\!:\!1$ relative Young--Laplace $L_1$ pressure error.}
\label{fig:wb3d}
\end{figure}

\begin{table}[t!]
\centering
\caption{Quantitative 2D and 3D verification of static well-balancedness and Rayleigh--Lamb capillary-wave dynamics across density ratios $\rho_+/\rho_-\in[1,1000]$ (parenthetical tags match the legends of \Cref{fig:wb2d,fig:wb3d}; in 2D rows $\rho_+/\rho_- = 100 / 1$ denotes static 3-droplet / Mode-2 wave runs, and in 3D rows $10 / 1000$ denotes the two static 3-sphere ratios paired with the $10\!:\!1$ $\ell=2$ wave; $\|\nabla_{\mathrm{MAC}}\times\bm{F}_{\mathrm{cap}}\|_\infty$ is evaluated on the settled initial VI profile and static $\|\bm{u}\|_\infty$ after $150$ coupled steps in 2D or $120$ in 3D). Every unforced $\sigma=0$ control yields $\|\bm{u}\|_\infty = 0.0$ and $|\Delta M|/M_0 = 0.0$.}
\label{tab:wb_2d3d}
\small
\setlength{\tabcolsep}{3.8pt}
\renewcommand{\arraystretch}{1.14}
\resizebox{\textwidth}{!}{%
\begin{tabular}{@{}llcccccc@{}}
\toprule
Dimension / Test & Capillary operator & Grid & $\rho_+/\rho_-$ & Static $\|\bm{u}\|_\infty$ & $\|\nabla_{\mathrm{MAC}}\times\bm{F}_{\mathrm{cap}}\|_\infty$ & YL $|\Delta p|/\Delta p_{\mathrm{ex}}$ & Lamb $Q$ ($|\Delta\omega|/\omega_{\mathrm{visc}}$) \\
\midrule
2D 3-drop + Mode 2 & Boolean-masked collar (\texttt{Masked-CSF}) & $96^2 / 64^2$ & $100 / 1$ & $1.31\times 10^{-4}$ & $4.25\times 10^{-17}$ & $2.60\%$ ($2.77\%$) & $4.43$ \\
2D 3-drop + Mode 2 & Eikonal curvature (\texttt{Raw-WB-CSDF}) & $96^2 / 64^2$ & $100 / 1$ & $1.41\times 10^{-5}$ & $9.37\times 10^{-5}$ & $1.80\%$ ($3.51\%$) & $18.05$ \\
2D 3-drop + Mode 2 & Ridge-blended eikonal (\texttt{Ridge-WB-CSDF}) & $96^2 / 64^2$ & $100 / 1$ & $1.41\times 10^{-5}$ & $9.37\times 10^{-5}$ & $1.80\%$ ($3.51\%$) & $18.04$ \\
2D 3-drop + Mode 2 & Ungated collar (\texttt{T21-Collar}) & $96^2 / 64^2$ & $100 / 1$ & $2.06\times 10^{-7}$ & $4.25\times 10^{-17}$ & $1.30\%$ ($2.50\%$) & $1.51$ (overdamped) \\
2D 3-drop + Mode 2 & \textbf{Gated VI force (\texttt{Exact-WB-CSDF})} & $96^2 / 64^2$ & $100 / 1$ & $\mathbf{8.16\times 10^{-18}}$ & $\mathbf{5.32\times 10^{-17}}$ & $\mathbf{0.46\%}$ ($\mathbf{0.24\%}$ on $128^2$) & $\mathbf{18.07}$ \\
\midrule
3D 1-sphere (9-ratio) & \textbf{Gated VI force (\texttt{Exact-WB-CSDF})} & $64^3$ & $1\text{--}1000$ & $\mathbf{5.62\times 10^{-17}}\text{--}\mathbf{1.98\times 10^{-16}}$ & $\mathbf{1.81\times 10^{-17}}$ & $\mathbf{0.397\%}$ & --- \\
3D 1-sphere (9-ratio) & \textbf{Gated VI force (\texttt{Exact-WB-CSDF})} & $96^3$ & $1\text{--}1000$ & $\mathbf{4.46\times 10^{-17}}\text{--}\mathbf{1.24\times 10^{-16}}$ & $\mathbf{1.86\times 10^{-17}}$ & $\mathbf{0.457\%}$ & --- \\
3D 3-sphere + $\ell=2$ & Eikonal curvature (\texttt{Raw-WB-CSDF}) & $64^3$ & $10 / 1000$ & $3.19\times 10^{-4} / 1.30\times 10^{-5}$ & $1.12\times 10^{-4}$ & $5.916\%$ & $167.77$ ($4.53\%$) \\
3D 3-sphere + $\ell=2$ & Ungated collar (\texttt{T21-Collar}) & $64^3$ & $10 / 1000$ & $1.48\times 10^{-16} / 9.99\times 10^{-17}$ & $8.65\times 10^{-17}$ & $1.193\%$ & $2.48$ (overdamped) \\
3D 3-sphere + $\ell=2$ & \textbf{Gated VI force (\texttt{Exact-WB-CSDF})} & $64^3$ & $10 / 1000$ & $\mathbf{1.09\times 10^{-16}} / \mathbf{8.98\times 10^{-17}}$ & $\mathbf{7.54\times 10^{-17}}$ & $\mathbf{0.794\%}$ & $\mathbf{167.77}$ ($4.53\%$) \\
3D 3-sphere + $\ell=2$ & \textbf{Gated VI force (\texttt{Exact-WB-CSDF})} & $96^3$ & $10 / 1000$ & $\mathbf{2.31\times 10^{-15}} / \mathbf{1.04\times 10^{-16}}$ & $\mathbf{8.53\times 10^{-16}}$ & $\mathbf{0.612\%}$ & $\mathbf{186.03}$ ($3.30\%$; $39.86$ at $1\!:\!1$) \\
\bottomrule
\end{tabular}%
}
\end{table}

The same complementarity gate extends without modification to three dimensions across density contrasts up to $1000\!:\!1$ (\Cref{fig:wb3d} and \Cref{tab:wb_2d3d}, $\epsilon=1.85\,\Delta x$, with linear curvature inversion $c_2^{(3)}=0$ per \Cref{eq:kappa_vi_d}). Across nine density ratios $\rho_+/\rho_-\in\{1,3,10,30,100,200,500,750,1000\}$ on $64^3$ and $96^3$ (\Cref{fig:wb3d}(a,c)), cold-start variable-density GMG-PCG projection holds single-sphere parasitic velocities between $4.46\times 10^{-17}$ and $1.98\times 10^{-16}$ with density-ratio-invariant Laplace accuracy ($0.397\%$ on $64^3$, $0.457\%$ on $96^3$, governed on the Cartesian mesh by spatial collar quadrature rather than continuum curvature error). When three unequal spheres ($R_1=8\Delta x$, $R_2=10.5\Delta x$, $R_3=13\Delta x$) coexist at $\rho_+/\rho_-=1000$, the gated force attains $\|\bm{u}\|_\infty = 8.98\times 10^{-17}$ on $64^3$ ($5.54\times 10^{-17}$ in $10$-step direct stasis) and $1.04\times 10^{-16}$ on $96^3$ ($0.612\%$ maximum Laplace error across all three spheres, zero droplet-wise volume drift), lowering spurious currents by $10^{11}\times$ relative to the eikonal force. On three-dimensional $\ell=2$ spheroidal Rayleigh--Lamb waves across $N\in\{32,48,64,96\}$ (\Cref{fig:wb3d}(b,d,e)), the same operator maintains underdamped oscillations at both $\rho_+/\rho_-=1$ ($Q=26.82\to 39.86$, frequency error converging from $13.53\%$ on $32^3$ to $4.59\%$ on $96^3$, or $8.89\%\to 6.75\%$ by uniform damped-cosine fit) and $\rho_+/\rho_-=10$ ($Q=18.68\to 76.96\to 167.77\to 186.03$, frequency error converging from $9.10\%$ to $3.30\%$, $|\Delta M|/M_0 = 2.30\times 10^{-15}$), whereas the ungated collar potential on $64^3$ collapses $Q$ to $1.91$ and $2.48$.

\subsection{Bound-preserving interface transport and coupled multi-droplet stasis}\label{subsec:bench_adv_enright}

\begin{figure}[t!]
\centering
\cmamefig{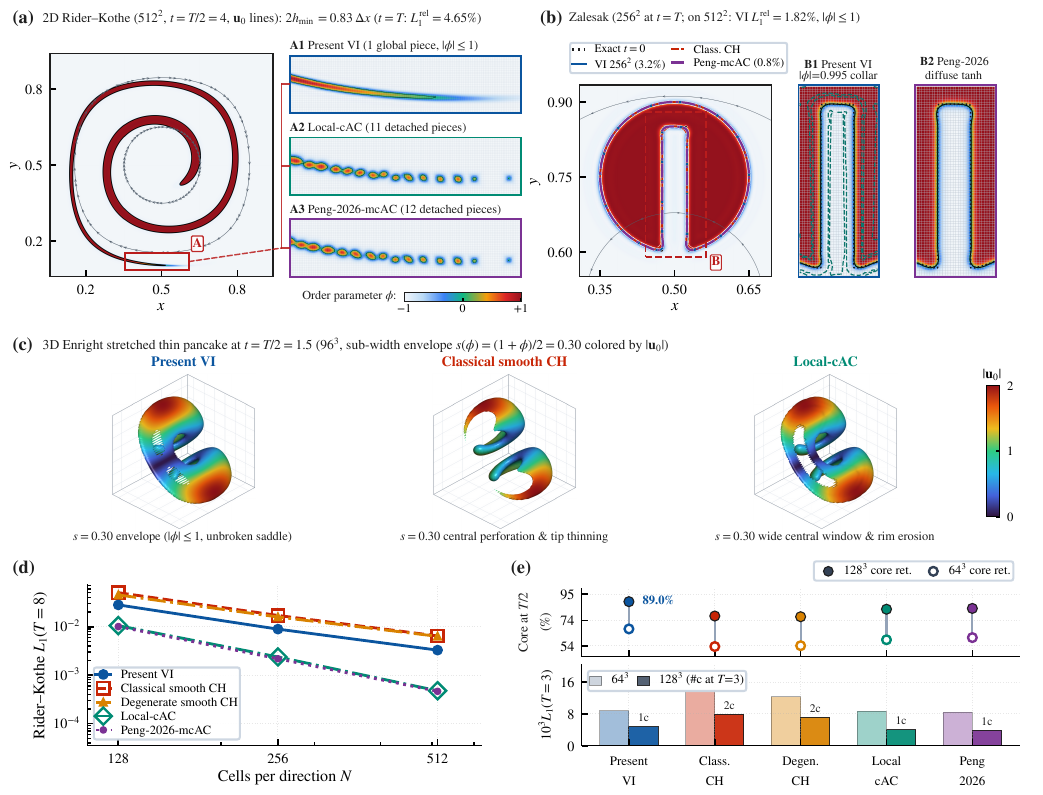}{width=\textwidth}
\caption{Matched-parameter comparison ($\epsilon=1.20\,\Delta x$, 9-point and 27-point isotropic Laplacians) across 2D and 3D kinematic stretching and rotation benchmarks against four diffuse-interface baselines. (a)~2D Rider--Kothe~\cite{Rider1998} reversible single-vortex spiral on $512^2$ at maximum stretching ($t=T/2$, $T=8.0$) with cell-registered Cartesian-mesh tip zooms: \texttt{A1}~present double-obstacle VI (labelled \texttt{Present VI}, $2h_{\min}=0.83\,\Delta x$ unbroken sub-cell ribbon, $1$ piece, zero bound violation); \texttt{A2}~conservative Allen--Cahn (\texttt{Local-cAC}, $11$ detached beads); \texttt{A3}~counter-diffusive cAC (\texttt{Peng-2026-mcAC}, $12$ detached beads). (b)~2D Zalesak~\cite{Zalesak1979} slotted-disk contours after one revolution ($256^2$ contours against the exact analytical $t=0$ slotted disk; $512^2$: $L_1^{\mathrm{rel}}=1.82\%$, zero bound violation) and Cartesian-mesh slot zooms contrasting the compact support collar (\texttt{B1}, $|\phi|\le 0.995$ span $<6\Delta x$, strict $\phi\equiv -1$ outside) against the unbounded $\tanh$ tail in \texttt{Peng-2026-mcAC} (\texttt{B2}). (c)~Three-dimensional Enright~\cite{Enright2002} stretched thin pancake at $t=1.5$ ($96^3$ sub-width sheet $s(\phi)=(1+\phi)/2=0.30$ colored by $|\bm{u}_0|\in[0,2]$): connected $s=0.30$ saddle bridge under the present VI (\texttt{Present VI + FCT4}) versus central perforation in classical smooth CH and window/rim erosion in \texttt{Local-cAC}. (d)~Three-grid ($128^2\to 512^2$) Rider--Kothe $L_1$ convergence. (e)~Paired 3D Enright ($64^3, 128^3$, continuous $\cos(\pi t/T_{\mathrm{rev}})$ clock) midpoint $\phi>0$ core volume retention (top, $88.96\%$ on $128^3$) and final $T=3.0$ $L_1$ shape error ($\times 10^{-3}$, bottom, annotated with $T=3$ connected-piece counts).}
\label{fig:adv_enright}
\end{figure}

To evaluate thin-filament survival and invariant-domain preservation under severe stretching, we compare the present formulation against four diffuse-interface models on identical MAC grids at matched parameters ($\epsilon = 1.20\,\Delta x$, second-order 9-point and 27-point isotropic Laplacians, and interface-localized $(1-\phi^2)_+$ regularization with $\alpha_c = 0.25/R_0$ for the counter-diffusive model of Peng et al.~\cite{Peng2026}; \Cref{fig:adv_enright} and \Cref{tab:advection_enright}): classical smooth CH~\cite{Cahn1958,Abels2012}, degenerate-mobility smooth CH~\cite{Gunnarsson2026_JCP}, the conservative Allen--Cahn model~\cite{Chiu2011,Jain2022} (labelled \texttt{Local-cAC}), and its counter-diffusive modification~\cite{Peng2026} (labelled \texttt{Peng-2026-mcAC}). In the reversible 2D Rider--Kothe single-vortex test ($T=8.0$, $128^2\to 512^2$, \Cref{fig:adv_enright}(a,d)) and Zalesak slotted-disk rotation (\Cref{fig:adv_enright}(b)), the two non-variational Allen--Cahn models yield smaller final $L_1$ shape errors after velocity reversal ($0.67\%$ and $0.66\%$ on the $512^2$ vortex, and $0.43\%$ and $0.41\%$ on Zalesak) than the present variational inequality ($4.65\%$ and $1.82\%$; on the $512^2$ vortex this improves by $1.9\times\text{--}2.0\times$ over classical and degenerate smooth CH at $9.12\%$ and $8.93\%$, whereas on Zalesak it lies between degenerate CH at $1.56\%$ and classical CH at $1.99\%$). Inspection at maximum stretching $t=T/2$ reveals the topological cost of that sharper reversed error: lacking variational capillary cohesion, their counter-diffusive fluxes shatter the stretched spiral into $11$ and $12$ detached beads on $512^2$ ($21$ and $20$ beads on $128^2$, \Cref{fig:adv_enright}(a, \texttt{A2}--\texttt{A3})) that re-merge only because the prescribed velocity reverses sign. The present formulation instead maintains a single unbroken sub-cell ribbon at $t=T/2$ ($2h_{\min}=0.83\,\Delta x$, $1$ connected piece, $0$ precipitates, exact $0.00$ bound violation), whereas leaving the counter-flux in~\cite{Peng2026} unlocalized or scaled as $\alpha_c\propto\epsilon^{-1}$ destabilizes pure-bulk modes and nucleates spurious precipitates as predicted by \Cref{thm:nogo}(d).

\begin{table}[t!]
\centering
\caption{Matched-parameter verification ($\epsilon=1.20\,\Delta x$, second-order 9-point and 27-point isotropic Laplacians, localized $\alpha_c=0.25/R_0$) across the 2D Rider--Kothe vortex ($T=8$), 2D Zalesak slotted disk ($1$ revolution), 3D Enright sphere deformation ($T=3$, continuous $\cos(\pi t/T_{\mathrm{rev}})$ flow), and coupled Navier--Stokes 2D unequal-droplet proximity stasis ($R_1=8\Delta x, R_2=16\Delta x$, gap $H=6\Delta x > \pi\epsilon$, $\mathcal{S}_0=0$, $2000$ coupled steps after single-droplet equilibrium relaxation; $n_{E\uparrow}$ counts steps with $\Delta\mathcal{E}_{\mathrm{tot},h}>0$).}
\label{tab:advection_enright}
\small
\setlength{\tabcolsep}{3.5pt}
\renewcommand{\arraystretch}{1.14}
\resizebox{\textwidth}{!}{%
\begin{tabular}{@{}lccccccc@{}}
\toprule
Formulation & 2D Vortex $512^2$ & Vortex $T/2$ & Zalesak $512^2$ & 3D Enright $128^3$ & Enright $T/2$ & Max bound viol. & Proximity loss $-\Delta V_1/V_{1,0}$ \\
 & $L_1$ ($L_1^{\mathrm{rel}}$) & pieces / precip. & $L_1^{\mathrm{rel}}$ & $L_1$ ($T=3$ pcs) & core ret. & $\max(|\phi|-1)_+$ & ($\|\bm{u}\|_\infty$, $n_{E\uparrow}$) \\
\midrule
\textbf{Present VI (\texttt{Present VI + FCT4})} & $\mathbf{3.28\times 10^{-3}}$ ($\mathbf{4.65\%}$) & $\mathbf{1}$ / $\mathbf{0}$ & $\mathbf{1.82\%}$ & $\mathbf{4.81\times 10^{-3}}$ ($\mathbf{1}$) & $\mathbf{88.96\%}$ & $\mathbf{0.00}$ & $\mathbf{0.000\%}$ ($\mathbf{1.7\times 10^{-16}}$, $\mathbf{0}$) \\
Classical smooth CH & $6.45\times 10^{-3}$ ($9.12\%$) & $1$ / $0$ & $1.99\%$ & $7.95\times 10^{-3}$ ($2$) & $77.62\%$ & $8.57\times 10^{-2}$ & $0.82\%$ ($9.8\times 10^{-3}$, $0$) \\
Degenerate smooth CH & $6.31\times 10^{-3}$ ($8.93\%$) & $1$ / $0$ & $1.56\%$ & $7.11\times 10^{-3}$ ($2$) & $77.06\%$ & $5.73\times 10^{-2}$ & $0.55\%$ ($8.4\times 10^{-3}$, $0$) \\
Conservative AC (\texttt{Local-cAC})~\cite{Chiu2011,Jain2022} & $4.77\times 10^{-4}$ ($0.67\%$) & $\mathbf{11}$ (beads) / $0$ & $0.43\%$ & $4.05\times 10^{-3}$ ($1$) & $82.99\%$ & $3.48\times 10^{-7}$ & $0.06\%$ ($5.1\times 10^{-3}$, $\mathbf{20}$) \\
Counter-diffusive cAC (\texttt{Peng-2026-mcAC})~\cite{Peng2026} & $4.63\times 10^{-4}$ ($0.66\%$) & $\mathbf{12}$ (beads) / $0$ & $0.41\%$ & $3.90\times 10^{-3}$ ($1$) & $83.62\%$ & $3.60\times 10^{-7}$ & $0.09\%$ ($3.4\times 10^{-3}$, $\mathbf{38}$) \\
\bottomrule
\end{tabular}%
}
\end{table}

Three-dimensional thin-sheet stretching in the Enright deformation benchmark~\cite{Enright2002} ($R_0=0.15$ sphere at $(0.35,0.35,0.35)$ stretched into a two-cell-thick curved sheet at $t=1.5$ and reversed continuously via $\cos(\pi t/T_{\mathrm{rev}})$ to $T=3.0$ across $64^3$, $96^3$, and $128^3$; \Cref{fig:adv_enright}(c,e) and \Cref{tab:advection_enright}) exhibits the same structural distinction. On $128^3$, the double-obstacle formulation retains $88.96\%$ of its $\phi>0$ core volume at $t=1.5$ ($99.86\%$ within the primary component and $1$ connected sub-width $s(\phi)\ge 0.15$ sheet, $80.32\%$ on $96^3$), exceeding classical and degenerate smooth CH ($77.62\%$ and $77.06\%$) as well as both Allen--Cahn variants ($82.99\%\text{--}83.62\%$). At $T=3.0$, it recovers a single sphere with $L_1 = 4.81\times 10^{-3}$ ($1.65\times$ and $1.48\times$ lower than classical and degenerate smooth CH, both of which leave $2$ detached fragments and $5.73\%\text{--}8.57\%$ phase overshoots) at exact $0.00$ bound violation. Controlled spatial ablations on $256^2$ confirm that fifth-order upwind reconstruction lowers $L_1$ error by $3.24\times$ (vortex) and $4.68\times$ (Zalesak) over second-order MUSCL-FCT and by $12.59\times$ and $25.71\times$ over first-order upwind, while increasing Zalesak FCT passes from $P=1$ to $P=4$ monotonically reduces the peak per-step obstacle-kink advection energy increment ($+5.84\times 10^{-4}\to +5.63\times 10^{-4}$; $80$-step cumulative drift $+7.83\times 10^{-3}\to +7.75\times 10^{-3}$) at $0.0$ bound violation (\Cref{thm:wb_fct}(c)).

Coupling phase transport to the Navier--Stokes equations in a quiescent unequal-droplet proximity test ($R_1=8\Delta x$, $R_2=16\Delta x$, gap $H=6\Delta x$, $\mathcal{S}_0=0$, \Cref{tab:advection_enright}) separates variational compact support from non-variational sharpening. At $\epsilon=1.20\,\Delta x$, where $H - \pi\epsilon = 2.23\,\Delta x > 2\Delta x$ leaves pure-bulk cells isolating the two active collars, smooth CH models drive continuous Ostwald-ripening drainage ($0.55\%\text{--}0.82\%$ volume loss over $2000$ steps, $\|\bm{u}\|_\infty = 8.4\text{--}9.8\times 10^{-3}$), while both non-variational Allen--Cahn models excite $\mathcal{O}(10^{-3})$ parasitic currents and violate discrete free-energy monotonicity ($n_{E\uparrow}=20$ and $n_{E\uparrow}=38$ uphill steps with $\Delta\mathcal{E}_{\mathrm{tot},h}>0$). The present formulation preserves $R_1$ to machine precision ($\Delta V_1/V_{1,0} = 0.000\%$, $\|\bm{u}\|_\infty = 1.72\times 10^{-16}$, Young--Laplace errors $3.65\%$ and $5.97\%$, $n_{E\uparrow}=0$, $\max(|\phi|-1)_+ = 0.00$). Widening $\epsilon$ to $1.80\,\Delta x$ at fixed $H=6.0\,\Delta x$ shrinks $H - \pi\epsilon$ to $0.35\,\Delta x$ (a single bulk cell), causing the multi-shell collar stencil to bridge the gap and induce coalescence across all models ($18.9\%$ volume transfer here versus $38.2\%\text{--}41.0\%$ in the smooth baselines), which confirms the disjoint-halo requirement $H > \pi\epsilon + 2\Delta x$ of \Cref{thm:wb_fct}(a).

\subsection{Buoyancy-driven benchmark flows}\label{subsec:bench_hysing_rt}

\begin{figure}[t!]
\centering
\cmamefig{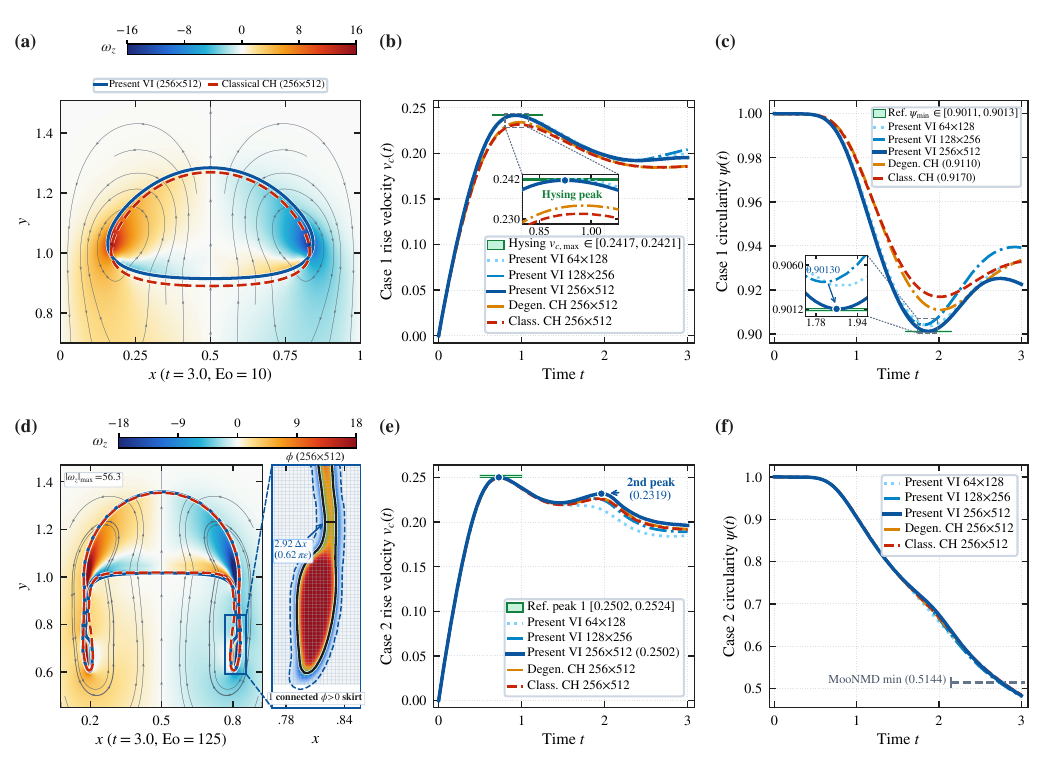}{width=\textwidth}
\caption{Quantitative rising-bubble verification against the sharp-interface finite-element benchmark suite of Hysing et al.~\cite{Hysing2009} on $[0,1]\times[0,2]$ ($64\times 128 \to 256\times 512$; bubble interior $\phi=+1$, ambient liquid $\phi=-1$). Top row (a--c), Case~1 (ambient-to-bubble ratios $\rho_-/\rho_+=10$, $\eta_-/\eta_+=10$, $\mathrm{Re}=35$, $\mathrm{Eo}=10$): (a)~$t=3$ wake vorticity $\omega_z\in[-16,+16]$, recirculation streamlines, and $256\times 512$ bubble contours (present VI labelled \texttt{Present VI} vs.\ classical CH); (b)~rise velocity $v_c(t)$ with zoomed consensus-band peak inset ($[0.2417,0.2421]$); (c)~circularity $\psi(t)$ with zoomed minimum inset ($[0.9011,0.9013]$). Bottom row (d--f), Case~2 ($\rho_-/\rho_+=1000$, $\eta_-/\eta_+=100$, $\mathrm{Re}=35$, $\mathrm{Eo}=125$): (d)~$t=3$ wake vorticity $\omega_z$ (peak $|\omega_z|_{\max}=56.3$), streamlines, and $256\times 512$ sub-cell Cartesian-mesh callout resolving the unbroken trailing skirt ($\phi=0$ shaft thickness $2.92\,\Delta x = 0.62\,\pi\epsilon$, $1$ connected $\phi>0$ component, zero satellite breakup); (e)~rise velocity $v_c(t)$ against consensus peak $[0.2502,0.2524]$; (f)~circularity $\psi(t)$ against the \texttt{MooNMD} sharp-interface reference minimum ($0.5144$).}
\label{fig:hysing}
\end{figure}

\begin{figure}[t!]
\centering
\cmamefig{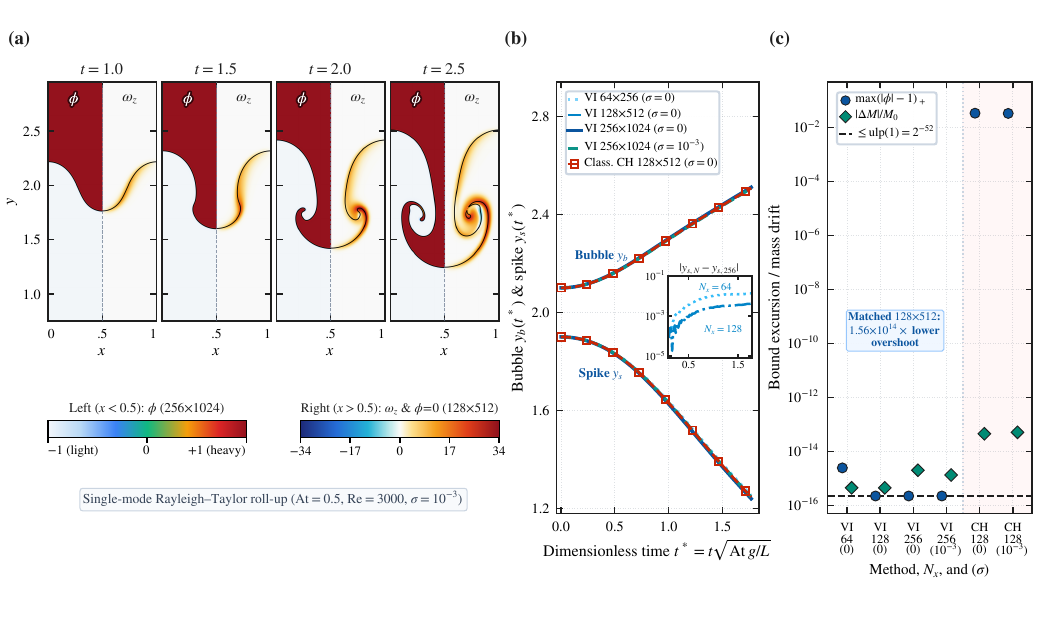}{width=\textwidth}
\caption{Single-mode Rayleigh--Taylor instability ($\mathrm{At}=0.5$, $\rho_+/\rho_-=3$, $\mathrm{Re}=3000$ on $[0,1]\times[0,4]$, $64\times 256 \to 256\times 1024$). (a)~Symmetry-split Kelvin--Helmholtz mushroom roll-up sequence ($t\in\{1.0,1.5,2.0,2.5\}$, $\sigma=10^{-3}$), pairing $256\times 1024$ phase field $\phi$ ($x<0.5$) with matched $128\times 512$ baroclinic vorticity $\omega_z\in[-34,+34]$ and $\phi=0$ interface contour ($x>0.5$). (b)~Three-grid convergence of bubble ($y_b$) and spike ($y_s$) tip trajectories versus dimensionless time $t^* = t\sqrt{\mathrm{At}\,g/L}$ (inset: semilogy spike-tip grid convergence $|y_s(N)-y_s(256\times 1024)|$). (c)~Logarithmic audit of phase-bound excursion $\max(|\phi|-1)_+$ and global mass drift $|\Delta M|/M_0$: the present VI holds bound excursions at or below $\mathrm{ulp}(1) = 2^{-52} \approx 2.22\times 10^{-16}$ across both $128\times 512$ and $256\times 1024$, outperforming matched-slope $128\times 512$ classical smooth CH ($3.46\times 10^{-2}$) by at least $1.56\times 10^{14}\times$.}
\label{fig:rt}
\end{figure}

Variable-density Abels--Garcke--Gr{\"u}n momentum coupling (\Cref{eq:gov_agg_do}) is tested next on the rising-bubble benchmarks of Hysing et al.~\cite{Hysing2009}, together with their diffuse-interface benchmark counterpart established by Aland and Voigt~\cite{Aland2012_IJNMF}, and on the single-mode Rayleigh--Taylor instability of Tryggvason~\cite{Tryggvason1988,Guermond2000} (\Cref{fig:hysing,fig:rt} and \Cref{tab:hysing_rt}). For a fair interface-thickness comparison between $\sin(z/\epsilon_{\mathrm{DO}})$ and $\tanh(z/(\sqrt{2}\epsilon_{\mathrm{GL}}))$ profiles, all models in \Cref{tab:hysing_rt} use the identical second-order 9-point isotropic Laplacian and matched zero-crossing slope thickness $1/|\phi'(0)| = 1.50\,\Delta x$ ($\epsilon_{\mathrm{DO}}=1.50\,\Delta x$, $\epsilon_{\mathrm{GL}}=\frac{1.50}{\sqrt{2}}\,\Delta x = 1.061\,\Delta x$; smooth CH baselines additionally require clamping $\operatorname{clip}(s(\phi),0,1)$ in \Cref{eq:rho_eta_laws} to prevent bulk overshoots from creating negative densities at $10^3$ contrast, whereas $\phi\in[-1,1]$ requires none). In Hysing Case~1 ($\rho_-/\rho_+=10$, \Cref{fig:hysing}(a--c)), refining the MAC grid from $64\times 128$ to $256\times 512$ brings all three metrics to within $0.05\%$ of the sharp-interface consensus band: $v_{c,\max}=0.24169$ at $t=0.924$ ($10^{-5}$ below $[0.2417,0.2421]$, after $0.24179$ on $128\times 256$), $\psi_{\min}=0.90130$ at $t=1.860$ (inside $[0.9011,0.9013]$), and $y_c(3)=1.08219$ ($0.045\%$ above $[1.0799,1.0817]$), with area drift $|\Delta A|/A_0 = 2.47\times 10^{-4}$ and bound excursion $5.55\times 10^{-15}$. On the same $256\times 512$ grid, classical and degenerate smooth CH incur $3.1\%\text{--}4.2\%$ peak-velocity deficits ($v_{c,\max}=0.23162$ and $0.23411$; $3.6\%\text{--}5.0\%$ at unscaled $\epsilon_{\mathrm{GL}}=1.50\,\Delta x$) and $1.6\%\text{--}1.8\%$ centroid lag ($y_c(3)=1.0621$ and $1.0643$) due to bulk chemical-potential shifts and tangential surface diffusion.

\begin{table}[t!]
\centering
\caption{Quantitative comparison on Hysing et al.~\cite{Hysing2009} rising-bubble Cases~1 ($\rho_-/\rho_+=10$) and~2 ($\rho_-/\rho_+=1000$) and 2D Rayleigh--Taylor instability ($\mathrm{At}=0.5, \rho_+/\rho_-=3, \mathrm{Re}=3000$) using second-order 9-point Laplacians with fourth-order directional isotropy and matched zero-crossing slope thickness $1/|\phi'(0)|=1.50\,\Delta x$ ($\epsilon_{\mathrm{DO}}=1.50\,\Delta x$, $\epsilon_{\mathrm{GL}}=\frac{1.50}{\sqrt{2}}\,\Delta x = 1.061\,\Delta x$). Reference intervals denote the published \texttt{TP2D}/\texttt{FreeLIFE}/\texttt{MooNMD} consensus bands~\cite{Hysing2009}.}
\label{tab:hysing_rt}
\small
\setlength{\tabcolsep}{3.8pt}
\renewcommand{\arraystretch}{1.14}
\resizebox{\textwidth}{!}{%
\begin{tabular}{@{}llccccccc@{}}
\toprule
Benchmark & Formulation / Grid & $v_{c,\max 1}$ ($t_{\max 1}$) & $\psi_{\min}$ ($t_{\psi_{\min}}$) & $y_c(3)$ or $y_s(2.5)$ & $|\Delta A|/A_0$ & $|\Delta M|/M_0$ & $\max(|\phi|-1)_+$ & Pcs \\
\midrule
Hysing Case 1 & Sharp ref.\ band~\cite{Hysing2009} & $[0.2417, 0.2421]$ ($0.922$) & $[0.9011, 0.9013]$ ($1.90$) & $[1.0799, 1.0817]$ & $0$ & $0$ & $0$ & $1$ \\
($10\!:\!1$, $\mathrm{Eo}=10$) & Present VI $128\times 256$ & $0.24179$ ($0.930$) & $0.90417$ ($1.820$) & $1.08433$ & $4.53\times 10^{-3}$ & $6.21\times 10^{-15}$ & $9.77\times 10^{-15}$ & $1$ \\
 & \textbf{Present VI $256\times 512$} & $\mathbf{0.24169}$ ($\mathbf{0.924}$) & $\mathbf{0.90130}$ ($\mathbf{1.860}$) & $\mathbf{1.08219}$ & $\mathbf{2.47\times 10^{-4}}$ & $\mathbf{1.51\times 10^{-14}}$ & $\mathbf{5.55\times 10^{-15}}$ & $\mathbf{1}$ \\
 & Degenerate CH $256\times 512$ & $0.23411$ ($0.984$) & $0.91100$ ($2.020$) & $1.06432$ & $2.15\times 10^{-3}$ & $4.10\times 10^{-13}$ & $1.24\times 10^{-14}$ & $1$ \\
 & Classical CH $256\times 512$ & $0.23162$ ($0.964$) & $0.91696$ ($2.020$) & $1.06209$ & $3.22\times 10^{-3}$ & $1.15\times 10^{-12}$ & $1.65\times 10^{-2}$ & $1$ \\
\midrule
Hysing Case 2 & Sharp ref.\ band~\cite{Hysing2009} & $[0.2502, 0.2524]$ ($0.73$) & $0.5144$ (\texttt{MooNMD}) & $[1.1249, 1.1380]$ & $0$ & $0$ & $0$ & $1\text{--}3$ \\
($1000\!:\!1$, $\mathrm{Eo}=125$) & Present VI $128\times 256$ & $0.25024$ ($0.730$) & $0.48107$ ($3.000$) & $1.11198$ & $2.75\times 10^{-3}$ & $5.65\times 10^{-15}$ & $3.98\times 10^{-13}$ & $1$ \\
 & \textbf{Present VI $256\times 512$} & $\mathbf{0.25022}$ ($\mathbf{0.725}$) & $\mathbf{0.48332}$ ($\mathbf{3.000}$) & $\mathbf{1.12182}$ & $\mathbf{9.18\times 10^{-5}}$ & $\mathbf{1.84\times 10^{-14}}$ & $\mathbf{3.56\times 10^{-13}}$ & $\mathbf{1}$ \\
 & Degenerate CH $256\times 512$ & $0.24977$ ($0.745$) & $0.48332$ ($3.000$) & $1.11518$ & $1.88\times 10^{-3}$ & $3.68\times 10^{-13}$ & $1.32\times 10^{-9}$ & $1$ \\
 & Classical CH $256\times 512$ & $0.24959$ ($0.725$) & $0.48554$ ($3.000$) & $1.11391$ & $3.82\times 10^{-3}$ & $7.00\times 10^{-13}$ & $1.92\times 10^{-2}$ & $1$ \\
\midrule
Rayleigh--Taylor & Present VI $256\times 1024$ ($\sigma=0$) & --- & $0.81655$ ($2.475$) & $y_s=1.2401, y_b=2.5101$ & $9.97\times 10^{-4}$ & $2.00\times 10^{-15}$ & $\mathbf{2.22\times 10^{-16}}$ & $5$ \\
($\mathrm{At}=0.5$, $3000$) & \textbf{Present VI $256\times 1024$ ($\sigma=10^{-3}$)} & --- & $\mathbf{0.85370}$ ($\mathbf{2.500}$) & $\mathbf{y_s=1.2470, y_b=2.5052}$ & $\mathbf{1.24\times 10^{-5}}$ & $\mathbf{1.33\times 10^{-15}}$ & $\mathbf{2.22\times 10^{-16}}$ & $\mathbf{1}$ \\
 & Classical CH $128\times 512$ ($\sigma=0$) & --- & $0.84052$ ($2.500$) & $y_s=1.2468, y_b=2.5069$ & $8.30\times 10^{-4}$ & $4.50\times 10^{-14}$ & $3.46\times 10^{-2}$ & $3$ \\
\bottomrule
\end{tabular}%
}
\end{table}

In the $\rho_-/\rho_+=1000$ Hysing Case~2 (\Cref{fig:hysing}(d--f)), the double-obstacle formulation captures both the primary rise-velocity peak $v_{c,\max 1}=0.25022$ at $t=0.725$ (inside $[0.2502,0.2524]$) and the secondary plateau peak $v_{c,\max 2}=0.23190$, while holding bubble area drift to $9.18\times 10^{-5}$ ($42\times$ tighter than classical CH and $21\times$ tighter than degenerate CH). Two unbroken trailing skirts form symmetrically behind the rim ($\phi=0$ shaft thickness $2.92\,\Delta x = 0.62\,\pi\epsilon$, tapering into a sub-width raised-cosine tip within a single connected $\phi>0$ component and matching fine-grid \texttt{FreeLIFE}/\texttt{MooNMD} skirt coherence without premature droplet snapping), and under Cahn scaling ($\epsilon_{\mathrm{DO}}=1.50\,\Delta x$) the terminal centroid height converges monotonically toward $[1.1249,1.1380]$ ($y_c(3)=1.1005 \to 1.1120 \to 1.1218$ across $64\times 128 \to 128\times 256 \to 256\times 512$, within $0.27\%$ of \texttt{FreeLIFE} and $1.39\%$ of \texttt{MooNMD}).

On the $\mathrm{Re}=3000$ single-mode Rayleigh--Taylor instability (\Cref{fig:rt} and \Cref{tab:hysing_rt}), spike and bubble tip trajectories collapse across $64\times 256$, $128\times 512$, and $256\times 1024$. Throughout secondary Kelvin--Helmholtz vortex roll-up, four-pass FCT with \Cref{eq:M_ONRO} keeps phase-bound excursions at or below $\mathrm{ulp}(1) = 2.22\times 10^{-16}$ on both $128\times 512$ and $256\times 1024$, improving on matched-slope $128\times 512$ classical smooth CH ($3.46\times 10^{-2}$) by more than fourteen orders of magnitude.

\subsection{Verification of Ostwald ripening fluxes and screened disjoining repulsion}\label{subsec:bench_rip_dlvo}

\begin{figure}[t!]
\centering
\cmamefig{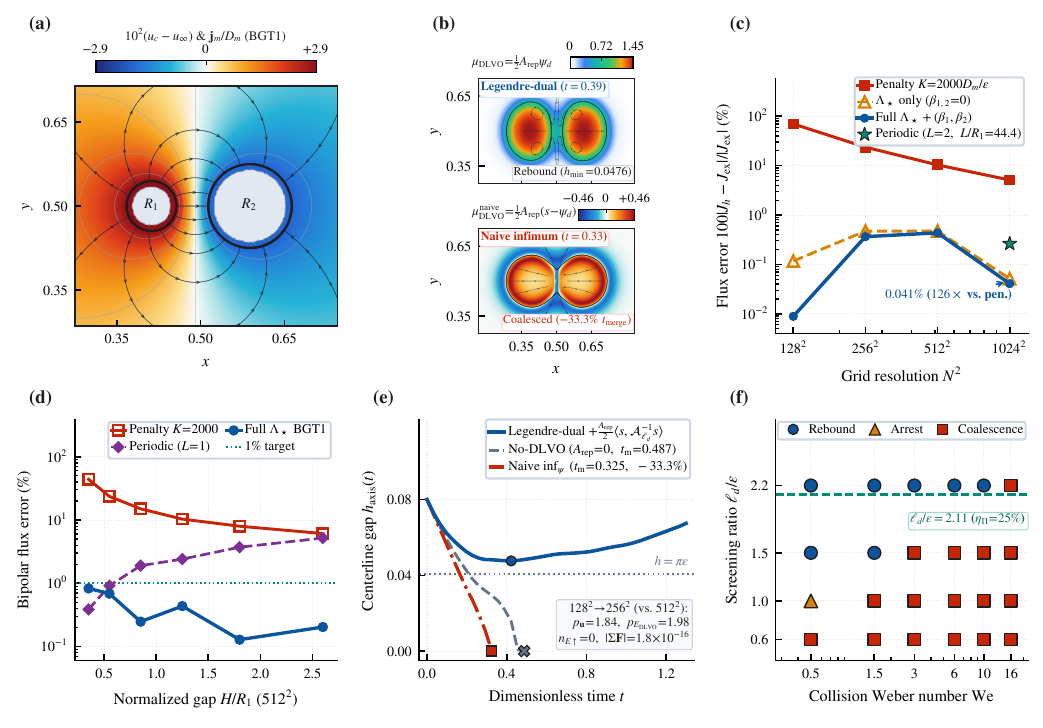}{width=\textwidth}
\caption{Verification of $\Lambda_\star$-renormalized Robin--Kelvin ripening and Legendre-dual Yukawa--DLVO repulsion (\Cref{prop:dlvo_cont,prop:ripening_sl,thm:dlvo_rip}). (a)~Two-dimensional BGT1 monomer supersaturation dipole $10^2(u_c - u_\infty)\in[-2.90,+2.90]$ and dimensionless diffusive flux lines $\bm{j}_m/D_m = -(1-s(\phi))\nabla u_c$ ($R_1=0.045, R_2=0.075, H/R_1=1.25$). (b)~Unmollified sign-duality collision snapshots ($\mathrm{We}=2.0, \ell_d/\epsilon=2.0$): top shows the Legendre-dual repulsive cushion $\mu_{\mathrm{DLVO}} = \frac{1}{2}A_{\mathrm{rep}}\psi_d\in[0,1.45]$ arresting approach at $t=0.39$ ($h_{\min}=0.048$); bottom shows the naive-infimum attractive potential $\mu_{\mathrm{DLVO}}^{\mathrm{naive}} = \frac{1}{2}A_{\mathrm{rep}}(s-\psi_d)\in[-0.46,+0.46]$ driving accelerated neck fusion at $t=0.33$. (c)~Four-grid Cahn-scaling convergence ($128^2\to 1024^2$, $\epsilon=1.85\Delta x$) against exact infinite-space bipolar Laplace flux: BGT1 reaches $-0.0407\%$ at $1024^2$ ($126\times$ more accurate than naive penalty $K=2000\,D_m/\epsilon$, $+5.12\%$; $L/R_1=44.4$ periodic box: $+0.263\%$). (d)~Six-gap bipolar sweep ($H/R_1\in[0.35,2.60]$ on $512^2$). (e)~Centerline film gap history $h_{\mathrm{axis}}(t)$ (terminated upon topological merge $N_{\mathrm{comp}}\!:\!2\to 1$, with collar-contact line $h=\pi\epsilon$) confirming elastic rebound ($n_{E\uparrow}=0$, four-grid $p_{\bm{u}}=1.84$, $p_{E_{\mathrm{DLVO}}}=1.98$) versus $33.3\%$ faster fusion under \Cref{eq:F_dlvo_naive}. (f)~Discrete $6\times 4$ $(\ell_d/\epsilon,\mathrm{We})$ bounce--arrest--coalescence outcome map against the analytical $\eta_\Pi=25\%$ ($\ell_d/\epsilon=2.11$) threshold (\Cref{eq:eta_pi_exact}).}
\label{fig:w3_rip_dlvo}
\end{figure}

\begin{figure}[tp!]
\centering
\cmamefig{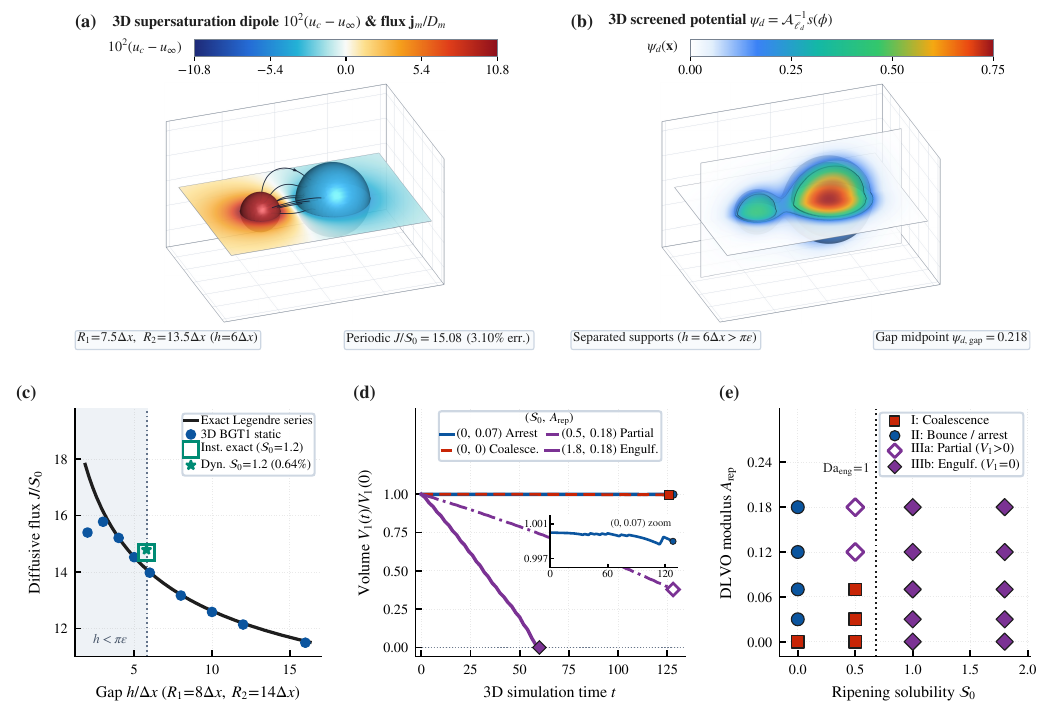}{width=0.98\textwidth}
\caption{Three-dimensional Jeffery--Stimson bispherical verification and non-reciprocal intact-film chasing phase diagram. (a)~Three-dimensional scaled monomer supersaturation dipole $10^2(u_c - u_\infty)\in[-10.8,+10.8]$ on the equatorial plane and 3D diffusive flux lines $\bm{j}_m/D_m$ arcing from $R_1=7.5\Delta x$ to $R_2=13.5\Delta x$ on a periodic $128\times 96^2$ box ($h=6\Delta x$, static $J/\mathcal{S}_0=15.08$, $3.10\%$ vs.\ infinite-space exact $14.62$). (b)~Quadrant cutaway of the Legendre-dual Yukawa--DLVO screened potential $\psi_d(\bm{x})\in[0,0.75]$ showing the positive repulsive saddle cushion ($\psi_{d,\mathrm{gap}}=0.218$) across $h=6\Delta x > \pi\epsilon$. (c)~Normalized 3D diffusive flux $J/\mathcal{S}_0$ versus gap $h/\Delta x\in[2,16]$ ($R_1=8\Delta x, R_2=14\Delta x$ on $128\times 96^2$ with BGT1 far-field radiation and curved $(\Lambda_\star,\beta_1,\beta_2^{(3)})$) against the exact bispherical Legendre series~\cite{Jeffery1912} ($0.02\%\text{--}0.49\%$ across $h\ge 4\Delta x$, shaded collar-overlap zone $h<\pi\epsilon$; dynamic chasing at $\mathcal{S}_0=1.2, \bar{h}=5.76\Delta x$: $0.64\%$). (d)~Small-droplet component volume histories $V_1(t)/V_1(0)$ across contact coalescence ($(0,0)$), intact-film DLVO arrest/rebound ($(0,0.07)$, $n_{E\uparrow}=0$), partial intact-film ripening ($(0.5,0.18)$), and complete intact-film engulfment ($(1.8,0.18)$). (e)~Twenty-case 3D $(\mathcal{S}_0,A_{\mathrm{rep}})$ regime map ($\mathcal{S}_0\in\{0,0.5,1.0,1.8\}\times A_{\mathrm{rep}}\in\{0,0.03,0.07,0.12,0.18\}$) across the vertical Damk{\"o}hler boundary $\mathrm{Da}_{\mathrm{eng}}=1$ ($\mathcal{S}_{0,c}=0.678$).}
\label{fig:chasing3d}
\end{figure}

Boundary-layer-renormalized Ostwald ripening (\Cref{prop:ripening_sl}) and discrete mass conservation (\Cref{thm:dlvo_rip}(c)) are verified against exact two-dimensional bipolar and three-dimensional bispherical Laplace solutions in \Cref{fig:w3_rip_dlvo}(a,c,d), \Cref{fig:chasing3d}(a,c), and \Cref{tab:ripening_w3a}. Coupling the curved-obstacle Sturm--Liouville constants $(\Lambda_\star,\beta_1,\beta_2^{(2)})$ (\Cref{eq:lambda_star_betas}) with self-consistent BGT1 dipole radiation boundaries holds the 2D bipolar ripening flux error within $0.44\%$ across all four grids and reaches $-0.0407\%$ at $1024^2$ ($+0.263\%$ on an $L=2.0$ periodic box with $L/R_1=44.4$), whereas a high-stiffness penalty $K=2000\,D_m/\epsilon$ incurs $+69.96\%$ error on $128^2$ and $+5.12\%$ at $1024^2$ ($126\times$ larger, and $+73.56\%$ under fixed-$\epsilon$ refinement due to outer-collar pinning). In three dimensions (\Cref{fig:chasing3d}(a,c)), GMG-PCG solves of \Cref{eq:ripening_pde} with BGT1 radiation boundaries and $(\Lambda_\star,\beta_1,\beta_2^{(3)})$ match the exact bispherical Legendre series of Jeffery~\cite{Jeffery1912} within $0.02\%\text{--}0.49\%$ across all seven resolved gaps $h\in[4,16]\Delta x$ ($0.43\%\text{--}0.90\%$ under periodic boundaries on $128\times 96^2$, $3.10\%$ on the smaller-sphere periodic visualization box of \Cref{fig:chasing3d}(a), and $2.26\%$ at $h=3\Delta x < \pi\epsilon=5.81\Delta x$ once diffuse collars overlap).

\begin{table}[t!]
\centering
\caption{Analytical verification of $\Lambda_\star$-renormalized Robin--Kelvin Ostwald ripening fluxes: four-grid 2D bipolar Laplace convergence ($R_1=0.045, R_2=0.075, H/R_1=1.25, \epsilon=1.85\Delta x$) with BGT1 radiation and $L=2.0$ periodic boundaries, and 3D bispherical gap sweep ($R_1=8\Delta x, R_2=14\Delta x$ on $128\times 96^2$ with BGT1 radiation) against the exact bispherical Legendre series of Jeffery~\cite{Jeffery1912}. At $\mathcal{S}_0=0$, $\max|q_\phi|\equiv 0.0$ identically ($^*$denotes collar overlap $h<\pi\epsilon=5.81\Delta x$).}
\label{tab:ripening_w3a}
\small
\setlength{\tabcolsep}{4.4pt}
\renewcommand{\arraystretch}{1.14}
\resizebox{\textwidth}{!}{%
\begin{tabular}{@{}lccccclccc@{}}
\toprule
\multicolumn{5}{c}{2D Bipolar Laplace Convergence ($\epsilon=1.85\Delta x$)} & & \multicolumn{4}{c}{3D Bispherical BGT1 Gap Sweep ($128\times 96^2$)} \\
\cmidrule(r){1-5}\cmidrule(l){7-10}
Grid / BC & Full $\Lambda_\star+(\beta_1,\beta_2^{(2)})$ & $\Lambda_\star$ only & Naive $K=2000$ & $\sum q_\phi\Delta x^2$ & & Surface gap $h$ & Present VI $J/\mathcal{S}_0$ & Exact series~\cite{Jeffery1912} & Rel.\ diff. \\
\midrule
$128^2$ BGT1 & $+0.009\%$ & $-0.117\%$ & $+69.958\%$ & $-1.28\times 10^{-15}$ & & $h=3.0\Delta x^*$ ($<\pi\epsilon$) & $15.784$ & $16.149$ & $2.26\%$ \\
$256^2$ BGT1 & $-0.365\%$ & $-0.472\%$ & $+23.680\%$ & $-3.20\times 10^{-15}$ & & $h=4.0\Delta x$ & $15.211$ & $15.217$ & $\mathbf{0.04\%}$ \\
$512^2$ BGT1 & $-0.436\%$ & $-0.473\%$ & $+10.353\%$ & $-5.67\times 10^{-15}$ & & $h=6.0\Delta x$ & $13.977$ & $13.993$ & $\mathbf{0.11\%}$ \\
$\mathbf{1024^2}$ \textbf{BGT1} & $\mathbf{-0.041\%}$ & $-0.051\%$ & $+5.121\%$ & $-1.27\times 10^{-14}$ & & $h=10.0\Delta x$ & $12.585$ & $12.623$ & $\mathbf{0.30\%}$ \\
$\mathbf{1024^2}$ \textbf{Per.\ ($L=2$)} & $\mathbf{+0.263\%}$ & --- & --- & $-2.99\times 10^{-15}$ & & $h=16.0\Delta x$ & $11.493$ & $11.550$ & $\mathbf{0.49\%}$ \\
\bottomrule
\end{tabular}%
}
\end{table}

\begin{table}[t!]
\centering
\caption{Verification of discrete Legendre-dual Yukawa--DLVO mechanics and 3D coupled chasing regimes: 2D four-grid rebound convergence ($\epsilon=0.024, \ell_d=0.048, \mathrm{We}=2.5$), $128^2$ variational sign-duality collision ($\epsilon=1.65/128\approx 0.0129, \ell_d=2\epsilon\approx 0.0258, \mathrm{We}=2.0, A_{\mathrm{rep}}=3.0$), and 3D 20-case $(\mathcal{S}_0,A_{\mathrm{rep}})$ intact-film chasing map ($96\times 64^2$, $\ell_d/\epsilon=2.70$).}
\label{tab:dlvo_chasing}
\small
\setlength{\tabcolsep}{3.8pt}
\renewcommand{\arraystretch}{1.14}
\resizebox{\textwidth}{!}{%
\begin{tabular}{@{}llcccccc@{}}
\toprule
Suite / Case & Formulation & Outcome & $n_{E\uparrow}$ & $\max\Delta\mathcal{E}$ & $|\sum\bm{f}_{\mathrm{DLVO}}|$ & MAC \eqref{eq:mac_product_rule} / Helm.\ res. & Convergence / Regime metric \\
\midrule
2D Conv.\ $64^2$ & Legendre-dual & Bounce & $0$ & $-1.01\times 10^{-5}$ & $1.25\times 10^{-16}$ & $1.29\times 10^{-15}$ & $\|\bm{u}-\bm{u}_{512}\|_{L_2}=6.68\times 10^{-3}$ \\
2D Conv.\ $128^2$ & Legendre-dual & Bounce & $0$ & $-5.27\times 10^{-6}$ & $1.46\times 10^{-16}$ & $2.62\times 10^{-15}$ & $2.55\times 10^{-3}$ ($p=1.39$) \\
2D Conv.\ $256^2$ & Legendre-dual & Bounce & $0$ & $-2.65\times 10^{-6}$ & $1.67\times 10^{-16}$ & $5.28\times 10^{-15}$ & $\mathbf{7.13\times 10^{-4}}$ ($\bm{p=1.84}$) \\
\midrule
2D Sign Duality & \textbf{Legendre-dual \eqref{eq:F_DLVO_dual}} & \textbf{Bounce ($2$ drops)} & $\mathbf{0}$ & $-1.80\times 10^{-6}$ & $1.80\times 10^{-16}$ & $4.14\times 10^{-15}$ & $h_{\min}=0.0476$ \\
($\mathrm{We}=2.0, 128^2$) & \textbf{Naive infimum \eqref{eq:F_dlvo_naive}} & \textbf{Fuse ($33.3\%$ faster)} & $0$ & $-1.52\times 10^{-5}$ & $9.28\times 10^{-17}$ & --- & $t_{\mathrm{merge}}=0.3250$ \\
 & No-DLVO ($A_{\mathrm{rep}}=0$) & Fuse ($1$ drop) & $0$ & $-8.61\times 10^{-6}$ & $0.00$ & --- & $t_{\mathrm{merge}}=0.4875$ \\
\midrule
3D Chasing & $(0,0)$ \& $\mathcal{S}_0=0.5, A_{\mathrm{rep}}\le 0.07$ ($4$ runs) & Regime I: Coalesce & --- & --- & $\le 2.89\times 10^{-15}$ & $\le 1.5\times 10^{-14}$ & $|\Delta M|/M_0 \le 4.52\times 10^{-13}$ \\
($4\times 5$ map) & $\mathcal{S}_0=0$, $A_{\mathrm{rep}}\in[0.03,0.18]$ ($4$ runs) & Regime II: Bounce & $\mathbf{0}$ & $\le -5.8\times 10^{-5}$ & $\le 7.11\times 10^{-15}$ & $\le 1.4\times 10^{-14}$ & $|\Delta M|/M_0 \le 4.74\times 10^{-16}$ \\
 & $\mathcal{S}_0=0.5$, $A_{\mathrm{rep}}\in\{0.12,0.18\}$ ($2$ runs) & Regime IIIa: Part.\ rip. & --- & --- & $\le 7.11\times 10^{-15}$ & $\le 1.5\times 10^{-14}$ & $\mathrm{Da}_{\mathrm{eng}}=0.74<1$ ($V_1/V_{1,0}\approx 0.37$) \\
 & $\mathcal{S}_0\in\{1.0,1.8\}$, $A_{\mathrm{rep}}\ge 0$ ($10$ runs) & Regime IIIb: Full engulf. & --- & --- & $\le 7.99\times 10^{-15}$ & $\le 1.5\times 10^{-14}$ & $\mathrm{Da}_{\mathrm{eng}} > 1$ ($V_1\to 0$, intact film) \\
\bottomrule
\end{tabular}%
}
\end{table}

Head-on droplet collisions in \Cref{fig:w3_rip_dlvo}(b,e,f) and \Cref{tab:dlvo_chasing} confirm the variational sign duality and barrier efficiency law of \Cref{prop:dlvo_cont,thm:dlvo_rip}(a,b). On matched unmollified $128^2$ collisions ($\mathrm{We}=2.0$, $\ell_d/\epsilon=2.0$, $A_{\mathrm{rep}}=3.0$), minimizing the naive infimum functional \Cref{eq:F_dlvo_naive} pulls the droplets into coalescence $33.3\%$ faster than the unstabilized $A_{\mathrm{rep}}=0$ control ($t_{\mathrm{merge}}=0.3250$ versus $0.4875$), whereas the Legendre-dual functional \Cref{eq:F_DLVO_dual} produces elastic rebound ($h_{\min}=0.0476 > \pi\epsilon = 0.0405$) with discrete MAC product-rule residual $4.14\times 10^{-15}$, net domain force $1.80\times 10^{-16}$, and monotonic total-energy decay ($n_{E\uparrow}=0$ across all four grids and all 24 runs of the $(\ell_d/\epsilon,\mathrm{We})$ map, converging at $p_{\bm{u}}=1.84$ and $p_{E_{\mathrm{DLVO}}}=1.98$). Across the $6\times 4$ $(\ell_d/\epsilon,\mathrm{We})$ map (\Cref{fig:w3_rip_dlvo}(f)), under-resolved screening at $\ell_d/\epsilon = 0.6$ ($\eta_\Pi = 1.77\%$ per \Cref{eq:eta_pi_exact}) coalesces across all $\mathrm{We}\in[0.5,16.0]$, $\ell_d/\epsilon = 1.0$ ($\eta_\Pi=6.80\%$) arrests approach only at $\mathrm{We}=0.5$, and resolved screening $\ell_d/\epsilon = 2.2 \ge 2.11$ ($\eta_\Pi = 26.39\%$) sustains rebound across $\mathrm{We}\in[0.5,10.0]$ prior to inertial breakthrough at $\mathrm{We}=16.0$.

Activating both closures simultaneously in three dimensions ($96\times 64^2$, $\ell_d/\epsilon=2.70 \implies \eta_\Pi=33.30\%$, \Cref{fig:chasing3d}(b,d,e) and \Cref{tab:dlvo_chasing}, with bispherical Stokes hydrodynamics benchmarked since Stimson and Jeffery~\cite{Jeffery1912,Stimson1926} and an independent $128\times 96^2$ dynamic chasing check at $\mathcal{S}_0=1.2, \bar{h}=5.76\Delta x$ matching the bispherical flux within $0.64\%$) maps a $4\times 5$ grid of 20 coupled $(\mathcal{S}_0,A_{\mathrm{rep}})$ trajectories across four regimes: contact coalescence (Regime~I, $4$ runs), elastic rebound (Regime~II, $4$ runs at $\mathcal{S}_0=0, A_{\mathrm{rep}}\ge 0.03$, all with $n_{E\uparrow}=0$), partial intact-film ripening (Regime~IIIa, $2$ runs at $\mathcal{S}_0=0.5, A_{\mathrm{rep}}\ge 0.12$, $\mathrm{Da}_{\mathrm{eng}}=0.74<1$, leaving $V_1(T_{\mathrm{obs}})/V_1(0)\approx 0.37$), and complete intact-film diffusional engulfment (Regime~IIIb, $10$ runs at $\mathcal{S}_0\in\{1.0,1.8\}$, $\mathrm{Da}_{\mathrm{eng}}>1$, driving $V_1\to 0$ smoothly across an unbroken carrier film). All 20 outcomes separate cleanly along the theoretical Damk{\"o}hler boundary $\mathrm{Da}_{\mathrm{eng}} = t_{\mathrm{chase}}/t_{\mathrm{rip}} = 1$ ($\mathcal{S}_{0,c}=0.678$). Whereas experimental predator--prey chasing between oil droplets in micellar solutions is driven by non-reciprocal Marangoni tractions~\cite{Meredith2020}, the present constant-$\sigma$ system isolates pure diffusive Stefan interface propulsion arrested by a conservative disjoining cushion.

\subsection{Experimental validation across emulsions, colloids, rising bubbles, and filament pinch-off}\label{subsec:bench_experiments}

\begin{figure}[tp!]
\centering
\cmamefig{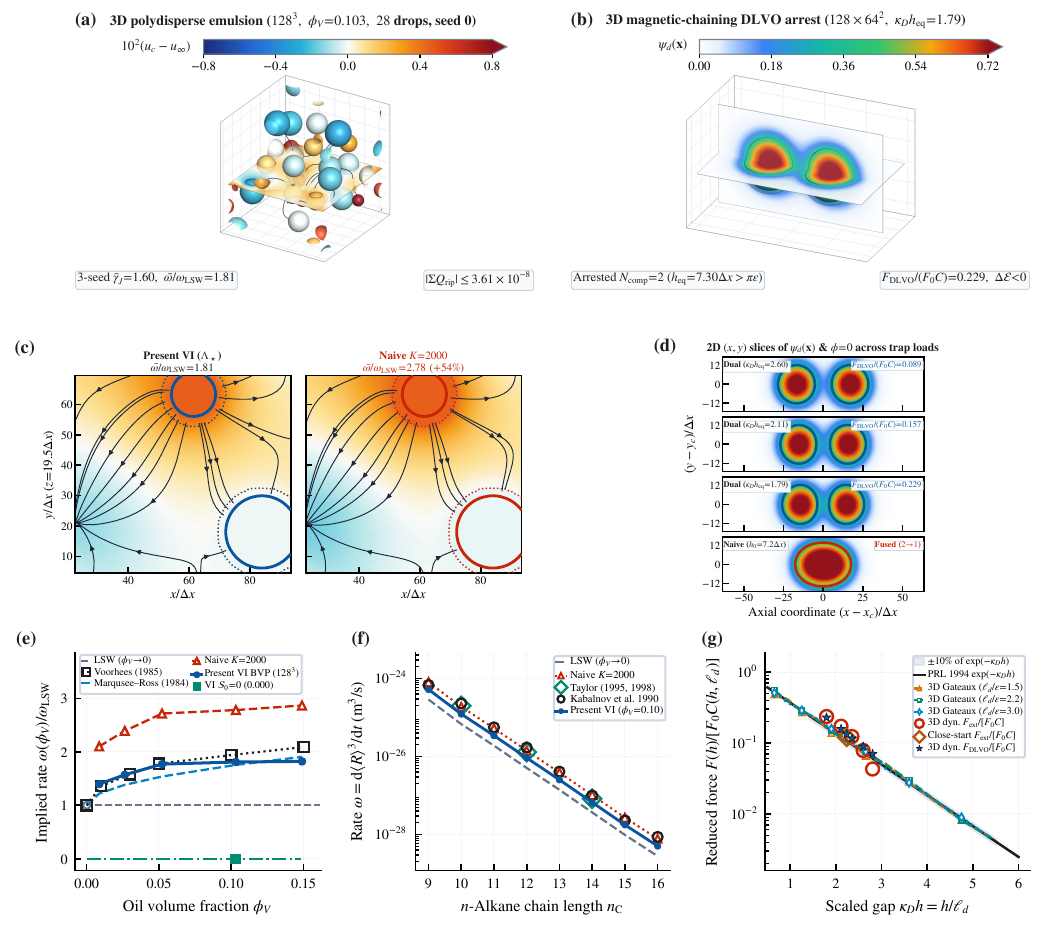}{width=0.90\textwidth}
\caption{Colloidal and polydisperse emulsion experimental validations of $\Lambda_\star$-renormalized ripening and Legendre-dual Yukawa--DLVO repulsion without fitted physical parameters. (a)~Three-dimensional octant cutaway of a representative 28-droplet polydisperse alkane-in-water emulsion solved as a stationary Robin--Kelvin BVP on $128^3$ ($\phi_V=0.103$, $h_{\min}>\pi\epsilon$), displaying $\phi=0$ interfaces colored by $10^2(u_c - u_\infty)\in[-0.8,+0.8]$ (upper tail to $+1.26$), equatorial supersaturation slice, and 3D diffusive flux lines $\bm{j}_m/D_m$ (three-realization mean $\bar{\gamma}_J=1.60$ vs.\ Marqusee--Ross~\cite{Marqusee1984} $1+\sqrt{3\phi_V}=1.56$; stationary implied rate $\bar{\omega}/\omega_{\mathrm{LSW}}=1.810$ vs.\ Voorhees~\cite{Voorhees1985} Table~I $1.94$). (b)~Three-dimensional cutaway of coupled colloidal Stokes + Model~H magnetic-chaining equilibrium on $128\times 64^2$ arrested at $\kappa_D h_{\mathrm{eq}}=1.79$ ($h_{\mathrm{eq}}=7.30\Delta x > \pi\epsilon$, $N_{\mathrm{comp}}=2$, $\Delta\mathcal{E}<0$), showing disjoint droplets and the positive $\psi_d(\bm{x})$ saddle cushion (displayed over $[0,0.72]$, core peak $0.79$). (c)~Paired 2D slice zoom ($z=19.5\Delta x$, identical $\phi<0$ carrier mask) of the $128^3$ emulsion contrasting the present VI (labelled \texttt{Present VI}, $\Lambda_\star$: smooth collar profile locking onto $u_K$ at solid $\phi=0$, $\bar{\omega}/\omega_{\mathrm{LSW}}=1.81$) against naive stiffness $K=2000\,D_m/\epsilon$ (premature pinning at the dotted outer support boundary $\phi=-0.97$, inflating $\bar{\omega}/\omega_{\mathrm{LSW}}$ by $+53.8\%$ to $2.78$). (d)~Central 2D $(x,y)$ slices of $\psi_d(\bm{x})$ and $\phi=0$ across axial cosine-trap loads ($\kappa_D h_{\mathrm{eq}}=2.60 \to 2.11 \to 1.79$) versus short-range attractive fusion ($N_{\mathrm{comp}}: 2\to 1$ at $g_{\mathrm{trap}}=4\times 10^{-3}, h_0=7.2\Delta x$) under the naive infimum functional \Cref{eq:F_dlvo_naive}. (e)~Stationary $128^3$ BVP implied ripening rate $\omega(\phi_V)/\omega_{\mathrm{LSW}}$ across $\phi_V\in[0,0.15]$ against Marqusee--Ross~\cite{Marqusee1984} and published multipole benchmarks of Voorhees~\cite{Voorhees1985} Table~I (with $0.000$ stasis at $\mathcal{S}_0=0$). (f)~Stationary 3D BVP prediction ($1.810\,\omega_{\mathrm{LSW}}$, without fitted physical parameters) for Ostwald ripening rates $\omega = \dd\langle R\rangle^3/\dd t$ across four orders of magnitude for homologous $n$-alkanes ($\mathrm{C_9}$--$\mathrm{C_{16}}$ at $\phi_V=0.10$) against photon correlation spectroscopy measurements of Kabalnov et al.~\cite{Kabalnov1990} and Taylor~\cite{Taylor1995,Taylor1998}. (g)~Finite-radius two-sphere Yukawa-reduced 3D disjoining force $F(h)/[F_0\,C(h,\ell_d)]$ ($C(h,\ell_d)=\frac{2R_0}{2R_0+h}\frac{\ell_d^{-1}+(2R_0+h)^{-1}}{\ell_d^{-1}+(2R_0)^{-1}}$) versus $\kappa_D h = h/\ell_d$ comparing static Gateaux isotherms ($\ell_d/\epsilon\in\{1.5,2.2,3.0\}$) and coupled 3D zero-drag Stokes states (five equilibrated loads labelled \texttt{F2}--\texttt{F5} including closer-gap start \texttt{F3-close}, plus relaxing far-tail load \texttt{F1}) against the optical Bragg-diffraction exponential fit $\exp(-\kappa_D h)$ ($\pm 10\%$ band) of Leal-Calderon et al.~\cite{LealCalderon1994}.}
\label{fig:colloid_exp}
\end{figure}

\begin{figure}[tp!]
\centering
\cmamefig{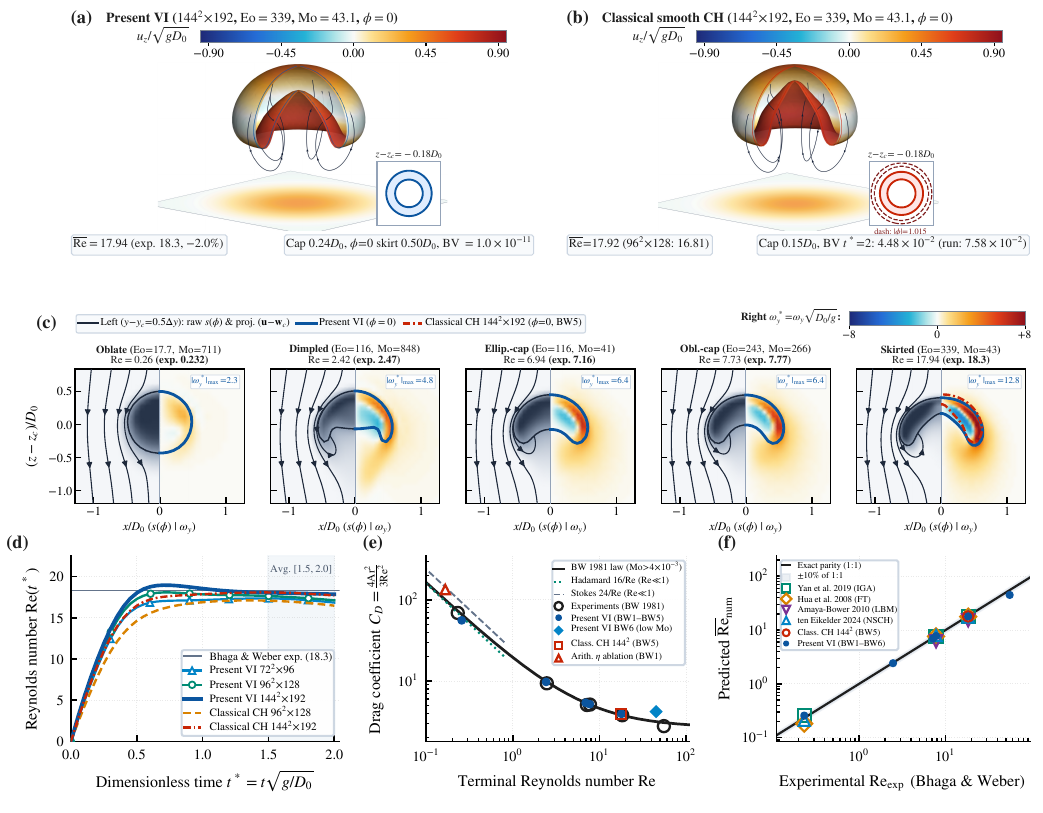}{width=\textwidth}
\caption{Three-dimensional experimental validation without fitted physical parameters against Bhaga and Weber~\cite{Bhaga1981} rising bubbles across six morphological regimes ($\rho_-/\rho_+=1000$, $\eta_-/\eta_+=100$). (a,b)~Matched-resolution ($144^2\times 192$, $D_0=24\Delta x$, $\epsilon=1.20\Delta x$) three-dimensional $270^\circ$ geometric cutaway ($\phi=0$ bubble shell colored by $u_z/\sqrt{gD_0}$, radial cut rims, co-moving toroidal wake streamlines, wake-velocity floor slice, and $z-z_c=-0.18\,D_0$ horizontal skirt-ring inset) on the annular thin-skirted bubble (BW5, $\mathrm{Eo}=339, \mathrm{Mo}=43.1, \mathrm{Re}_{\exp}=18.3$): present VI (labelled \texttt{Present VI}, $\overline{\mathrm{Re}}=17.94$, $-2.0\%$, $\phi=0$ cap crown $0.24\,D_0$, $\phi=0$ skirt length $0.50\,D_0$ tapering across $s(\phi)\ge 0.08$ to $0.58\,D_0$, bound excursion $1.0\times 10^{-11}$) versus classical smooth CH ($\overline{\mathrm{Re}}=17.92$ on $144^2\times 192$ vs.\ $16.81$ on $96^2\times 128$, bulk-thinned cap crown $0.15\,D_0$, and dynamic bulk overshoot $\max(|\phi|-1)_+ = 7.58\times 10^{-2}$ with dashed $|\phi|>1.015$ ring halos). (c)~Five-regime meridional split-mirror gallery pairing simulated near-midplane volume-fraction light-sheet exposures $s(\phi)$ and co-moving wake streamlines $(u_x,u_z-w_c)$ (left half $x\le 0$) with dimensionless azimuthal vorticity $\omega_y^* = \omega_y\sqrt{D_0/g}\in[-8,+8]$ and $\phi=0$ contours (right half $x\ge 0$: solid \texttt{Present VI} on BW1--BW5 and dash-dotted $144^2\times 192$ classical CH on BW5). (d)~Three-grid $\mathrm{Re}(t^*)$ convergence ($72^2\times 96 \to 96^2\times 128 \to 144^2\times 192$) against classical smooth CH on BW5. (e)~Drag coefficient $C_D = \frac{4\mathrm{Ar}^2}{3\mathrm{Re}^2}$ across three decades of $\mathrm{Re}$ against the Bhaga--Weber correlation, Hadamard--Rybczynski ($16/\mathrm{Re}$), and Stokes rigid-sphere ($24/\mathrm{Re}$) limits. (f)~Terminal Reynolds number parity across BW1--BW6 benchmarked against isogeometric analysis~\cite{Yan2019}, front-tracking~\cite{Hua2008}, kinetic diffuse-interface~\cite{AmayaBower2010}, unified-conservation 3D phase-field~\cite{TenEikelder2024_JCP}, and $144^2\times 192$ classical CH.}
\label{fig:bhaga_weber}
\end{figure}

\begin{figure}[tp!]
\centering
\cmamefig{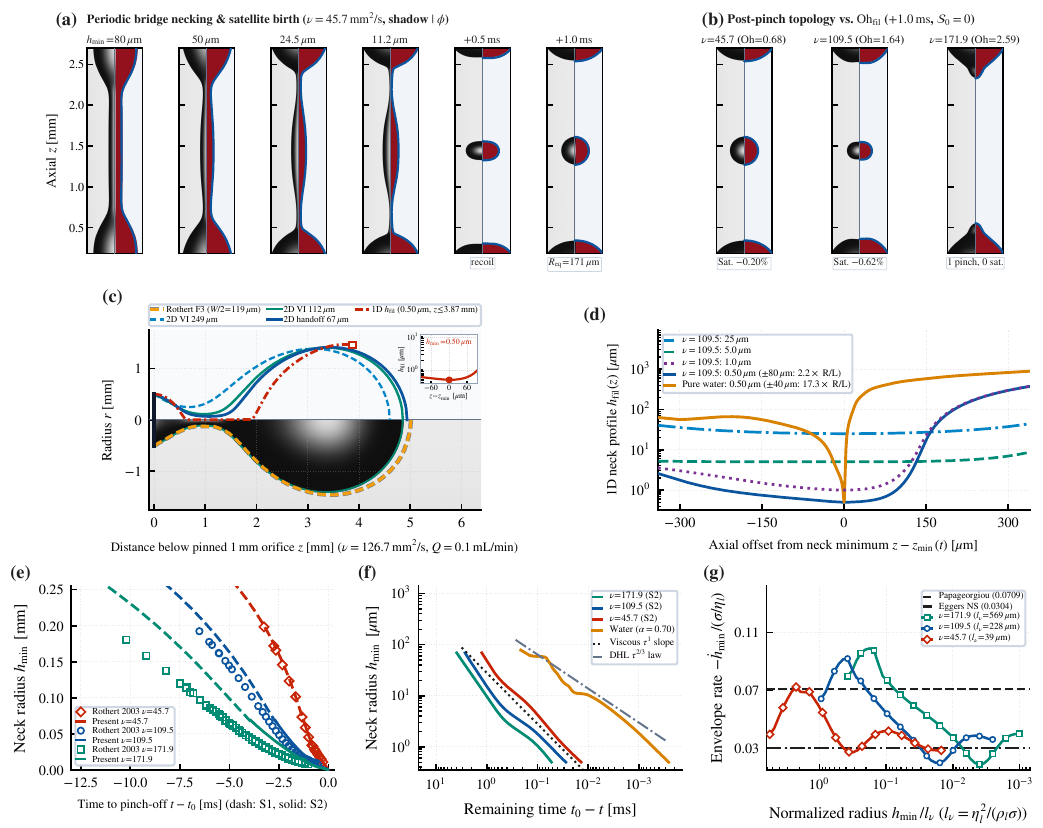}{width=\textwidth}
\caption{Experimental validation without fitted physical parameters of Newtonian viscous pendant-drop and filament pinch-off against Rothert, Richter, and Rehberg~\cite{Rothert2003} ($1\,\mathrm{mm}$ outer-diameter orifice, $\mathcal{S}_0=0$, zero artificial damping or interface smoothing). (a)~Six-frame axially periodic capillary-bridge split-mirror sequence at $\nu=45.7\,\mathrm{mm}^2/\mathrm{s}$ ($\mathrm{Oh}_{\mathrm{fil}}=0.68$, $\Delta x=3.33\,\mu\mathrm{m}$) pairing simulated backlit optical shadowgraphs ($r\le 0$) against the phase field $\phi$ ($r\ge 0$) across neck thinning ($h_{\min}: 80 \to 50 \to 24.5 \to 11.2\,\mu\mathrm{m}$), double end-pinching recoil ($+0.5\,\mathrm{ms}$), and steady spherical satellite retention ($+1.0\,\mathrm{ms}$, $R_{\mathrm{eq}}=171\,\mu\mathrm{m}$). (b)~Viscosity-controlled post-pinch topological transition ($+1.0\,\mathrm{ms}$) from satellite trapping at $\mathrm{Oh}_{\mathrm{fil}}\in\{0.68, 1.64\}$ to single central pinch-off without satellites at $\mathrm{Oh}_{\mathrm{fil}}=2.59$. (c)~Viscosity-matched gravitational pendant drop below a hinged Gibbs-pinned $1\,\mathrm{mm}$ orifice ($\nu=126.7\,\mathrm{mm}^2/\mathrm{s}$, experimental syringe feed $Q=0.1\,\mathrm{mL/min}$): axisymmetric VI simulated backlit optical shadowgraph ($r_{\min}=112\,\mu\mathrm{m}$ vs.\ experimental half-width $W/2=119\,\mu\mathrm{m}$ of Rothert et al.~\cite{Rothert2003} Fig.~2 frame F3, $r\le 0$) paired with axisymmetric VI contours down to $h_{\min}=67\,\mu\mathrm{m}$ (Stage~1) and 1D slender-jet continuation profiles $h_{\mathrm{fil}}(z,t)$ down to $h_{\min}=0.50\,\mu\mathrm{m}$ with neck zoom inset ($r\ge 0$, Stage~2; \Cref{app:slender_jet}). (d)~Sub-micron radial neck profiles $h(z-z_{\min})$ on $[-340,+340]\,\mu\mathrm{m}$ contrasting the viscous--inertial asymmetric Eggers double-cone ($2.2\times$ slope asymmetry at $\nu=109.5\,\mathrm{mm}^2/\mathrm{s}$) against the pure-water inertial slender-jet cone ($17.3\times$ slope asymmetry). (e)~Direct overlay of $h_{\min}(t-t_0)$ below a $1\,\mathrm{mm}$ orifice against calibrated shadowgraph data from Rothert et al.~\cite{Rothert2003} Fig.~7 ($\mathrm{RMSE}=2.05\,\mu\mathrm{m}$, median error $2.85\%$ at $\nu=45.7\,\mathrm{mm}^2/\mathrm{s}$ and late-stage $\mathrm{RMSE}=7.97\,\mu\mathrm{m}$ at $\nu=109.5\,\mathrm{mm}^2/\mathrm{s}$, while $\nu=171.9\,\mathrm{mm}^2/\mathrm{s}$ exhibits $+0.8\,\mathrm{ms}$ longer noise-free symmetric Stokes persistence prior to inertial crossover). (f)~Log--log scaling trajectories from spline-resampled handoff $h_{\min}\in[68.5,78.3]\,\mu\mathrm{m}$ down to $0.50\,\mu\mathrm{m}$ contrasting viscous ($\tau^1$) and pure-water inertial ($\tau^{2/3}$; fit $\alpha=0.699$, $R^2=0.973$ over $[2.5,65]\,\mu\mathrm{m}$) necks. (g)~Dimensionless thinning rate $-\dot{h}_{\min}/(\sigma/\eta_l)$ versus $h_{\min}/l_\nu$ showing crossover from the symmetric Papageorgiou Stokes regime ($0.0709$) into the viscous--inertial Eggers band ($0.0304$).}
\label{fig:pinchoff_exp}
\end{figure}

Agreement with analytical solutions and sharp-interface benchmarks confirms that the governing equations are solved accurately. We now test whether the formulation reproduces laboratory measurements without fitted physical parameters, progressing in dynamical complexity from quasi-steady emulsion ripening and colloidal Stokes equilibria (\Cref{fig:colloid_exp} and \Cref{tab:colloid_emulsion_exp}) to three-dimensional rising bubbles (\Cref{fig:bhaga_weber} and \Cref{tab:bhaga_weber}) and viscous filament pinch-off (\Cref{fig:pinchoff_exp} and \Cref{tab:pinchoff_exp}). On $128^3$ periodic boxes initialized with Lifshitz--Slyozov--Wagner polydisperse droplet ensembles across volume fractions $\phi_V\in[0.0087, 0.1489]$ ($10\text{--}32$ disjoint droplets with $h_{\min}> \pi\epsilon$, \Cref{fig:colloid_exp}(a,c,e)), solving the stationary 3D Robin--Kelvin boundary-value problem \Cref{eq:ripening_pde} yields the diffusive flux screening factor $\gamma_J(\phi_V) \coloneqq (\sum_m |J_m|)/(\sum_m |J_{m,\mathrm{LSW}}|)$. At $\phi_V=0.103$ across three 28-droplet realizations, the $\Lambda_\star$-renormalized closure gives $\gamma_J = 1.596 \pm 0.091$ ($+2.58\%$ relative to the Marqusee--Ross~\cite{Marqusee1984} effective-medium value $1+\sqrt{3\phi_V}=1.556$, $|\sum Q_{\mathrm{rip}}|\le 3.61\times 10^{-8}$) and implied coarsening rates $\omega(\phi_V)/\omega_{\mathrm{LSW}} = 1.398 \to 1.571 \to 1.771 \to 1.810 \to 1.822$ ($+3.05\%$ vs.\ Marqusee--Ross $1.756$ and $-6.7\%$ vs.\ the multipole benchmark $1.94$ of Voorhees~\cite{Voorhees1985} at $\phi_V=0.10$). By contrast, naive penalty pinning $K=2000\,D_m/\epsilon$ clamps $u_c$ onto $R_k + \frac{\pi}{2}\epsilon$ (\Cref{fig:colloid_exp}(c)) and inflates $\omega/\omega_{\mathrm{LSW}}$ by $+53.8\%$ to $2.783$. Transient $96^3$ 20-droplet evolutions confirm pre-asymptotic coarsening ($K_{3\mathrm{D}}/\omega_{\mathrm{LSW}}=1.034$ over the initial $4\%$ $\langle R^3\rangle$ growth) alongside exact stasis at $\mathcal{S}_0=0$ ($\langle R^3\rangle/\langle R_0^3\rangle = 1.000000$, zero bound violation, $n_{E\uparrow}=0$, versus $-2.74\%$ bulk-shift shrinkage and $7.98\times 10^{-2}$ bound overshoot in smooth CH). Mapped directly onto the homologous $n$-alkane series ($\mathrm{C_9H_{20}}$ through $\mathrm{C_{16}H_{34}}$ in $0.1\,\mathrm{M}$ aqueous SDS at $298.15\,\mathrm{K}$, $\phi_V=0.10$) measured via photon correlation spectroscopy by Kabalnov et al.~\cite{Kabalnov1990} and Taylor~\cite{Taylor1995,Taylor1998} across four orders of magnitude ($8.7\times 10^{-29}$ to $6.8\times 10^{-25}\,\mathrm{m^3/s}$, \Cref{fig:colloid_exp}(f)), the finite-volume-fraction factor $1.810$ accounts for $\log_{10}1.810 = 0.258$ of the $0.457$-decade mean offset between dilute LSW theory and experiment (lowering the mean logarithmic deviation to $0.199$ decades), while the remaining $\sim 1.6\times$ experimental excess lies outside pure monomer diffusion and is attributed in~\cite{Kabalnov1990,Taylor1998} to supramicellar solubilization transport above the $8\,\mathrm{mM}$ critical micelle concentration. Accordingly, the $1.810$ prefactor is tested against the effective-medium and multipole benchmarks above, and the alkane series against the four-decade carbon-chain scaling of $\omega$.

\begin{table}[t!]
\centering
\caption{Three-dimensional experimental validation on polydisperse alkane emulsion ripening ($\phi_V=0.10$) against Marqusee--Ross~\cite{Marqusee1984}, Voorhees~\cite{Voorhees1985}, Kabalnov et al.~\cite{Kabalnov1990}, and Taylor~\cite{Taylor1995,Taylor1998}, and on colloidal magnetic-chaining disjoining forces ($128\times 64^2$) against Leal-Calderon et al.~\cite{LealCalderon1994}, without fitted physical parameters.}
\label{tab:colloid_emulsion_exp}
\small
\setlength{\tabcolsep}{3.8pt}
\renewcommand{\arraystretch}{1.14}
\resizebox{\textwidth}{!}{%
\begin{tabular}{@{}llccccc@{}}
\toprule
Validation system & Configuration / Formulation & Grid & Simulated metric & Reference / Experiment & Agreement / Deviation & Conservation / Bound audit \\
\midrule
Insoluble emulsion ($\mathcal{S}_0=0$) & \textbf{Present VI (20 drops, transient)} & $96^3$ & $\Delta\langle R^3\rangle/\langle R_0^3\rangle=\mathbf{0.0000}$ & Exact stasis ($0.0000$) & $\mathbf{0.00\%}$ & $\max(|\phi|-1)_+=\mathbf{0.00}$, $n_{E\uparrow}=0$ \\
Insoluble emulsion ($\mathcal{S}_0=0$) & Classical smooth CH (20 drops) & $96^3$ & $\Delta\langle R^3\rangle/\langle R_0^3\rangle=-0.0274$ & Exact stasis ($0.0000$) & $-2.74\%$ shrinkage & $\max(|\phi|-1)_+=7.98\times 10^{-2}$ \\
Polydisperse BVP ($\phi_V=0.103$) & \textbf{Present VI ($\Lambda_\star$ closure, 3 seeds)} & $128^3$ & $\gamma_J=\mathbf{1.596}$, $\omega/\omega_{\mathrm{LSW}}=\mathbf{1.810}$ & $1.556$ ($1.756$~\cite{Marqusee1984}, $1.94$~\cite{Voorhees1985}) & $\mathbf{+2.58\%}$ ($+3.05\%, -6.7\%$) & $|\sum Q_{\mathrm{rip}}|\le 3.61\times 10^{-8}$ \\
Polydisperse BVP ($\phi_V=0.103$) & Naive penalty ($K=2000\,D_m/\epsilon$) & $128^3$ & $\gamma_J=2.333$, $\omega/\omega_{\mathrm{LSW}}=2.783$ & $1.556$ ($1.756$~\cite{Marqusee1984}) & $+53.8\%$ pinning bias & Outer-collar pinned \\
$\mathrm{C_9}$--$\mathrm{C_{16}}$ alkanes ($\phi_V=0.10$) & \textbf{Present VI ($1.810\,\omega_{\mathrm{LSW}}$ mapping)} & $128^3$ & Mean $|\log_{10}(\omega_{\mathrm{num}}/\omega_{\exp})|=\mathbf{0.199}$ & PCS data~\cite{Kabalnov1990,Taylor1995,Taylor1998} & $\mathbf{0.199}$ vs.\ $0.457$ dec.\ (LSW) & Consistent with $\sim 1.6\times$ micellar shift \\
\midrule
Colloidal Yukawa force & \textbf{Static Gateaux ($36$ states, $\ell_d/\epsilon\in[1.5,3.0]$)} & $128\times 64^2$ & $\exp(-\kappa_D h)$ scaling ($h\ge\pi\epsilon$) & Analytical Yukawa / Bragg~\cite{LealCalderon1994} & $\mathbf{3.27\%\text{--}5.80\%}$ mean err. & $|\sum\bm{f}_{\mathrm{DLVO}}|\le 4.68\times 10^{-13}$ \\
Colloidal magnetic chain & \textbf{Coupled Stokes + VI ($h_{\mathrm{eq}}/\ell_d\in[1.79,2.60]$)} & $128\times 64^2$ & Disjoint arrest ($2$ drops, \texttt{F2}--\texttt{F5}) & Bragg force curve~\cite{LealCalderon1994} & $|F_{\mathrm{DLVO}}-F_{\mathrm{ext}}|/F_{\mathrm{ext}}\in\mathbf{[3.15\%,13.67\%]}$ & $\Delta\mathcal{E}_{\mathrm{tot}}\in[-3.59,-0.16]$, viol.\ $\mathbf{0.00}$ \\
Colloidal magnetic chain & Naive infimum \eqref{eq:F_dlvo_naive} ($g_{\mathrm{trap}}=4\times 10^{-3}$) & $128\times 64^2$ & Attractive fusion ($N_{\mathrm{comp}}\!:\!2\to 1$) & Disjoint arrest~\cite{LealCalderon1994} & Coalesces in $180\text{--}200$ steps & Spurious sign reversal \\
\bottomrule
\end{tabular}%
}
\end{table}

Direct colloidal disjoining-force measurements~\cite{Bibette1992,LealCalderon1994} are tested in \Cref{fig:colloid_exp}(b,d,g) and \Cref{tab:colloid_emulsion_exp}. Thirty-six static 3D Gateaux evaluations of the finite-radius Yukawa-reduced force $F(h)/[F_0\,C(h,\ell_d)]$ across $\ell_d/\epsilon\in\{1.5,2.2,3.0\}$ on $128\times 64^2$ match the analytical non-overlapping two-sphere Yukawa law within mean errors of $5.80\%$, $4.79\%$, and $3.27\%$ for $h\ge \pi\epsilon$ ($|\sum\bm{f}_{\mathrm{DLVO}}|\le 4.68\times 10^{-13}$, Helmholtz residual $\le 6.25\times 10^{-15}$). Six coupled 3D inertialess ($\mathrm{Re}=0$) Stokes + Model~H compressive chaining simulations ($\ell_d/\epsilon=2.20$, $A_{\mathrm{rep}}=0.25$, $\sigma=0.35 > |\Delta\sigma_{\mathrm{DLVO}}|\approx 0.16$ per \Cref{rem:dlvo_scope}) are driven by the periodic axial cosine potential $V_{\mathrm{ext}}(x) = -g_{\mathrm{trap}}\cos\bigl(2\pi(x-x_c)/L_x\bigr)$ across $g_{\mathrm{trap}}\in\{1.25,2.40,4.00,6.00,9.00\}\times 10^{-3}$ from $h_0=9.2\Delta x$, plus an independent closer-gap run from $h_0=7.2\Delta x$ at $g_{\mathrm{trap}}=4.00\times 10^{-3}$ (labelled \texttt{F3-close}). All six coupled runs maintain disjoint droplets with $0.00$ bound violation and net free-energy dissipation $\Delta\mathcal{E}_{\mathrm{tot}}\in[-3.59,-0.16] < 0$ (up to $\mathcal{O}(10^{-5})$ late-time operator-splitting ripples), whereas matched runs under the naive infimum functional \Cref{eq:F_dlvo_naive} coalesce within $180\text{--}200$ steps (\Cref{fig:colloid_exp}(d)). Moderate-to-high trap loads (\texttt{F2}--\texttt{F5}) equilibrate across $h_{\mathrm{eq}}/\ell_d\in[1.79,2.60]$ with internal-to-external force balance $|F_{\mathrm{DLVO}}-F_{\mathrm{ext}}|/F_{\mathrm{ext}}\in[3.15\%,13.67\%]$ along the $\exp(-\kappa_D h)$ scaling of Leal-Calderon et al.~\cite{LealCalderon1994}; at the highest loads, deformable contact-facet flattening at finite $\sigma=0.35$ places the points $19.5\%\text{--}37.3\%$ above the rigid-sphere curve, while the weakest far-tail load (\texttt{F1}, $h_{\mathrm{eq}}/\ell_d=2.80$) is still relaxing slowly with a $62.7\%$ force residual after $7200$ Stokes steps (\Cref{fig:colloid_exp}(g)).

\begin{table}[t!]
\centering
\caption{Three-dimensional rising-bubble validation without fitted physical parameters against the experiments of Bhaga and Weber~\cite{Bhaga1981} ($\rho_-/\rho_+=1000, \eta_-/\eta_+=100$, $\mathrm{Ar}=(\mathrm{Eo}^3/\mathrm{Mo})^{1/4}$). The unforced $\sigma=g=0$ null control yields $\|\bm{u}\|_\infty = 0.0$ and $n_{E\uparrow}=0$.}
\label{tab:bhaga_weber}
\small
\setlength{\tabcolsep}{3.8pt}
\renewcommand{\arraystretch}{1.13}
\resizebox{\textwidth}{!}{%
\begin{tabular}{@{}llccccccc@{}}
\toprule
Regime / Case & $(\mathrm{Eo}, \mathrm{Mo})$ & Grid ($D_0/\Delta x$) & $\mathrm{Re}_{\exp}$ (3D refs.) & Present $\overline{\mathrm{Re}}$ & Rel.\ diff. & $C_{D,\mathrm{num}}$ ($C_{D,\mathrm{BW}}$) & $|\Delta V|/V_0$ & $\max(|\phi|-1)_+$ \\
\midrule
Null control ($\sigma=g=0$) & $\sigma=g=0$ & $72^2\times 96$ ($12$) & $0$ & $0.0000$ & --- & --- & $8.48\times 10^{-16}$ & $0.00$ \\
BW1 Oblate creeping (harm.\ $\eta$) & $(17.7,\, 711)$ & $96^3$ ($12$) & $0.232$ ($0.259^{\mathrm{a}}, 0.211^{\mathrm{d}}$) & $\mathbf{0.2578}$ & $+11.1\%$ ($-0.6\%^{\mathrm{a}}$) & $56.04$ ($73.08$) & $5.69\times 10^{-13}$ & $5.02\times 10^{-13}$ \\
BW1 Ablation (arith.\ $\eta$) & $(17.7,\, 711)$ & $96^3$ ($12$) & $0.232$ & $0.1665$ & $-28.2\%$ & $134.28$ ($73.08$) & $3.28\times 10^{-13}$ & $2.62\times 10^{-12}$ \\
BW2 Dimpled cap & $(116,\, 848)$ & $96^2\times 128$ ($16$) & $2.47$ ($2.38^{\mathrm{d}}$) & $\mathbf{2.4216}$ & $\mathbf{-1.96\%}$ & $\mathbf{9.755}$ ($9.791$) & $1.34\times 10^{-12}$ & $4.56\times 10^{-12}$ \\
BW3 Ellipsoidal cap & $(116,\, 41.1)$ & $96^2\times 128$ ($16$) & $7.16$ ($6.93^{\mathrm{d}}$) & $\mathbf{6.9379}$ & $\mathbf{-3.10\%}$ & $\mathbf{5.398}$ ($5.295$) & $1.09\times 10^{-13}$ & $1.11\times 10^{-11}$ \\
BW4 Spherical cap & $(243,\, 266)$ & $96^2\times 128$ ($16$) & $7.77$ ($7.59^{\mathrm{a}}, 6.20^{\mathrm{b}}, 7.90^{\mathrm{d}}$) & $\mathbf{7.7272}$ & $\mathbf{-0.55\%}$ & $\mathbf{5.186}$ ($5.104$) & $1.14\times 10^{-13}$ & $2.28\times 10^{-12}$ \\
BW5 Skirted (coarse) & $(339,\, 43.1)$ & $72^2\times 96$ ($12$) & $18.3$ & $17.136$ & $-6.36\%$ & $4.317$ ($3.776$) & $8.86\times 10^{-14}$ & $1.31\times 10^{-12}$ \\
BW5 Skirted (medium) & $(339,\, 43.1)$ & $96^2\times 128$ ($16$) & $18.3$ ($15.20^{\mathrm{b}}, 17.5^{\mathrm{d}}$) & $17.393$ & $-4.96\%$ & $4.190$ ($3.776$) & $1.28\times 10^{-13}$ & $1.74\times 10^{-11}$ \\
BW5 Skirted (Classical CH) & $(339,\, 43.1)$ & $96^2\times 128$ ($16$) & $18.3$ & $16.810$ & $-8.14\%$ & $4.486$ ($3.776$) & $2.56\times 10^{-12}$ & $\mathbf{4.40\times 10^{-2}}$ \\
BW5 Skirted (Classical CH) & $(339,\, 43.1)$ & $144^2\times 192$ ($24$) & $18.3$ & $17.919$ & $-2.08\%$ & $3.948$ ($3.776$) & $2.98\times 10^{-12}$ & $\mathbf{7.58\times 10^{-2}}$ \\
BW5 \textbf{Skirted (fine)} & $(339,\, 43.1)$ & $\mathbf{144^2\times 192}$ ($\mathbf{24}$) & $18.3$ ($18.17^{\mathrm{a}}, 17.76^{\mathrm{c}}$) & $\mathbf{17.939}$ & $\mathbf{-1.98\%}$ & $\mathbf{3.939}$ ($3.776$) & $\mathbf{5.34\times 10^{-14}}$ & $\mathbf{1.04\times 10^{-11}}$ \\
BW6 Inertial ellipsoid & $(32.2,\, 8.2\times 10^{-4})$ & $128^3$ ($16$) & $55.3$ & $44.982$ (pk $52.1$) & $-18.7\%$ & $4.205$ ($3.074$) & $2.33\times 10^{-13}$ & $7.40\times 10^{-12}$ \\
\bottomrule
\multicolumn{9}{@{}l}{\footnotesize 3D published benchmarks on BW1--BW5: $^{\mathrm{a}}$Yan et al.~\cite{Yan2019} (IGA: $0.259$ on BW1, $7.59$ on BW4, $18.17$ on BW5); $^{\mathrm{b}}$Amaya-Bower and Lee~\cite{AmayaBower2010} ($6.20$ on BW4, $15.20$ on BW5); $^{\mathrm{c}}$Hua et al.~\cite{Hua2008} (front-tracking: $17.76$ on BW5); $^{\mathrm{d}}$ten Eikelder et al.~\cite{TenEikelder2024_JCP} (3D PF: $0.211, 2.38, 6.93, 7.90, 17.5$ on BW1--BW5).}
\end{tabular}%
}
\end{table}

We next test three-dimensional air--liquid buoyancy dynamics across six morphological regimes (BW1--BW6) of Bhaga and Weber~\cite{Bhaga1981} (\Cref{fig:bhaga_weber} and \Cref{tab:bhaga_weber}). An initially spherical gas bubble of diameter $D_0=1$ ($\phi=+1$) rises from rest in a quiescent liquid column ($\phi=-1$, $\rho_-/\rho_+=1000$, $\eta_-/\eta_+=100$, periodic lateral boundaries and no-slip top and bottom walls) without symmetry planes; specifying the experimental pair $(\mathrm{Eo},\mathrm{Mo})$ fixes $\sigma = \Delta\rho\,g D_0^2/\mathrm{Eo}$, $\eta_- = (\mathrm{Mo}\,\rho_-^2\sigma^3/(g\Delta\rho))^{1/4}$, and $\mathrm{Ar} = (\mathrm{Eo}^3/\mathrm{Mo})^{1/4}$ without adjustable physical constants, with terminal diagnostics averaged over $t^* = t\sqrt{g/D_0} \in [1.5,2.0]$ ($[2.7,3.2]$ for high-viscosity BW2) following~\cite{Yan2019,TenEikelder2024_JCP}. In creeping and low-$\mathrm{Re}$ regimes (BW1--BW2), mechanical-jump harmonic shear viscosity~\cite{Reder2024_CMAME} preserves internal gas circulation, yielding $\overline{\mathrm{Re}}=0.2578$ on BW1 ($\mathrm{Eo}=17.7,\mathrm{Mo}=711,\mathrm{Re}_{\exp}=0.232$, within $0.6\%$ of the coarse-mesh isogeometric diffuse-interface value $0.2593$~\cite{Yan2019}). On this $D_0=12\Delta x$ grid, harmonic collar viscosity ($+11.1\%$) and arithmetic collar viscosity \Cref{eq:rho_eta_laws} ($\overline{\mathrm{Re}}=0.1665$, $-28.2\%$, $C_D=134.3$) bracket the sharp-interface Hadamard--Rybczynski and experimental rise velocity from above and below across the finite-thickness collar. In finite-inertia regimes (BW3--BW6; numerical $(\epsilon/\Delta x, \gamma_0, \beta)$ settings listed in \Cref{app:params}), cell-centered geometric mixture viscosity $\eta(\phi)=\eta_-^{1-s(\phi)}\eta_+^{s(\phi)}$ paired with edge-harmonic shear viscosity retains gas-core mobility while preventing kinematic-viscosity troughs in the liquid wake. Across dimpled-cap (BW2: $\overline{\mathrm{Re}}=2.422$, $-1.96\%$, vs.\ $2.38$~\cite{TenEikelder2024_JCP}), ellipsoidal-cap (BW3: $\overline{\mathrm{Re}}=6.938$, $-3.10\%$, vs.\ $6.93$~\cite{TenEikelder2024_JCP}), and indented spherical-cap (BW4: $\overline{\mathrm{Re}}=7.727$, $-0.55\%$, vs.\ $7.59$~\cite{Yan2019} and $7.90$~\cite{TenEikelder2024_JCP}) regimes, the simulations reproduce experimental shapes, terminal Reynolds numbers, and drag coefficients $C_D = \frac{4\mathrm{Ar}^2}{3\mathrm{Re}^2}$, while on the higher-inertia ellipsoid BW6 ($\mathrm{Eo}=32.2,\mathrm{Mo}=8.2\times 10^{-4},\mathrm{Re}_{\exp}=55.3$) $128^3$ resolution captures the closed toroidal wake (peak $\mathrm{Re}=52.06$, window mean $44.98$ as the thin boundary layer $\delta_{\mathrm{BL}}\sim D_0/\sqrt{\mathrm{Re}}\approx 2.1\Delta x$ approaches the diffuse collar thickness).

On the annular thin-skirted bubble BW5 ($\mathrm{Eo}=339,\mathrm{Mo}=43.1,\mathrm{Re}_{\exp}=18.3$, \Cref{fig:bhaga_weber}(a,b,d) and \Cref{tab:bhaga_weber}), refining the grid across $72^2\times 96 \to 96^2\times 128 \to 144^2\times 192$ ($D_0/\Delta x \in \{12,16,24\}$) increases $\overline{\mathrm{Re}}$ monotonically toward the experimental value via $17.136 \to 17.393 \to 17.939$ ($-1.98\%$ vs.\ experiment on $144^2\times 192$, compared with $18.17$~\cite{Yan2019}, $17.76$~\cite{Hua2008}, $17.5$~\cite{TenEikelder2024_JCP}, and $15.20$~\cite{AmayaBower2010}). The $144^2\times 192$ run sustains a single connected off-axis ($r\ge 0.35\,D_0$) three-dimensional annular skirt of length $L_{\mathrm{skirt}}=0.583\,D_0$ ($\phi=0$ core skirt length $0.50\,D_0$ at thickness $2.55\,\Delta x$, tapering across the sub-width $s(\phi)\ge 0.08$ tip to median thickness $h_{s\ge 0.08}\approx 1.25\,\Delta x$, while unskirted BW1--BW4 register zero off-axis skirt) with $|\Delta V|/V_0 = 5.34\times 10^{-14}$ and bound excursion $1.04\times 10^{-11}$. Matched comparisons against classical smooth CH at $\epsilon=1.20\,\Delta x$ on $96^2\times 128$ ($\overline{\mathrm{Re}}=17.393$ vs.\ $16.810$, bound excursion $1.74\times 10^{-11}$ vs.\ $4.40\times 10^{-2}$) and $144^2\times 192$ ($\overline{\mathrm{Re}}=17.939$ vs.\ $17.919$, sub-cell $\phi=0$ cap crown $0.24\,D_0$ [$0.208\,D_0$ cell-center] vs.\ bulk-thinned $0.15\,D_0$ [$0.125\,D_0$ cell-center], and bound excursion $1.04\times 10^{-11}$ vs.\ $7.58\times 10^{-2}$) demonstrate a $7.3\times 10^9\times$ reduction in dynamic bound violation together with preserved cap thickness.

\begin{table}[t!]
\centering
\caption{Axisymmetric validation of Newtonian liquid-bridge and pendant-drop capillary pinch-off against the high-speed shadowgraph experiments of Rothert et al.~\cite{Rothert2003} and analytical similarity laws~\cite{Eggers1993,Papageorgiou1995,Day1998} ($\mathcal{S}_0=0$, without fitted physical parameters; axisymmetric operators and 1D slender-jet continuation equations are detailed in \Cref{app:slender_jet}).}
\label{tab:pinchoff_exp}
\small
\setlength{\tabcolsep}{3.6pt}
\renewcommand{\arraystretch}{1.13}
\resizebox{\textwidth}{!}{%
\begin{tabular}{@{}llccccccc@{}}
\toprule
Configuration & $\nu$ ($\mathrm{mm}^2/\mathrm{s}$) & Grid / Continuation & Reference / Experiment & Simulated & Rel.\ diff. & Satellite / Trajectory metric & $|\Delta V|/V_0$ & $\max(|\phi|-1)_+$ \\
\midrule
Liquid bridge~\cite{Rothert2003} & $45.7$ & $128\times 864$ ($3.33\,\mu\mathrm{m}$) & $v_{\exp}=72.32\,\mathrm{mm/s}$ & $\mathbf{70.14\,\mathrm{mm/s}}$ ($[15,30]$) & $\mathbf{-3.02\%}$ & sat.\ $\Delta V=-0.20\%$ ($R_{\mathrm{eq}}=171\,\mu\mathrm{m}$) & $6.82\times 10^{-14}$ & $4.91\times 10^{-12}$ ($n_{E\uparrow}=0$) \\
Liquid bridge~\cite{Rothert2003} & $109.5$ & $128\times 864$ ($3.33\,\mu\mathrm{m}$) & $u_s=0.0709$~\cite{Papageorgiou1995} & $\mathbf{0.07033}$ & $\mathbf{-0.81\%}$ & sat.\ $\Delta V=-0.62\%$ ($R_{\mathrm{eq}}=138\,\mu\mathrm{m}$) & $5.55\times 10^{-14}$ & $1.94\times 10^{-12}$ ($n_{E\uparrow}=0$) \\
Liquid bridge~\cite{Rothert2003} & $171.9$ & $128\times 864$ ($3.33\,\mu\mathrm{m}$) & $u_s=0.0709$~\cite{Papageorgiou1995} & $\mathbf{0.07020}$ & $\mathbf{-0.99\%}$ & $1$ central pinch, $0$ sat. & $7.42\times 10^{-14}$ & $2.42\times 10^{-12}$ ($n_{E\uparrow}=0$) \\
\midrule
Pendant 2D VI $\to$ 1D jet & $45.7$ & $128\times 384 \to 0.50\,\mu\mathrm{m}$ & $u_{as}=0.0304$~\cite{Eggers1993} & $\mathbf{0.02994}$ ($[0.52,1.5]$; $0.0337$ on $[0.52,3]$) & $\mathbf{-1.50\%}$ & Fig.\,7 RMSE $\mathbf{2.05\,\mu\mathrm{m}}$ ($2.85\%$) & $4.72\times 10^{-16}$ (1D jet) & $1.40\times 10^{-12}$ (2D VI) \\
Pendant 2D VI $\to$ 1D jet & $109.5$ & $128\times 384 \to 0.50\,\mu\mathrm{m}$ & $u_{as}=0.0304$~\cite{Eggers1993} & $\mathbf{0.02838}$ ($[1.5,3]$; $0.0334$ on $[0.52,3]$) & $-6.63\%$ ($+9.86\%$) & $2.2\times$ cone; late RMSE $7.97\,\mu\mathrm{m}$ & $1.33\times 10^{-15}$ (1D jet) & $2.41\times 10^{-12}$ (2D VI) \\
Pendant 2D VI $\to$ 1D jet & $126.7$ & $128\times 384 \to 0.50\,\mu\mathrm{m}$ & $u_{as}=0.0304$~\cite{Eggers1993} & $\mathbf{0.02968}$ ($[0.52,3]$; $u_s=0.0725$) & $\mathbf{-2.36\%}$ & $3.0\times$ Eggers cone (Fig.\,2b~\cite{Rothert2003}) & $1.94\times 10^{-15}$ (1D jet) & $3.24\times 10^{-12}$ (2D VI) \\
Pendant 2D VI $\to$ 1D jet & $171.9$ & $128\times 384 \to 0.50\,\mu\mathrm{m}$ & $u_{as}=0.0304$~\cite{Eggers1993} & $\mathbf{0.03473}$ ($[0.52,1.5]$; $0.0236$ on $[0.52,3]$) & $+14.2\%$ ($-22.3\%$) & $4.0\times$ cone; late RMSE $11.01\,\mu\mathrm{m}$ & $5.01\times 10^{-15}$ (1D jet) & $4.57\times 10^{-12}$ (2D VI) \\
Pendant 2D VI $\to$ 1D jet & $1.00$ & $128\times 352 \to 0.50\,\mu\mathrm{m}$ & $\alpha=2/3$~\cite{Day1998} & $\mathbf{0.6990}$ ($C_{\mathrm{DHL}}=0.768$) & $\mathbf{+4.86\%}$ & $17.3\times$ cone ($R^2=0.973$) & $4.47\times 10^{-16}$ (1D jet) & $1.51\times 10^{-12}$ (2D VI) \\
\bottomrule
\end{tabular}%
}
\end{table}

Capillary thinning and breakup of viscous liquid bridges and gravitational pendant drops test interfacial mechanics across more than two decades of radial length scale (\Cref{fig:pinchoff_exp} and \Cref{tab:pinchoff_exp}, benchmarked against the high-speed shadowgraph experiments of Rothert, Richter, and Rehberg~\cite{Rothert2003} below a $1\,\mathrm{mm}$ outer-diameter brass nozzle at $\mathcal{S}_0=0$ without artificial damping). On fixed-grid glycerin--water liquid bridges ($r_{\mathrm{fil},0}=85\,\mu\mathrm{m}$, $\nu\in\{45.7, 109.5, 171.9\}\,\mathrm{mm}^2/\mathrm{s}$ in air with $\rho_l/\rho_g \in [1008,1030]$, $\eta_l/\eta_g \in [3.07\times 10^3, 1.18\times 10^4]$, $\mathrm{Oh}_{\mathrm{fil}}\in\{0.68, 1.64, 2.59\}$, $128\times 864$ mesh at $\Delta x = 3.33\,\mu\mathrm{m}$, $\epsilon=1.85\,\Delta x$; \Cref{fig:pinchoff_exp}(a,b)), the axisymmetric phase-field solver resolves neck thinning down to $h_{\min}=12\,\mu\mathrm{m}$. Over $h_{\min}\in[15,35]\,\mu\mathrm{m}$, peak Stokes necking rates match Papageorgiou's similarity law $v_{\mathrm{P}} = 0.0709\,\sigma/\eta_l$~\cite{Papageorgiou1995} within $-0.81\%$ ($u_s = 0.07033$ at $\nu=109.5\,\mathrm{mm}^2/\mathrm{s}$) and $-0.99\%$ ($u_s = 0.07020$ at $\nu=171.9\,\mathrm{mm}^2/\mathrm{s}$), and match the macroscopic experimental fit on $[60,150]\,\mu\mathrm{m}$ at $\nu=45.7\,\mathrm{mm}^2/\mathrm{s}$ within $-3.02\%$ ($70.14\,\mathrm{mm/s}$ peak over $[15,30]\,\mu\mathrm{m}$ vs.\ $72.32\,\mathrm{mm/s}$, following initial acceleration through $51.84\,\mathrm{mm/s}$ on $[30,65]\,\mu\mathrm{m}$). Post-pinch evolutions capture double end-pinching satellite formation at $\mathrm{Oh}_{\mathrm{fil}}\in\{0.68,1.64\}$ ($R_{\mathrm{eq}} = 171.0$ and $137.7\,\mu\mathrm{m}$, surviving $1.0\,\mathrm{ms}$ after breakup with $|\Delta V_{\mathrm{sat}}|/V_{\mathrm{sat},0} \le 0.62\%$) and the viscous transition to single central pinch-off without satellites at $\mathrm{Oh}_{\mathrm{fil}}=2.59$, maintaining $n_{E\uparrow}=0$, $\max(|\phi|-1)_+ \le 4.91\times 10^{-12}$, and $|\Delta V|/V_0 \le 7.42\times 10^{-14}$.

For full gravitational pendant drops below a hinged Gibbs-pinned $1\,\mathrm{mm}$ orifice ($\mathrm{Bo}=0.41$, experimental syringe feed $Q=0.1\,\mathrm{mL/min}$~\cite{Rothert2003}, \Cref{fig:pinchoff_exp}(c--g)), physical scales must be attributed cleanly between the two coupled stages detailed in \Cref{app:slender_jet}. Stage~1 evolves the axisymmetric double-obstacle phase-field solver ($128\times 384$, $\Delta x = 16\,\mu\mathrm{m}$ for glycerin--water with $r_{\mathrm{cl}}\in[0.4997,0.5004]\,\mathrm{mm}$, and $128\times 352$, $\Delta x = 20\,\mu\mathrm{m}$ for pure water with $r_{\mathrm{cl}}\in[0.4993,0.5017]\,\mathrm{mm}$; $\max(|\phi|-1)_+ \le 4.57\times 10^{-12}$) down to the Eulerian handoff neck radius $h_{\min}\in[66.7,76.0]\,\mu\mathrm{m}$, matching the frame-F3 experimental half-width $W/2=119\,\mu\mathrm{m}$ at $\nu=126.7\,\mathrm{mm}^2/\mathrm{s}$ with $r_{\min}=112\,\mu\mathrm{m}$ (\Cref{fig:pinchoff_exp}(c)). Stage~2 then projects the filament via a cell-volume-preserving $C^2$ cumulative-volume spline ($C^1$ cross-sectional area $\mathcal{A}(z)$, starting from resampled smooth neck radii $h_{\min}\in[68.5,78.3]\,\mu\mathrm{m}$ with handoff volume jump $\le 4.84\times 10^{-16}$) onto an open-loop 1D sharp-interface slender-jet continuation ($\Delta s = 0.50\,\mu\mathrm{m}$, synchronized volume drift $\le 5.01\times 10^{-15}$) down to $0.50\,\mu\mathrm{m}$. Direct comparison of $h_{\min}(t-t_0)$ against the $1\,\mathrm{mm}$-orifice data of Rothert et al.~\cite{Rothert2003} Fig.~7 (\Cref{fig:pinchoff_exp}(e)) yields $\mathrm{RMSE}=2.05\,\mu\mathrm{m}$ ($2.85\%$ median error) at $\nu=45.7\,\mathrm{mm}^2/\mathrm{s}$ and late-stage $\mathrm{RMSE}=7.97\,\mu\mathrm{m}$ ($12.61\,\mu\mathrm{m}$ full window) at $\nu=109.5\,\mathrm{mm}^2/\mathrm{s}$. At the highest viscosity $\nu=171.9\,\mathrm{mm}^2/\mathrm{s}$ (viscocapillary length $l_\nu \coloneqq \rho_l\nu^2/\sigma = 569\,\mu\mathrm{m}$, late-stage $\mathrm{RMSE}=11.01\,\mu\mathrm{m}$, full window $30.31\,\mu\mathrm{m}$), the noise-free continuum calculation remains on the linearly stable symmetric Stokes branch ($u_s\in[0.0721,0.0734]$ over $[35,65]\,\mu\mathrm{m}$) about $0.8\,\mathrm{ms}$ longer than in experiment, consistent with finite-amplitude laboratory perturbations triggering subcritical symmetry breaking near $h_T\approx 28.5\,\mu\mathrm{m}$~\cite{Rothert2003}.

Across all four glycerin--water viscosities $\nu\in\{45.7,109.5,126.7,171.9\}\,\mathrm{mm}^2/\mathrm{s}$ (\Cref{fig:pinchoff_exp}(d,g) and \Cref{tab:pinchoff_exp}), the continued filaments cross over from symmetric Stokes thinning into the asymmetric viscous--inertial Eggers double-cone regime~\cite{Eggers1993} ($u_{as}=0.02994$ on $[0.52,1.5]\,\mu\mathrm{m}$ at $\nu=45.7$; $0.02838$ on $[1.5,3.0]\,\mu\mathrm{m}$ and $0.03340$ on $[0.52,3.0]\,\mu\mathrm{m}$ at $\nu=109.5$; $0.02968$ on $[0.52,3.0]\,\mu\mathrm{m}$, $-2.36\%$ vs.\ $0.0304$, at $\nu=126.7$; and $0.03473$ on $[0.52,1.5]\,\mu\mathrm{m}$ at $\nu=171.9$, with $2.2\times\text{--}4.0\times$ steep-to-slender slope asymmetry across $\pm 80\,\mu\mathrm{m}$). In pure water ($\nu=1.00\,\mathrm{mm}^2/\mathrm{s}$) with zero artificial axial viscosity, eliminating coarse-grid slope kinks through the $C^1$-area volume-preserving handoff carries the single-valued 1D continuation to $h_{\min}=0.50\,\mu\mathrm{m}$, capturing a $17.3\times$ steep-to-slender cone slope asymmetry across $\pm 40\,\mu\mathrm{m}$ and a terminal inertial exponent $\alpha=0.6990$ ($+4.86\%$ relative to the potential-flow exponent $\alpha=2/3$~\cite{Day1998}, prefactor $0.768$, $R^2=0.973$ over $h_{\min}\in[2.5,65]\,\mu\mathrm{m}$; \Cref{fig:pinchoff_exp}(d,f)).


\section{Conclusions}\label{sec:conclusions}\label{sec:fem_concl}

This work has developed a single-scalar diffuse-interface formulation for immiscible and weakly soluble incompressible two-phase flows based on a deep-quench double-obstacle variational inequality. Starting from an exact one-dimensional Siggia cancellation identity and smooth single-scalar obstruction results (\Cref{thm:nogo}), we showed that smooth polynomial potentials, quadratic Bogomolnyi and functionalized Cahn--Hilliard defect penalties, and well-weighted tail-stiffened functionals remain trapped between attractive exponential tail overlap and an empirical $\sim 3.7\text{--}4.6\,\Delta x$ lattice-pinning wall on uniform Eulerian meshes, whereas non-variational counter-diffusive fluxes forfeit Lyapunov energy dissipation or induce outward droplet drift and bulk nucleation. Replacing smooth polynomial wells with the non-smooth indicator $I_{[-1,1]}(\phi)$ resolves this conflict within a single order parameter: matched asymptotics (\Cref{thm:asymptotics}) establishes exact outer bulk isolation to all algebraic orders at zero solubility, sixth-order three-dimensional spherical Young--Laplace superconvergence ($-\frac{\pi^2}{12}\delta^6$ at the collar midpoint radius $R_{\mathrm{mid}}$, and $\mathcal{O}(\delta^4)$ with respect to the conserved mass radius $R_{\mathrm{mass}}$), and non-evaporating raised-cosine liquid sheets thinner than the interfacial collar ($\pi\epsilon$) whose $90^\circ$ medial-axis normal flip is suppressed to $\mathcal{O}((\partial_x h)^2)$ by the orthonormal ridge-safe operator (ONRO) for normal mobility.

At the level of near-field physical closures (\Cref{prop:dlvo_cont,prop:ripening_sl}), the strictly convex Legendre-dual Yukawa formulation eliminates the short-range sign reversal of naive infimum convolutions and characterizes pre-contact steric and electrostatic rebound through an analytical barrier-efficiency law $\eta_\Pi(\ell_d/\epsilon)$, while boundary-layer matching across the compact sinusoidal collar determines the universal renormalization constant $\Lambda_\star = 3.880110546586$ and curvature shifts required to recover sharp-interface Ostwald ripening through third order in $\epsilon/R_{\mathrm{mid}}$. On staggered finite-volume grids (\Cref{lem:collar_coarea,thm:wb_fct,thm:dlvo_rip}), pairing fifth-order conservative upwind reconstruction with four-pass Zalesak flux-corrected limiting preserves the invariant domain $\phi\in[-1,1]$ without state clipping, and gating an interface-centered curvature estimator against a complementarity equilibrium indicator in the well-balanced continuum surface-delta force (WB-CSDF) reconciles machine-precision multi-droplet hydrostatics ($\mathcal{O}(10^{-17})$ spurious currents across $1\!:\!1$ to $1000\!:\!1$ density ratios) with underdamped two- and three-dimensional Rayleigh--Lamb capillary oscillations.

Across the verification suites of \Cref{sec:benchmarks}, these continuum and discrete properties sustain unbroken sub-cell ribbons ($2h_{\min} = 0.83\,\Delta x$) under Rider--Kothe vortex reversal, retain $89.0\%$ of the resolved liquid core during three-dimensional Enright sheet deformation, preserve unbroken $1000\!:\!1$ rising-bubble skirts of $\phi=0$ shaft thickness $2.92\,\Delta x = 0.62\,\pi\epsilon$, and achieve $0.041\%$ bipolar ripening flux accuracy on a $1024^2$ mesh alongside sub-percent three-dimensional Stimson--Jeffery bispherical chasing rates. Without fitted physical parameters, the same framework reproduces laboratory measurements of homologous alkane emulsion coarsening, superparamagnetic colloidal force--distance isotherms, three-dimensional Bhaga--Weber rising-bubble terminal Reynolds numbers and annular thin-skirted wakes, and millimeter-to-sub-micron pendant-drop filament pinch-off trajectories (coupled at Eulerian neck radius $h_{\min}\in[66.7,76.0]\,\mu\mathrm{m}$ to a volume-preserving one-dimensional slender-jet continuation) down to $0.50\,\mu\mathrm{m}$ viscous-inertial and inertial cones.

Several scope boundaries of the present study point directly to natural extensions. First, our implementation relies on uniform Cartesian and axisymmetric staggered meshes; pairing the continuum variational inequality of \Cref{sec:continuum_vi} with pressure-jump enrichment or unfitted cut-cell ghost-penalty stabilization~\cite{Burman2016_CMAME} and differentiable nonlinear stabilization~\cite{Badia2017_CMAME,Bonilla2019_CMAME} or monolithic convex limiting~\cite{Guermond2018_CMAME,Kuzmin2020_CMAME,Kuzmin2020b_CMAME} offers a direct path to transfer $1000\!:\!1$ hydrostatic balance and primal--dual active-set sparsity~\cite{Blowey1992,Graser2009_SINUM,Barrett2013_CMAME,Gerasimov2019_CMAME} to adaptive octree finite-element~\cite{Khanwale2020_JCP,Khanwale2023_JCP} and multi-patch isogeometric discretizations~\cite{Gomez2010_CMAME,Bracco2023_CMAME}. Second, the screened Yukawa disjoining closure requires resolving the screening length ($\ell_d/\epsilon \ge 2.11$ for $\eta_\Pi \ge 0.25$) and models equilibrium colloidal or surfactant stabilization rather than hydrodynamic gas cushioning; capturing uncharged sub-millimeter drop bouncing in ambient gas calls for coupling the compact obstacle collar to a rate-dependent Stefan--Reynolds--Knudsen lubrication layer, while elevating the Eulerian-to-slender-jet handoff to a monolithic two-way coupling would remove the open-loop shoulder approximation below $h_{\min}\approx 67\,\mu\mathrm{m}$. Coupling this compact variational inequality to complex rheological constitutive laws, and extending the obstacle indicator from the scalar interval $[-1,1]$ to the Gibbs $N$-simplex, provides a consistent variational route toward structured multiphase emulsions, foams, and wetting flows in which thin-film survival, capillary statics, and diffusive coarsening must be governed simultaneously on a fixed Eulerian grid.


\appendix
\crefalias{section}{appendix}
\setcounter{table}{0}
\setcounter{figure}{0}

\section{Proofs of the smooth single-scalar no-go results (\texorpdfstring{\Cref{thm:nogo}}{Theorem 1})}\label{app:nogo}

\subsection{Proof of \texorpdfstring{\Cref{thm:nogo}(a)}{Theorem 1(a)}: Hamiltonian non-existence of stationary smooth thin films}
At reservoir chemical potential $\mu=0$, any stationary one-dimensional critical point $\phi\in C^2(\R)$ of $\mathcal{F}[\phi]=\int_\R e_{\mathrm{loc}}(\phi,\phi')\,\dd x$ ($e_{\mathrm{loc}}\in C^3(\R^2)$) satisfies the Euler--Lagrange equation $\frac{\dd}{\dd x}\partial_{\phi'}e_{\mathrm{loc}}(\phi,\phi') - \partial_\phi e_{\mathrm{loc}}(\phi,\phi') = 0$. Multiplying by $\phi'(x)$ yields conservation of the Beltrami--Noether Hamiltonian:
\begin{equation}\label{eq:app_beltrami}
\frac{\dd}{\dd x}\Bigl[\phi'\partial_{\phi'}e_{\mathrm{loc}}(\phi,\phi') - e_{\mathrm{loc}}(\phi,\phi')\Bigr] = \phi'\left(\frac{\dd}{\dd x}\partial_{\phi'}e_{\mathrm{loc}} - \partial_\phi e_{\mathrm{loc}}\right) \equiv 0.
\end{equation}
Suppose $\phi\in C^2(\R)$ is a homoclinic thin-film profile satisfying $\lim_{x\to\pm\infty}\phi(x) = -1$ and $\lim_{x\to\pm\infty}\phi'(x)=0$. Evaluating \Cref{eq:app_beltrami} as $|x|\to\infty$ using $e_{\mathrm{loc}}(-1,0)=0$ fixes the first integral identically to zero: $\mathcal{H}(\phi,\phi') \coloneqq \phi'\partial_{\phi'}e_{\mathrm{loc}}(\phi,\phi') - e_{\mathrm{loc}}(\phi,\phi') \equiv 0$ for all $x\in\R$. By Rolle's theorem, any non-constant homoclinic orbit must attain an interior extremum at some $x_0\in\R$ where $\phi'(x_0)=0$ and $\phi(x_0)=\phi_{\mathrm{mid}}\neq -1$. At $x=x_0$, $\mathcal{H}(\phi_{\mathrm{mid}},0)=0$ implies $e_{\mathrm{loc}}(\phi_{\mathrm{mid}},0)=0$. By strict double-well positivity ($e_{\mathrm{loc}}(\phi,0)>0$ for $\phi\in(-1,1)$ and $e_{\mathrm{loc}}(\pm 1,0)=0$), $\phi_{\mathrm{mid}}$ cannot lie in $(-1,1)$. If $\phi_{\mathrm{mid}}>1$, continuity gives an intermediate point $x_1$ where $\phi(x_1)=+1$; at $x_1$, $\mathcal{H}(+1,\phi'(x_1)) = \phi'(x_1)\partial_{\phi'}e_{\mathrm{loc}}(+1,\phi'(x_1)) - e_{\mathrm{loc}}(+1,\phi'(x_1)) = 0$. However, uniform ellipticity $\partial_{pp}^2 e_{\mathrm{loc}} \ge c_0 > 0$ implies $\frac{\partial}{\partial p}\bigl(p\partial_p e_{\mathrm{loc}}(+1,p) - e_{\mathrm{loc}}(+1,p)\bigr) = p\,\partial_{pp}^2 e_{\mathrm{loc}}(+1,p)$, which has the strict sign of $p$, so $\mathcal{H}(+1,p) > \mathcal{H}(+1,0) = -e_{\mathrm{loc}}(+1,0) = 0$ for every $p\neq 0$. Thus $\phi'(x_1)=0$ and $(\phi,\phi')(x_1)=(+1,0)$. Since $e_{\mathrm{loc}}\in C^3(\R^2)$ and $\partial_{pp}^2 e_{\mathrm{loc}}\ge c_0>0$, the normal-form ODE $\phi'' = (\partial_\phi e_{\mathrm{loc}} - \partial_{\phi p}^2 e_{\mathrm{loc}}\,\phi')/\partial_{pp}^2 e_{\mathrm{loc}}$ has a locally Lipschitz right-hand side, so Picard--Lindel{\"o}f uniqueness forces $\phi\equiv +1$, contradicting $\lim_{x\to\pm\infty}\phi(x)=-1$. If instead $\phi_{\mathrm{mid}}<-1$, global nonnegativity $e_{\mathrm{loc}}(\cdot,0)\ge 0$ combined with $e_{\mathrm{loc}}(\phi_{\mathrm{mid}},0)=0$ implies $\partial_\phi e_{\mathrm{loc}}(\phi_{\mathrm{mid}},0)=0$ at the interior minimum $x_0$ ($\phi'(x_0)=0$), so Picard--Lindel{\"o}f uniqueness at $x_0$ forces $\phi\equiv\phi_{\mathrm{mid}}\neq -1$, again contradicting $\lim_{x\to\pm\infty}\phi(x)=-1$. Hence $\phi\equiv -1$ is the sole homoclinic solution. \qed

\subsection{Proof of \texorpdfstring{\Cref{thm:nogo}(b)}{Theorem 1(b)}: quadratic Bogomolnyi and FCH defect invariance}
On the isolated planar GL kink $\phi_K(z) = \tanh\bigl(z/(\sqrt{2}\epsilon)\bigr)$, equipartition $\frac{\epsilon}{2}(\phi_K')^2 = \frac{1}{\epsilon}W_{\mathrm{GL}}(\phi_K)$ gives $\mathcal{Q}(\phi_K,\phi_K')\equiv 0$ and $\mathcal{G}(\phi_K,\phi_K'')\equiv 0$ identically. Since the additional integrand $e_{\mathrm{def}} = \gamma_{\mathrm{GL}}\bigl[\frac{\beta_B\epsilon}{2}\omega(\phi)\mathcal{Q}^2 + \frac{\tilde{\gamma}_F\epsilon}{2}\mathcal{G}^2\bigr]$ vanishes quadratically on $\{\mathcal{Q}=0,\mathcal{G}=0\}$, both $e_{\mathrm{def}}$ and its first Fr{\'e}chet derivative $\delta\int e_{\mathrm{def}}/\delta\phi$ vanish identically at $\phi=\phi_K$, leaving $\phi_K$ and $\sigma = \mathcal{F}_{\beta,\gamma}[\phi_K]$ unchanged for all $(\beta_B,\tilde{\gamma}_F)$.

For a symmetric kink--antikink pair separated by $h_f \gg \epsilon$, write $\phi(x) = \phi_0(x;h_f) + \psi(x)$ with superposition ansatz $\phi_0(x;h_f) = \phi_K(x + h_f/2) - \phi_K(x - h_f/2) - 1$ and tail correction $\|\psi\|_{H^2} = \mathcal{O}(e^{-\sqrt{2}h_f/\epsilon})$. Define the small overlap parameter $E \coloneqq e^{-\sqrt{2}h_f/\epsilon} \ll 1$. Around each kink core $x \approx \mp h_f/2$, $\phi(x) = \pm\phi_K(x\pm h_f/2) + \mathcal{O}(E)$. Evaluating $\mathcal{Q}(\phi_0,\phi_0')$ and $\mathcal{G}(\phi_0,\phi_0'')$ pointwise yields $|\mathcal{Q}(\phi_0,\phi_0')| = \mathcal{O}(E)$ and $|\mathcal{G}(\phi_0,\phi_0'')| = \mathcal{O}(E)$ uniformly in $x$, supported over a length $\mathcal{O}(h_f)$. Consequently,
\begin{equation}\label{eq:app_q2_g2_order}
\int_\R \omega(\phi)\mathcal{Q}(\phi,\phi')^2\,\dd x = \mathcal{O}(h_f\,E^2), \qquad \int_\R \mathcal{G}(\phi,\phi'')^2\,\dd x = \mathcal{O}(h_f\,E^2),
\end{equation}
because the linearized kink operator $\mathcal{L}_K\psi = \epsilon\psi'' - \epsilon^{-1}W_{\mathrm{GL}}''(\phi_K)\psi$ satisfies $\mathcal{L}_K(e^{\mp\sqrt{2}x/\epsilon}) = \mathcal{O}(e^{-2\sqrt{2}|x|/\epsilon})$ in the tail. Thus $U_{\beta,\gamma}(h_f) - U_{\mathrm{GL}}(h_f) = \mathcal{O}\bigl(\sigma(\beta_B+\tilde{\gamma}_F)(h_f/\epsilon)\,e^{-2\sqrt{2}h_f/\epsilon}\bigr)$ for any fixed $(\beta_B,\tilde{\gamma}_F)$ as $h_f/\epsilon\to\infty$, proving \Cref{eq:nogo_b_asymptotic}. On a smooth curved interface of total curvature $\kappa = -\nabla\cdot(\nabla\phi/|\nabla\phi|)$, normal expansion $\phi(n) = \phi_K(n) + \epsilon\kappa\phi_1(n) + \mathcal{O}(\epsilon^2\kappa^2)$ gives $\mathcal{G}(\phi,\Delta\phi) = \epsilon\kappa\phi_K'(n) + \mathcal{O}(\epsilon^2\kappa^2)$ on $\phi_K$ ($\gamma_{\mathrm{GL}}\frac{\tilde{\gamma}_F\epsilon}{2}\int_\R (\epsilon\kappa\phi_K')^2\,\dd n = \frac{1}{2}\sigma\tilde{\gamma}_F\epsilon^2\kappa^2$) and $\epsilon\mathcal{Q}(\phi,\nabla\phi) = \mathcal{O}(\epsilon\kappa)$, so for well-localized weights $\omega(\pm 1)=0$ the defect integral contributes a Helfrich bending rigidity $\propto \sigma(\beta_B\mathcal{C}_\omega + \tilde{\gamma}_F\mathcal{C}_F)\epsilon^2\int_\Gamma \kappa^2\,\dd S$ that stabilizes long-wave Rayleigh--Plateau modes once $(\beta_B,\tilde{\gamma}_F)(\epsilon/R_0)^2 = \mathcal{O}(1)$, whereas $\beta_B < 0$ or $\tilde{\gamma}_F < 0$ renders $\mathcal{F}_{\beta,\gamma}$ unbounded below on high-wavenumber perturbations. \qed

Regarding \Cref{rem:nogo_num}, around the pure bulk $\phi=\pm 1$ the fourth-order FCH operator $(1 - \tilde{\gamma}_F(\epsilon^2\partial_x^2 - 2))(\epsilon^2\partial_x^2 - 2)(\phi\mp 1) = 0$ retains four real spatial eigenvalues $\pm\sqrt{2}, \pm\sqrt{2 + 1/\tilde{\gamma}_F}$ for all $\tilde{\gamma}_F > 0$; inside the inverted nonlinear core ($W_{\mathrm{GL}}''(0) = -1 < 0$), oscillatory spatial modes couple to these real exponential tails so numerical continuation past $\tilde{\gamma}_c \in (1.6,1.8)$ produces locked homoclinic ghost bilayers.

\subsection{Proof of \texorpdfstring{\Cref{thm:nogo}(c) and (d)}{Theorem 1(c) and (d)}: tail stiffening, Peierls--Nabarro wall, and counter-diffusive outward drift}
For \Cref{eq:F_tail_stiff}, the gradient coefficient $1 - a S(\phi)$ satisfies $\min_{\phi\in[-1,1]}(1-aS(\phi)) = 1 - a$ because $S(\phi)=\phi^2(2-\phi^2)$ attains its maximum $S(\pm 1)=1$ at the wells. Hence uniform ellipticity requires $a < 1$ (whereas in the degenerate limit $a\to 1^-$, the first integral $\frac{\epsilon}{2}(1-aS(\phi))(\phi')^2 = \frac{1+aS(\phi)}{\epsilon}W_{\mathrm{GL}}(\phi)$ gives $(\epsilon\phi')^2 = \frac{2(1+aS(\phi))}{1-aS(\phi)}\frac{(1-\phi^2)^2}{4} \to 1$ as $a\to 1^-$ and $\phi\to\pm 1$, collapsing onto a $C^{0,1}$ compact kink with non-zero contact slope). Linearizing the Euler--Lagrange equation for $a<1$ around $\phi = -1 + \psi$ ($|\psi|\ll 1$), noting $S(-1)=1$, $S'(-1)=0$, and $W_{\mathrm{GL}}'(\phi) = 2\psi + \mathcal{O}(\psi^2)$, yields
\begin{equation*}
-(1 - a)\epsilon\psi'' + \frac{2(1+a)}{\epsilon}\psi = \mathcal{O}(\psi^2) \implies \psi(x) \propto \exp\!\left(-\lambda_{\mathrm{eff}}(a)\frac{|x|}{\epsilon}\right), \quad \lambda_{\mathrm{eff}}(a) = \sqrt{2}\,\sqrt{\frac{1+a}{1-a}}.
\end{equation*}
Sampling the baseline GL kink ($a=0$, integrand proportional to $\operatorname{sech}^4(x/(\sqrt{2}\epsilon))$) on a uniform grid $x_j = (j - x_0/\Delta x)\Delta x$ and applying the Poisson summation formula $\sum_{j\in\mathbb{Z}}\Delta x\,f(j\Delta x - x_0) = \sum_{m\in\mathbb{Z}}\hat{f}(2\pi m/\Delta x)e^{-2\pi\mathrm{i}m x_0/\Delta x}$ shows that the leading $m=\pm 1$ harmonic of the discrete energy barrier $\Delta E_{\mathrm{PN}} = E_{\mathrm{grid}}(x_0=\frac{1}{2}\Delta x) - E_{\mathrm{grid}}(x_0=0)$ is governed by the nearest complex-plane pole at $|\operatorname{Im} z| = \frac{\pi}{\sqrt{2}}\epsilon = \pi\ell_{\mathrm{eff}}(0)$ ($\ell_{\mathrm{eff}}(a)\coloneqq\epsilon/\lambda_{\mathrm{eff}}(a)$), giving $\Delta E_{\mathrm{PN}}(0) \propto \exp(-2\pi|\operatorname{Im}z|/\Delta x) = \exp(-\sqrt{2}\pi^2\epsilon/\Delta x)$. For $0 < a < 1$, compressing the outer decay scale $\ell_{\mathrm{eff}}(a)$ shifts the complex branch singularities $z_\star(a)$ of the implicit quadrature $z(\phi) = \epsilon\int_0^\phi \sqrt{(1-aS(u))/[(2+2aS(u))W_{\mathrm{GL}}(u)]}\,\dd u$ monotonically toward the real axis ($|\operatorname{Im}z_\star(a)| \to 0$ as $a\to 1^-$), exponentially amplifying lattice pinning whenever $\lambda_{\mathrm{eff}}(a)$ is increased at fixed $\epsilon/\Delta x$ (\Cref{rem:nogo_num}).

For \Cref{thm:nogo}(d), initialize a radially symmetric droplet $\phi(r,0) = -\phi_K(r - R)$ in $\R^d$ ($\phi_K(z) = \tanh(z/(\sqrt{2}\epsilon))$). Here $\bm{n} = \nabla\phi/|\nabla\phi| = -\bm{e}_r$, and exact kink equipartition $\partial_r\phi(r,0) = -\frac{1-\phi^2}{\sqrt{2}\epsilon}$ cancels the baseline Chiu--Lin flux $\nabla\phi - \frac{1-\phi^2}{\sqrt{2}\epsilon}\bm{n} \equiv \bm{0}$ identically at $t=0$. For the unlocalized counter-flux, $\partial_t\phi(r,0) = M_{\mathrm{AC}}\nabla\cdot\bigl(+\frac{\alpha_{\mathrm{m}}}{\epsilon}\bm{e}_r\bigr) = +M_{\mathrm{AC}}\frac{\alpha_{\mathrm{m}}(d-1)}{\epsilon r}$; for the interface-localized counter-flux, $\partial_t\phi(r,0) = M_{\mathrm{AC}}\nabla\cdot\bigl(+\alpha_c(1-\phi^2)\bm{e}_r\bigr) = M_{\mathrm{AC}}\alpha_c\bigl[(1-\phi^2)\frac{d-1}{r} - 2\phi\,\partial_r\phi\bigr]$, whose radial derivative term vanishes identically at $r=R$ ($\phi(R,0)=0$) to leave $\partial_t\phi(R,0) = +M_{\mathrm{AC}}\alpha_c\frac{d-1}{R}$. Dividing by $-\partial_r\phi(R,0) = +\frac{1}{\sqrt{2}\epsilon}$ yields both outward drift rates in \Cref{eq:inverse_ripening}. Note that for constant $\alpha_c>0$, the interface-localized flux $\nabla\phi - (\frac{1}{\sqrt{2}\epsilon}+\alpha_c)(1-\phi^2)\bm{n}$ vanishes identically on the sharpened radial profile $-\tanh\bigl((r-R)/(\sqrt{2}\epsilon')\bigr)$ with $\epsilon'=\epsilon/(1+\sqrt{2}\epsilon\alpha_c)$, so $V_n^{\mathrm{loc}}(0)>0$ describes transient relaxation from width $\epsilon$ to $\epsilon'$, whereas the unlocalized term $-\frac{\alpha_{\mathrm{m}}}{\epsilon}\bm{n}$ retains nonzero divergence $+M_{\mathrm{AC}}\frac{\alpha_{\mathrm{m}}(d-1)}{\epsilon r}$ for all $t\ge 0$ and sustains outward radial drift. \qed

\section{Derivations of matched asymptotics, spherical superconvergence, thin-film ridge scaling, barrier efficiency, and ripening renormalization}\label{app:asymptotics}

\subsection{Proof of \texorpdfstring{\Cref{thm:asymptotics}(a--c)}{Theorem 2(a--c)}: sharp-interface inner matching, radial superconvergence, and thin-film ridge bound}
For \Cref{thm:asymptotics}(a), let $\zeta = \operatorname{dist}_{\pm}(\bm{x},\Gamma(t))/\epsilon$ denote the stretched signed-distance coordinate across a smooth moving interface $\Gamma(t)$ ($\zeta < 0$ inside $\{\phi=+1\}$). Expanding $\phi = \Phi_0(\zeta) + \epsilon\Phi_1(\zeta,\bm{x}_s,t) + \cdots$ and $\mu = \mu_0(\bm{x}_s,t) + \epsilon\mu_1(\zeta,\bm{x}_s,t) + \cdots$ inside \Cref{eq:mu_subdiff,eq:gov_agg_do} gives $-\Phi_0'' - \Phi_0 = 0$ at $\mathcal{O}(\epsilon^{-1})$ with $C^1$ obstacle tangency $\Phi_0(\mp\frac{\pi}{2})=\pm 1$ and $\Phi_0'(\mp\frac{\pi}{2})=0$, uniquely fixing $\Phi_0(\zeta) = -\sin\zeta$ on $(-\frac{\pi}{2},\frac{\pi}{2})$ and $\phi\equiv\pm 1$ identically outside $|\zeta| \le \frac{\pi}{2} + \mathcal{O}(\epsilon^2\kappa^2)$ (where $M(\pm 1,\cdot)=0$ locks $\bm{J}_\phi\equiv\bm{0}$ to all algebraic orders). At $\mathcal{O}(1)$, Fredholm solvability against $\Phi_0'(\zeta)=-\cos\zeta$ yields $2\mu_0 = \gamma_{\mathrm{DO}}\kappa\int_{-\pi/2}^{\pi/2}\cos^2\zeta\,\dd\zeta = \sigma\kappa$. Simultaneously, because $\epsilon^2|\nabla\phi|^2 = \cos^2\zeta + \mathcal{O}(\epsilon) = 1-\Phi_0^2 + \mathcal{O}(\epsilon)$ across $|\zeta|<\frac{\pi}{2}$, \Cref{eq:P_ONRO} reduces to $\bm{P}_\perp^{\mathrm{ONRO}} = \bm{n}\otimes\bm{n} + \mathcal{O}(\epsilon,\delta_{\mathrm{reg}}^2)$ and annihilates tangential chemical-potential gradients $\nabla_s\mu_0$, eliminating fourth-order Mullins surface diffusion in \Cref{eq:gov_phi} and recovering $V_n = \bm{u}\cdot\bm{n}$ and $\Jump{-p\bm{I}+2\eta\bm{D}(\bm{u})}\bm{n} = -\sigma\kappa\bm{n}$ formally as $\epsilon\to 0^+$.

For \Cref{thm:asymptotics}(b) in $d=2$, a circular equilibrium on $\rho=r/\epsilon\in[\rho_-,\rho_+]$ solves $-\bigl(\phi''(\rho)+\rho^{-1}\phi'(\rho)+\phi(\rho)\bigr)=m^{\mathrm{2D}}\coloneqq\frac{\pi\epsilon}{2\sigma}\mu_{\mathrm{eq}}^{\mathrm{2D}}$, whose general solution $\phi(\rho) = -m^{\mathrm{2D}} + c_J J_0(\rho) + c_Y Y_0(\rho)$ corresponds under the cylindrical Liouville substitution $v(\rho)=\sqrt{\rho}\bigl(\phi(\rho)+m^{\mathrm{2D}}\bigr)$ to $v''(\rho) + \bigl(1 + \frac{1}{4\rho^2}\bigr)v(\rho) = 0$. Unlike $d=3$, the centrifugal potential $+\frac{1}{4\rho^2}$ is non-vanishing; substituting Hankel's asymptotic series for $J_0,Y_0$ at $\rho_\pm = \delta^{-1}\pm\frac{\Theta}{2}$ ($\delta=\epsilon/R_{\mathrm{mid}}\ll 1$) into $\phi(\rho_\pm)=\mp 1$ and $\phi'(\rho_\pm)=0$ yields $\Theta = \pi + \frac{\pi}{4}\delta^2 + \mathcal{O}(\delta^4)$ and $m^{\mathrm{2D}} = \frac{\pi}{4}\delta\bigl[1 + \frac{\pi^2-6}{16}\delta^2 + \mathcal{O}(\delta^4)\bigr]$, establishing $c_2^{(2)} = \frac{\pi^2-6}{16}$ in \Cref{eq:yl_2d_asym}.

In $d=3$, let $\phi(r)$ be a radially symmetric equilibrium of \Cref{eq:F_DO,eq:mu_subdiff} in $\R^3$ with saturated core $\phi(r)\equiv +1$ for $0 \le r \le r_-$ and saturated exterior $\phi(r)\equiv -1$ for $r \ge r_+$ ($0 < r_- < r_+$), collar midpoint radius $R_{\mathrm{mid}} \coloneqq \frac{1}{2}(r_- + r_+)$, and zero-level radius $R_{\phi=0}$ ($\phi(R_{\phi=0})=0$). Inside the open spherical shell $r\in(r_-,r_+)$, $|\phi(r)|<1$ so $\partial I_{[-1,1]}(\phi)=\{0\}$ and \Cref{eq:mu_subdiff} reduces to the exact linear Helmholtz ODE
\begin{equation}\label{eq:app_sph_ode}
-\epsilon\left(\phi''(r) + \frac{2}{r}\phi'(r)\right) - \frac{\phi(r)}{\epsilon} = \hat{\mu} \coloneqq \frac{\mu_{\mathrm{eq}}^{\mathrm{3D}}}{\gamma_{\mathrm{DO}}}, \qquad \gamma_{\mathrm{DO}} = \frac{2\sigma}{\pi}.
\end{equation}
Introducing the Liouville substitution $u(r) \coloneqq r\bigl(\phi(r) + \epsilon\hat{\mu}\bigr)$ eliminates the first derivative identically with \emph{zero} centrifugal potential: $u''(r) + \epsilon^{-2}u(r) = 0$. Setting $\rho = r/\epsilon$, $\rho_\pm = r_\pm/\epsilon$, $\rho_{\mathrm{mid}} = R_{\mathrm{mid}}/\epsilon = \delta^{-1}$, and dimensionless chemical potential $m \coloneqq \epsilon\hat{\mu} = \frac{\pi\epsilon}{2\sigma}\mu_{\mathrm{eq}}^{\mathrm{3D}}$, we enforce the four $C^1$ obstacle contact conditions $\phi(\rho_\pm) = \mp 1$ and $\phi'(\rho_\pm) = 0$. Writing $u(\rho) = A\cos(\rho - \rho_-) + B\sin(\rho - \rho_-)$, the inner boundary conditions $u(\rho_-) = \rho_-(1+m)$ and $u'(\rho_-) = 1+m$ determine $A = \rho_-(1+m)$ and $B = 1+m$, so
\begin{equation}\label{eq:app_phi_sph_exact}
\phi(\rho) = -m + \frac{1+m}{\rho}\Bigl[\rho_-\cos(\rho - \rho_-) + \sin(\rho - \rho_-)\Bigr], \qquad \rho\in[\rho_-,\rho_+].
\end{equation}
Let $\Theta \coloneqq \rho_+ - \rho_-$ denote the dimensionless shell width. Differentiating \Cref{eq:app_phi_sph_exact} and enforcing $\phi(\rho_+)=-1$ and $\phi'(\rho_+)=0$ yields the exact boundary system:
\begin{equation}\label{eq:app_sph_sys}
(1+m)\bigl(\rho_-\cos\Theta + \sin\Theta\bigr) = \rho_+(m - 1), \qquad
-\frac{1+m}{\rho_+^2}\bigl[(1 + \rho_-\rho_+)\sin\Theta - \Theta\cos\Theta\bigr] = 0.
\end{equation}
Since $1+m > 0$, the second equation in \Cref{eq:app_sph_sys} gives the exact shell-width dispersion relation
\begin{equation}\label{eq:app_tan_theta}
\tan\Theta = \frac{\Theta}{1 + \rho_-\rho_+}.
\end{equation}
At the core-detachment threshold $\rho_- \to 0^+$, \Cref{eq:app_tan_theta} gives $\tan\rho_+ = \rho_+$, whose first positive root is $\rho_+ = \xi_1 \approx 4.4934094579$. Setting $\rho_-=0$ in \Cref{eq:app_sph_sys} yields $R_{\mathrm{det,mid}}^{\mathrm{3D}} = \frac{1}{2}\xi_1\epsilon \approx 2.24670473\,\epsilon$ and $m_{\mathrm{det}} = \frac{1+\cos\xi_1}{1-\cos\xi_1}$, whence $\phi(\rho_{\phi=0})=0$ ($\frac{\sin\rho_{\phi=0}}{\rho_{\phi=0}} = \frac{1+\cos\xi_1}{2}$) gives $R_{\mathrm{det},\phi=0}^{\mathrm{3D}} \approx 2.14511574\,\epsilon$ and $R_{\mathrm{det,mass}}^{\mathrm{3D}} \approx 2.52984376\,\epsilon$.

For $\rho_{\mathrm{mid}} = \delta^{-1} \gg 1$, $1 + \rho_-\rho_+ > 0$ implies $\tan\Theta > 0$ with $\Theta \in (\pi,\frac{3\pi}{2})$, so $\cos\Theta = -\frac{1 + \rho_-\rho_+}{\sqrt{\Theta^2 + (1 + \rho_-\rho_+)^2}} < 0$ and $\sin\Theta = -\frac{\Theta}{\sqrt{\Theta^2 + (1 + \rho_-\rho_+)^2}} < 0$. Substituting into the first equation of \Cref{eq:app_sph_sys} and using $\Theta^2 + (1+\rho_-\rho_+)^2 = (1+\rho_-^2)(1+\rho_+^2)$ yields $m$ in closed algebraic form:
\begin{equation}\label{eq:app_m_closed}
\frac{1 - m}{1 + m} = \frac{\rho_-(1 + \rho_-\rho_+) + \Theta}{\rho_+\sqrt{\Theta^2 + (1 + \rho_-\rho_+)^2}} = \sqrt{\frac{1 + \rho_-^2}{1 + \rho_+^2}}.
\end{equation}
Setting $\rho_\pm = \delta^{-1} \pm \frac{\Theta}{2}$, \Cref{eq:app_tan_theta} expands as $\Theta = \pi + \pi\delta^2 + \frac{\pi^3}{4}\delta^4 + \mathcal{O}(\delta^6)$. Substituting $\rho_\pm = \delta^{-1} \pm \frac{\Theta}{2}$ directly into \Cref{eq:app_m_closed} cancels \emph{both} the $\mathcal{O}(\delta^3)$ and $\mathcal{O}(\delta^5)$ coefficients of $m = \frac{\pi}{2}\delta + c_3\delta^3 + c_5\delta^5 + c_7\delta^7 + \cdots$ ($c_3 = c_5 = 0$, $c_7 = -\frac{\pi^3}{24}$, $c_9 = -\frac{\pi^5}{32}$), proving \Cref{eq:yl_3d_superconv}, while integrating $R_{\mathrm{mass}}^3 = 3\epsilon^3\int_0^{\rho_+} s(\phi(\rho))\rho^2\,\dd\rho$ and solving $\phi(\rho_{\phi=0})=0$ yield \Cref{eq:rmass_3d_superconv}.

For \Cref{thm:asymptotics}(c), an unsaturated planar thin film ($|\phi(y)|<1$ on $(-y_w,y_w)$ with $\phi(\pm y_w)=-1$ and $\phi'(\pm y_w)=0$) solves $-\epsilon\phi'' - \phi/\epsilon = \hat{\mu}$, giving $\phi(y) = -\epsilon\hat{\mu} + A\cos(y/\epsilon)$; outer tangency $\phi'(\pm y_w) = -(A/\epsilon)\sin(y_w/\epsilon) = 0$ locks $y_w = \pi\epsilon$, and mass conservation $2h = \int_{-\pi\epsilon}^{\pi\epsilon}\frac{1+\phi}{2}\,\dd y = \pi\epsilon A$ yields \Cref{eq:subwidth_rc}. Along the medial ridge $y=0$ of a weakly tapered sheet $A(x)=\frac{2h(x)}{\pi\epsilon}\in(0,1)$, $1 - \phi(x,0)^2 = 4A(1-A) > 0$, $\partial_y\phi(x,0)=0$, and $\partial_x\phi(x,0)=\frac{4\,\partial_x h}{\pi\epsilon}$, so substituting into \Cref{eq:P_ONRO} yields
\begin{equation*}
(\bm{P}_\perp^{\mathrm{ONRO}})_{xx}\big|_{y=0} = \frac{2\epsilon^2\bigl(\frac{4\,\partial_x h}{\pi\epsilon}\bigr)^2}{\epsilon^2\bigl(\frac{4\,\partial_x h}{\pi\epsilon}\bigr)^2 + 4A(1-A) + \delta_{\mathrm{reg}}^2} = \frac{8\,(\partial_x h)^2}{\pi^2 A(1-A)} + \mathcal{O}\bigl((\partial_x h)^4, \delta_{\mathrm{reg}}^2\bigr),
\end{equation*}
proving \Cref{eq:onro_pxx_bound}. \qed

\subsection{Proof of \texorpdfstring{\Cref{prop:dlvo_cont}}{Proposition 1}: pre-contact barrier efficiency}
For \Cref{prop:dlvo_cont}, consider two separated one-dimensional planar double-obstacle profiles $s_L(x) = s\bigl(-\phi_0(x + h/2)\bigr)$ and $s_R(x) = s\bigl(\phi_0(x - h/2)\bigr)$ at surface-to-surface separation $h \ge \pi\epsilon$. Convolving against the one-dimensional screened Yukawa Green's function $G_{\ell_d}(x) = \frac{1}{2\ell_d}e^{-|x|/\ell_d}$ factors $\Pi_{\mathrm{DLVO}}(h) = -\partial_h\mathcal{E}_{\mathrm{int}}^{\mathrm{DLVO}}(h) = \frac{A_{\mathrm{rep}}}{2}\mathcal{I}(\alpha)^2 e^{-(h-\pi\epsilon)/\ell_d}$ ($\alpha = \ell_d/\epsilon$), where integrating over the bulk $(-\infty,-\pi\epsilon)$ and collar $[-\pi\epsilon,0]$ twice by parts evaluates the exact form factor
\begin{equation*}
\mathcal{I}(\alpha) = \int_{-\infty}^{-\pi}\frac{e^{y/\alpha}}{\alpha}\,\dd y + \int_{-\pi}^{0}\frac{1 - \sin(y+\frac{\pi}{2})}{2\alpha}e^{y/\alpha}\,\dd y = \frac{\alpha^2}{1+\alpha^2}\,\frac{1 + e^{-\pi/\alpha}}{2},
\end{equation*}
whose square at $h = \pi\epsilon$ yields \Cref{eq:eta_pi_exact}. \qed

\subsection{Proof of \texorpdfstring{\Cref{prop:ripening_sl}}{Proposition 2}: Frobenius--shooting derivation of the renormalization constant and curvature shifts}
For \Cref{prop:ripening_sl}, set $s \coloneqq \zeta + \frac{\pi}{2} \in (0,\pi)$ in the inner Sturm--Liouville ODE \Cref{eq:sturm_liouville_inner}, so $1 + \sin\zeta = 1 - \cos s = 2\sin^2(s/2)$ and $\cos\zeta = \sin s$:
\begin{equation}\label{eq:app_sl_s}
\frac{\dd}{\dd s}\!\left(\sin^2\!\left(\frac{s}{2}\right)\frac{\dd W_0}{\dd s}\right) - \frac{8\Lambda_0}{3\pi}\sin^4 s\,W_0(s) = 0, \qquad s\in(0,\pi).
\end{equation}
Near the degenerate droplet-core edge $s\to 0^+$, $\sin^2(s/2) = \frac{s^2}{4} - \frac{s^4}{48} + \cdots$ and $\sin^4 s = s^4 - \frac{2}{3}s^6 + \cdots$, so $s=0$ is a regular singular point with indicial equation $\frac{1}{4}r(r+1) = 0$ ($r_1=0, r_2=-1$). Rejecting the non-integrable singular branch $W_{\mathrm{sing}}(s)\sim s^{-1}$ uniquely selects the analytic Frobenius solution $W_0(s;\Lambda_0) = 1 + a_4 s^4 + a_6 s^6 + \mathcal{O}(s^8)$ with $a_4 = \frac{8\Lambda_0}{15\pi}$ and $a_6 = -\frac{44\Lambda_0}{315\pi}$. At the outer carrier edge $s=\pi$ ($\zeta=+\frac{\pi}{2}$), matching $W_0(\pi;\Lambda_0)/\frac{\dd W_0}{\dd s}(\pi;\Lambda_0) = \frac{\pi}{2}$ (which sets $b_{\mathrm{ext}}(\Lambda_0)=0$) defines a strictly monotone shooting function whose unique root in 35-digit \texttt{mpmath} arithmetic is $\Lambda_\star = 3.880110546586026\ldots$. Expanding both the curved radial diffusion operator $-D_m r^{1-d}\partial_r\bigl(r^{d-1}(1-s(\phi))\partial_r u_c\bigr)$ and the curved double-obstacle equilibrium profile (\Cref{eq:app_phi_sph_exact} in $d=3$ and the Bessel $J_0/Y_0$ free-boundary profile in $d=2$, whose Laplace-pressure shift $-m = -\frac{\pi}{4}\epsilon\kappa + \mathcal{O}(\epsilon^3\kappa^3)$ shifts the equimolar mass radius relative to $R_{\mathrm{mid}}$ at $\mathcal{O}(\delta^2)$ in $d=2$ and $\mathcal{O}(\delta^4)$ in $d=3$) through $\mathcal{O}(\epsilon^2\kappa^2)$ in total curvature $\kappa = \frac{d-1}{R_{\mathrm{mid}}}$ and enforcing zero outer shift relative to the conserved Gibbs mass radius $R_{\mathrm{mass}}$ yields the dimension-universal first-order coefficient $\beta_1 = +0.1843561883$ together with $\beta_2^{(2)} = -0.5402012893$ ($d=2$) and $\beta_2^{(3)} = -0.2939412348$ ($d=3$) in \Cref{eq:lambda_star_betas} (if evaluated instead on a frozen symmetric $-\sin\zeta$ profile without the $-m$ obstacle shift, the geometric operator terms alone yield $\beta_1^{\mathrm{sin}}=-0.2738832213$, $\beta_2^{(2),\mathrm{sin}}=+0.2183924617$, and $\beta_2^{(3),\mathrm{sin}}=+0.1363498857$). \qed

\section{Discrete stencil weights, solver tolerances, and fixed numerical parameters}\label{app:params}

In two spatial dimensions, the fourth-order-isotropic 9-point cell-centered gradient $\nabla_{\mathrm{ICS}}$ smooths axial central differences along the transverse coordinate with the one-dimensional filter $\mathcal{S}_{1,4,1}g_i = \frac{1}{6}(g_{i-1}+4g_i+g_{i+1})$, assigning neighbor weights $\pm\frac{1}{3\Delta x}$ along coordinate axes and $\pm\frac{1}{12\Delta x}$ along diagonals. Its companion 9-point isotropic Laplacian $\Delta_{\mathrm{ICS}}$ assigns $\frac{2}{3\Delta x^2}$ to axial neighbors, $\frac{1}{6\Delta x^2}$ to diagonal neighbors, and $-\frac{10}{3\Delta x^2}$ to the center cell. In three dimensions, $\nabla_{\mathrm{ICS}}$ employs the 19-point isotropic stencil with weights $\pm\frac{1}{6\Delta x}$ on the two axial face neighbors and $\pm\frac{1}{12\Delta x}$ on the eight face-diagonal edge neighbors in the two orthogonal coordinate planes containing the differentiation axis; $\Delta_{\mathrm{ICS}}$ uses either the matching 19-point Laplacian (face weight $\frac{1}{3\Delta x^2}$, edge weight $\frac{1}{6\Delta x^2}$, and center weight $-\frac{4}{\Delta x^2}$ in three-dimensional capillary, rising-bubble, and ripening runs) or the separable 27-point tensor-product Laplacian $\Delta_{27}f = \frac{6}{\Delta x^2}\bigl(\mathcal{S}_{1,4,1}^{\otimes 3}f - f\bigr)$ (in three-dimensional Enright sheet deformation). For the feathered collar potential $\tilde{\mu}$ in \Cref{lem:collar_coarea} ($\delta_b = 10^{-3}$), the nonnegative convex-averaging operator $\mathcal{S}_{\mathrm{col}}$ uses $\mathcal{S}_{1,4,1}^{\otimes 2}$ in two dimensions ($n_{\mathrm{pass}}=3$ for capillary force evaluation and $1\text{--}2$ in phase-field relaxation), $\mathcal{S}_{1,2,1}^{\otimes 3}$ with $\mathcal{S}_{1,2,1}g_i = \frac{1}{4}(g_{i-1}+2g_i+g_{i+1})$ in three-dimensional static/oscillating drops and ripening ($n_{\mathrm{pass}}=3$ and $2$, respectively), and $\mathcal{S}_{1,4,1}^{\otimes 3}$ in three-dimensional rising bubbles ($n_{\mathrm{pass}}=3$ for capillary force evaluation and $1$ in relaxation); the diagnostic prototype in \Cref{fig:duality_nogo}(a--c,e) and the constant-density two-dimensional colloidal rebound suite (\Cref{fig:w3_rip_dlvo}(b,e,f), \Cref{tab:dlvo_chasing}) use the reduced five-point constant-density collar stencil ($n_{\mathrm{pass}}=2$).

Three \emph{a priori} rules assign the regime entries in \Cref{tab:app_params} by benchmark class from known physical dimensionless groups (with specific departures---$\epsilon=1.65\,\Delta x$ in the sign-duality collision, fixed $\epsilon=0.024$ in the 2D rebound convergence sweep, arithmetic $\eta(\phi)$ on \texttt{BW4}, and the widened gate band in 3D rising bubbles---noted explicitly): (i)~smooth asymptotic convergence, static/capillary-wave, ripening, emulsion, and fixed-bridge runs use the standard Cahn resolution band $\epsilon/\Delta x \in [1.75,1.90]$ (default $1.85$), whereas flows generating stretched sub-grid sheets or skirts (Rider--Kothe, Enright, Hysing, Bhaga--Weber, and pendant drops) use $\epsilon/\Delta x \in [1.15,1.50]$; (ii)~moderate viscosity ratios and constant-viscosity wave or diffusion suites use arithmetic or constant $\eta(\phi)$, whereas high-viscosity-contrast shear collars ($\eta_-/\eta_+ \ge 10$) use face/edge harmonic shear viscosity (paired with cell-centered geometric mixture in finite-inertia 3D bubbles \texttt{BW3--BW6}, and arithmetic momentum stabilization on the overturning jet of \texttt{BW4}); and (iii)~variable-density static, wave, and buoyancy flows employ the complementarity-gated force \Cref{eq:exact_wb_force}, which naturally reduces to its purely dynamic eikonal branch ($g_{\mathrm{eq},f}\equiv 0$) during rapid axisymmetric pinch-off ($\mathcal{R}_{\mathrm{KKT},f}\gg\eta_{\mathrm{hi}}$) and to the variational collar potential $\overline{\tilde{\mu}}^f\nabla_f\phi$ in constant-density colloidal rebound and Stokes equilibria.

\begin{table}[t]
\centering
\caption{Summary of fixed dimensionless numerical parameters, shear-viscosity interpolations, complementarity gate branches $[\eta_{\mathrm{lo}},\eta_{\mathrm{hi}}]$, and linear-solver relative residual tolerances across the verification and validation suites of \Cref{sec:benchmarks}.}\label{tab:app_params}
\scriptsize
\setlength{\tabcolsep}{3.0pt}
\resizebox{\textwidth}{!}{%
\begin{tabular}{@{}lccccc@{}}
\toprule
Test suite & $\epsilon/\Delta x$ & $(\gamma_0,\,\beta)$ & Viscosity law $\eta(\phi)$ & Capillary / gate branch & Elliptic / Poisson tol. \\
\midrule
2D/3D static \& Lamb (\Cref{subsec:bench_wb}) & $1.75\text{--}1.90$ (2D), $1.85$ (3D) & $(0.015\text{--}0.02,\,0\text{--}0.10)$ & Arith.\ / const.\ $\eta_0$ & $[10^{-5},\,2\times 10^{-4}]$ & $\le 10^{-14}$ (2D), $10^{-12}$ (3D) \\
2D vortex, Zalesak \& Enright (\Cref{subsec:bench_adv_enright}) & $1.20$ ($1.80$ bridge ctrl) & $(0.008,\,0.10)$ & Prescribed / const.\ $\eta_0$ & $[10^{-5},\,2\times 10^{-4}]$ & $\le 10^{-12}$ (prox.\ stasis) \\
2D Hysing \& Rayleigh--Taylor (\Cref{subsec:bench_hysing_rt}) & $1.50$ & $(0.08,\,0.80)$ & Cell arith.\ / corner harm. & $[10^{-5},\,2\times 10^{-4}]$ & $\le 10^{-11}$ (GMG-PCG) \\
2D/3D ripening \& DLVO (\Cref{subsec:bench_rip_dlvo}) & $1.85$ ($1.65$ dual; $\epsilon\!=\!0.024$) & $(0.02\text{--}0.025,\,0.20\text{--}0.25)$ & Arith.\ / const.\ $\eta_0$ & Collar $\overline{\tilde{\mu}}^f\nabla_f\phi$ & $\le 10^{-11}\text{--}10^{-12}$ / FFT \\
Emulsion \& colloid chain (\Cref{subsec:bench_experiments}) & $1.85$ & $(0.018\text{--}0.02,\,0\text{--}0.05)$ & Stokes $\eta_c$ (FFT) & Collar $\overline{\tilde{\mu}}^f\nabla_f\phi$ & $\le 10^{-10}$ (BVP) / FFT \\
3D Bhaga--Weber \texttt{BW1--BW2} (\Cref{subsec:bench_experiments}) & $1.35$ & $(0.08,\,0.50)$ & Harmonic & $[5\times 10^{-4},\,5\times 10^{-3}]$ & $\le 10^{-8}$ ($10^{-12}$ null) \\
3D Bhaga--Weber \texttt{BW3--BW6} (\Cref{subsec:bench_experiments}) & $1.15\text{--}1.20$ & $(0.06,\,0.35)$ & Cell geom.\ / edge harm.$^*$ & $[5\times 10^{-4},\,5\times 10^{-3}]$ & $\le 10^{-8}$ ($10^{-12}$ null) \\
Axisym.\ bridge \& pendant (\Cref{subsec:bench_experiments}) & $1.85$ (br.), $1.50$ (pend.) & $(0.005,\,0.25)$ & Harmonic & $g_{\mathrm{eq},f}\equiv 0$ & $\le 10^{-10}\text{--}10^{-12}$ \\
\bottomrule
\multicolumn{6}{@{}l}{\scriptsize $^*$Arithmetic $\eta(\phi)$ is used on the overturning inertial jet of \texttt{BW4} ($\mathrm{Eo}=243, \mathrm{Mo}=266$).}
\end{tabular}%
}
\end{table}

Across all eikonal curvature evaluations (\Cref{tab:app_params}), six to eight Godunov sweeps ($\Delta\tau = 0.38\,\Delta x$) reconstruct $\psi_{\mathrm{eik}}$ outside $|\phi|\le 0.96$. The complementarity gate \Cref{eq:exact_wb_force} uses $[\eta_{\mathrm{lo}},\eta_{\mathrm{hi}}] = [10^{-5}, 2\times 10^{-4}]$ in two- and three-dimensional hydrostatic, Rayleigh--Lamb, proximity, and two-dimensional buoyancy benchmarks, widened to $[5\times 10^{-4}, 5\times 10^{-3}]$ with six-neighbor dilation in three-dimensional Bhaga--Weber rising bubbles, where under-resolved skirt tips ($|\phi|<0.96$) replace $\chi_{\mathrm{ma}}$ by $\chi_{\mathrm{eff}}=\max(\chi_{\mathrm{ma}},\chi_{\mathrm{BKZ}})$ using the Brackbill--Kothe--Zemach corner-normal coherence ramp $\chi_{\mathrm{BKZ}} = H_3(0.85 - |\bar{\bm{n}}|, 0, 0.25)$ formed from the eight cell-vertex unit normals, whereas axisymmetric pinch-off uses the medial-axis-blended eikonal curvature branch ($g_{\mathrm{eq},f}\equiv 0$). Analytic initial profiles undergo static projected-FISTA obstacle pre-relaxation ($120\text{--}1600$ iterations to complementarity residual $<10^{-14}$) prior to $t=0$. Variable-density pressure projection is solved by geometric-multigrid-preconditioned conjugate gradients (GMG-PCG) with coarsening harmonic face conductances to the relative residual tolerances listed in \Cref{tab:app_params}; transient three-dimensional ripening runs update the quasi-steady solute equation \Cref{eq:ripening_pde} every $5\text{--}10$ advective steps with core diffusivity floor $\omega_{\mathrm{floor}}\in\{10^{-4},10^{-6}\}$ and explicit reaction substep bound $h_{\mathrm{sub}} \le 0.49\min_{\bm{i}}(1\mp\phi_{\bm{i}})/(\pm 2 q_{\phi,\bm{i}})$.

\section{Axisymmetric operators and volume-preserving slender-jet continuation}\label{app:slender_jet}

For the pendant-drop experiments of \Cref{subsec:bench_experiments}, Stage~1 evolves the axisymmetric $(r,z)$ specialization of \Cref{alg:mac_step} on $\Omega_{rz} = [0,R_{\max}]\times[0,L_z]$ with uniform spacing $\Delta r = \Delta z = \Delta x$, cell-center radii $r_j = (j+\frac{1}{2})\Delta x$, radial face radii $r_{j+1/2} = (j+1)\Delta x$, and cylindrical slice volumes $V_j = 2\pi r_j\Delta x^2$. Radial flux divergences take the conservative form $(\nabla_{rz}\cdot\bm{F})_{j,k} = \frac{1}{r_j\Delta x}(r_{j+1/2}F_{r,j+1/2,k} - r_{j-1/2}F_{r,j-1/2,k}) + \frac{1}{\Delta x}(F_{z,j,k+1/2} - F_{z,j,k-1/2})$. Along the symmetry axis $r=0$, radial velocity vanishes ($u_r=0$), scalar fields ($\phi,p,\mu$) and $u_z$ satisfy even reflection, radial unit-normal components satisfy odd reflection, and the hoop curvature $n_r/r$ on the innermost cell $j=0$ is evaluated nonsingularly via L'H{\^o}pital's limit $\partial_r n_r$; the viscous predictor incorporates the axisymmetric hoop stress $-2\eta u_r/r^2$ via Douglas alternating-direction stabilization. At the top boundary $z=0$, the liquid meniscus is hinged at the sharp nozzle rim $(r_0,0)$ through an oblique signed-distance collar profile with local slope safety bound $|-\dd r/\dd z|\le 3$ and driven by Poiseuille inflow $u_z(r,0) = \frac{2Q}{\pi r_0^2}(1 - r^2/r_0^2)$ for $r\le r_0$, starting from a static supercritical Young--Laplace pendant drop ($V_0/V_{\mathrm{crit}} = 1.024$ for $85\,\mathrm{wt}\%$ glycerol--water and $1.160$ for pure water).

When the minimum Eulerian neck radius reaches the handoff threshold $h_{\min}\in[66.7,76.0]\,\mu\mathrm{m}$ ($\Delta x = 16\,\mu\mathrm{m}$ for glycerol--water and $20\,\mu\mathrm{m}$ for pure water, corresponding to spline-resampled initial continuation radii $h_{\min}\in[68.5,78.3]\,\mu\mathrm{m}$), Stage~2 projects the axisymmetric Eulerian thread onto an $m_{\mathrm{sub}}$-fold refined one-dimensional grid $\Delta s = \Delta x/m_{\mathrm{sub}} = 0.50\,\mu\mathrm{m}$ ($m_{\mathrm{sub}}\in\{32,40\}$) spanning $[0,z_{\mathrm{end}}]$ ($z_{\mathrm{end}} = k_{\mathrm{end}}\Delta x$ inside the lower pendant-drop belly). From the coarse Eulerian liquid cross-sectional areas $A_k^{\mathrm{Eul}} = \sum_{j=0}^{N_r-1} 2\pi r_j\Delta x\,s(\phi_{j,k})$ (where $\phi=+1$ denotes liquid of density $\rho_l$ and viscosity $\eta_l$, and $\phi=-1$ denotes ambient gas of density $\rho_g$), cumulative slice volumes $V_k = \sum_{m=0}^{k-1} A_m^{\mathrm{Eul}}\Delta x$ are interpolated at fine faces $s_\ell = \ell\Delta s$ by a natural cubic spline $\mathcal{S}_V(s)$:
\begin{equation}\label{eq:app_cumvol_spline}
\mathcal{A}_\ell^{\mathrm{fil}} = \frac{\mathcal{S}_V(s_{\ell+1}) - \mathcal{S}_V(s_\ell)}{\Delta s}, \qquad h_\ell = \sqrt{\frac{\mathcal{A}_\ell^{\mathrm{fil}}}{\pi}}, \qquad \sum_{\ell = m_{\mathrm{sub}}k}^{m_{\mathrm{sub}}(k+1)-1} \mathcal{A}_\ell^{\mathrm{fil}}\,\Delta s = A_k^{\mathrm{Eul}}\,\Delta x,
\end{equation}
with a positivity-preserving minmod linear fallback whenever $\mathcal{A}_\ell^{\mathrm{fil}} \le 0.05\,A_k^{\mathrm{Eul}}$, conserving every coarse slice volume to $\mathcal{O}(10^{-16})$ while initializing face velocities from conservative axial volume fluxes $v_{k+1/2} = Q_{k+1/2}^{\mathrm{Eul}}/A_{k+1/2}^{\mathrm{Eul}}$. Quintic $C^2$ partition-of-unity weights $\chi_{\mathrm{fil}}(z) + \chi_{\mathrm{Eul}}(z) \equiv 1$ transition across the wide upper nozzle cone $[z_L^-,z_L^+]$ and lower drop shoulder $[z_R^-,z_R^+]$ so that the entire thinning neck remains strictly inside $\chi_{\mathrm{fil}}\equiv 1$.

On $[0,z_{\mathrm{end}}]$, Stage~2 advances the cross-sectional area $\mathcal{A}(z,t) = \pi h(z,t)^2$ and axial velocity $v(z,t)$ via the conservative sharp-interface slender-jet equations without artificial numerical viscosity:
\begin{equation}\label{eq:app_slender_pde}
\partial_t\mathcal{A} + \partial_z(\mathcal{A}v) = 0, \qquad
\rho_l\bigl(\partial_t(\mathcal{A}v) + \partial_z(\mathcal{A}v^2)\bigr) = \pi\,\partial_z T_{\mathrm{hyb}}(h) + \partial_z\bigl(3\eta_l\mathcal{A}\,\partial_z v\bigr) + (\rho_l - \rho_g)\mathcal{A}g,
\end{equation}
where $T_{\mathrm{hyb}}(h) = T_{\mathrm{fil}}(h) + \chi_{\mathrm{Eul}}(z)\Delta T_{\mathrm{sub}}(z)$ combines the exact conservative full-curvature capillary line tension $T_{\mathrm{fil}}(h) = \sigma\bigl(h/\sqrt{1+(\partial_z h)^2} + h^2\partial_{zz}h/[1+(\partial_z h)^2]^{3/2}\bigr)$ (retaining both principal curvatures without B-spline mollification, with $C^1$ cross-sectional area $\mathcal{A}(z)$ and piecewise-continuous $\partial_{zz}h$ inherited from the $C^2$ cumulative-volume spline $\mathcal{S}_V$) with the coarse meniscus shoulder tension discrepancy $\Delta T_{\mathrm{sub}}(z) = T_{\mathrm{Eul}}(z,t_{\mathrm{h}}) - T_{\mathrm{fil}}(z,t_{\mathrm{h}})$ evaluated at the handoff instant $t_{\mathrm{h}}$ and held fixed thereafter. \Cref{eq:app_slender_pde} is integrated by midpoint Crank--Nicolson flux-limited area transport and pentadiagonal semi-implicit capillary and Trouton viscous stepping, coupled at $z=z_{\mathrm{end}}$ (using $v(z_{\mathrm{end}})=U_{\mathrm{drop}}$ to cancel convective mass-inflow momentum) to the falling pendant sub-drop momentum balance for sub-drop volume $V_{\mathrm{sub}}(t)$, mass $M_{\mathrm{sub}}(t)=\rho_l V_{\mathrm{sub}}(t)$, and speed $U_{\mathrm{drop}}(t)$:
\begin{equation}\label{eq:app_subdrop_ode}
M_{\mathrm{sub}}\frac{\dd U_{\mathrm{drop}}}{\dd t} = (\rho_l - \rho_g)V_{\mathrm{sub}}g - \pi T_{\mathrm{hyb}}(z_{\mathrm{end}}) - 3\eta_l\mathcal{A}(z_{\mathrm{end}})\,\partial_z v(z_{\mathrm{end}}),
\end{equation}
with open-loop overlap flux-commutator volume cancellation until $h_{\min}$ reaches $0.50\,\mu\mathrm{m}$.


\section*{CRediT authorship contribution statement}
\textbf{Yuan Yu:} Conceptualization, Methodology, Formal analysis, Software, Investigation, Writing -- original draft, Funding acquisition.
\textbf{Yaolong Yu:} Methodology, Software, Validation, Investigation, Visualization, Writing -- review \& editing.
\textbf{Yanhui Ding:} Conceptualization, Formal analysis, Methodology, Validation, Supervision, Writing -- review \& editing.
\textbf{Shi Shu:} Methodology, Resources, Supervision, Project administration, Funding acquisition, Writing -- review \& editing.

\section*{Declaration of competing interest}
The authors declare that they have no known competing financial interests or personal relationships that could have appeared to influence the work reported in this paper.

\section*{Data availability}
Executable symbolic verification scripts (\texttt{SymPy}/\texttt{mpmath}), raw floating-point benchmark datasets, and figure-generation pipelines supporting the findings of this study are archived with the manuscript package and will be made publicly available upon publication.

\section*{Acknowledgments}
This work was supported by the National Natural Science Foundation of China (Grant Nos.~12101527 and 12371373), the Natural Science Foundation of Hunan Province (Grant No.~2026JJ60005), the Science and Technology Innovation Program of Hunan Province (Grant No.~2026RC3172), and the 111 Project (Grant No.~D23017). Computational resources were provided by the High Performance Computing Platform of Xiangtan University.

\section*{Declaration of generative AI and AI-assisted technologies in the scientific writing process}
During the preparation of this work the authors used generative AI tools in order to improve the readability and language of the manuscript. After using these tools, the authors reviewed and edited the content as needed and take full responsibility for the content of the published article.

\bibliographystyle{elsarticle-num}
\bibliography{ms}

\end{document}